\documentclass[fleqn]{aa}

\usepackage{graphicx}
\usepackage[fleqn]{amsmath}
\usepackage{amssymb}
\usepackage{amsfonts}
\usepackage{fixltx2e}
\usepackage{natbib}
\usepackage{txfonts}

\usepackage[hyperindex]{hyperref}
\hypersetup{
breaklinks = {true},
colorlinks = {false},
linkcolor={black},
pdfpagemode = {None}, 
pdfborder = {0 0 1},
pdftitle = {Cosmic shear covariance: The log-normal approximation},
pdfsubject = {the log-normal approximation to cosmic shear covariances},
pdfauthor = {Stefan Hilbert, Jan Hartlap, and Peter Schneider},
pdfkeywords = {gravitational lensing: weak -- large-scale structure of the Universe -- cosmological parameters -- cosmology: theory -- methods: analytical -- methods: numerical}
}

\bibpunct{(}{)}{;}{a}{}{,}

\newcommand{\satellitename}[1]{\textit{#1}}
\newcommand{\softwarename}[1]{\textsc{#1}}

\newcommand{\vect}[1]{\boldsymbol{#1}}
\newcommand{\matr}[1]{\mathsf{#1}}
\newcommand{\matrb}[1]{\boldsymbol{\mathsf{#1}}}

\newcommand{\transposed}[1]{#1^\mathsf{t}}

\newcommand{\ii}{\mathrm{i}}
\newcommand{\ee}{\mathrm{e}}

\newcommand{\uvect}[1]{\vect{u}_{#1}}

\newcommand{\R}{\mathbb{R}}

\newcommand{\parder}[3][]{\frac{\partial^{#1} {#2}}{\partial {#3}^{#1}}}
\newcommand{\dparder}[3][]{\dfrac{\partial^{#1} {#2}}{\partial {#3}^{#1}}}

\newcommand{\diff}[2][]{\mathrm{d}^{#1}{#2}}
\newcommand{\idiff}[2][]{\!\!\mathrm{d}^{#1}{#2}}
\newcommand{\iidiff}[2][]{\!\!\!\!\mathrm{d}^{#1}{#2}}

\newcommand{\EV}[1]{\left\langle{#1}\right\rangle}
\newcommand{\bEV}[1]{\bigl\langle{#1}\bigr\rangle}

\newcommand{\BiggEV}[1]{\Biggl\langle{#1}\Biggr\rangle}

\newcommand{\est}[1]{\hat{#1}}
\newcommand{\ft}[1]{\tilde{#1}}

\newcommand{\DiracDelta}{\delta_\mathrm{D}}

\newcommand{\kpc}{\ensuremath{\mathrm{kpc}}}
\newcommand{\Mpc}{\ensuremath{\mathrm{Mpc}}}

\newcommand{\Msolar}{\ensuremath{\mathrm{M}_\odot}}

\newcommand{\arcsect}{\ensuremath{\mathrm{arcsec}}}
\newcommand{\arcmint}{\ensuremath{\mathrm{arcmin}}}
\newcommand{\degt}{\ensuremath{\mathrm{deg}}}

\newcommand{\vtheta}{\vect{\theta}}
\newcommand{\vvartheta}{\vect{\vartheta}}
\newcommand{\vell}{\vect{\ell}}

\newcommand{\gammat}{\gamma_{\mathsc{t}}}
\newcommand{\gammax}{\gamma_{\times}}

\newcommand{\spavxi}{\bar{\xi}}
\newcommand{\spavKerP}{\bar{G}_{+}}
\newcommand{\spavKerM}{\bar{G}_{-}}
\newcommand{\spavKerPM}{\bar{G}_{\pm}}
\newcommand{\KerP}{G_{+}}
\newcommand{\KerM}{G_{-}}
\newcommand{\KerPM}{G_{\pm}}

\newcommand{\Ngal}{N_{\text{g}}}
\newcommand{\ngal}{n_{\text{g}}}
\newcommand{\meanngal}{\bEV{n_{\text{g}}}}
\newcommand{\deltagal}{\delta_{\text{g}}}

\newcommand{\zgal}{z}
\newcommand{\wgal}{w}

\newcommand{\thetagal}{\vect{\theta}}
\newcommand{\varthetagal}{\vartheta}
\newcommand{\varphigal}{\varphi}

\newcommand{\epsilongal}{\epsilon}
\newcommand{\epsilongali}{\epsilon_{1}}
\newcommand{\epsilongalii}{\epsilon_{2}}
\newcommand{\epsilongalx}{\epsilon_{\times}}
\newcommand{\epsilongalt}{\epsilon_{\mathsc{t}}}
\newcommand{\sigmaepsilongal}{\sigma_{\epsilon}}

\newcommand{\gammagal}{\gamma}

\newcommand{\gammagalt}{\gamma_{\mathsc{t}}}
\newcommand{\gammagalx}{\gamma_{\times}}
\newcommand{\gammagalobs}{\est{\gamma}}

\newcommand{\gammagalobst}{\est{\gamma}_{\mathsc{t}}}
\newcommand{\gammagalobsx}{\est{\gamma}_{\times}}

\newcommand{\sigmapixel}{\sigma_{\text{pix}}}
\newcommand{\Apixel}{A_{\text{pix}}}
\newcommand{\Npixel}{N_{\text{pix}}}

\newcommand{\FOV}{\mathbb{F}}
\newcommand{\AFOV}{A_{\FOV}}
\newcommand{\rFOV}{\theta_{\FOV}}

\newcommand{\ABin}{A_{\Delta}}

\newcommand{\NBins}{N_{\text{bins}}}

\newcommand{\likelihood}{L}
\newcommand{\loglikelihood}{\ln L}

\newcommand{\ShearSumPM}{\est{S}_{\pm}}

\newcommand{\NormSum}{\est{N}_\mathrm{p}}
\newcommand{\MeanNormSum}{N_\mathrm{p}}
\newcommand{\PairSum}{N_\mathrm{p}}

\DeclareMathOperator{\cov}{Cov}

\allowdisplaybreaks[4]
\begin{document}

\title{Cosmic shear covariance: The log-normal approximation}
\titlerunning{The log-normal approximation to cosmic shear covariances}
\author{
Stefan Hilbert\inst{1,2}\thanks{\texttt{shilbert@astro.uni-bonn.de}} \and
Jan Hartlap\inst{1} \and
Peter Schneider\inst{1}
}
\institute{
$^1$Argelander-Institut f{\"u}r Astronomie, Universit{\"a}t Bonn, Auf dem H{\"u}gel 71, 53121 Bonn, Germany\\
$^2$Max-Planck-Institut f{\"u}r Astrophysik, Karl-Schwarzschild-Stra{\ss}e 1, 85741 Garching, Germany
}
\date{Received  / Accepted }


\abstract{
Accurate estimates of the errors on the cosmological parameters inferred from cosmic shear surveys require accurate estimates of the covariance of the cosmic shear correlation functions.
}{
We seek approximations to the cosmic shear covariance that are as easy to use as the common approximations based on normal (Gaussian) statistics, but yield more accurate covariance matrices and parameter errors. 
}{
We derive expressions for the cosmic shear covariance under the assumption that the underlying convergence field follows log-normal statistics. We also derive a simplified version of this log-normal approximation by only retaining the most important terms beyond normal statistics. We use numerical simulations of weak lensing to study how well the normal, log-normal, and simplified log-normal approximations as well as empirical corrections to the normal approximation proposed in the literature reproduce shear covariances for cosmic shear surveys. We also investigate the resulting confidence regions for cosmological parameters inferred from such surveys.
}{
We find that the normal approximation substantially underestimates the cosmic shear covariances and the inferred parameter confidence regions, in particular for surveys with small fields of view and large galaxy densities, but also for very wide surveys. In contrast, the log-normal approximation yields more realistic covariances and confidence regions, but also requires evaluating slightly more complicated expressions. However, the simplified log-normal approximation, although as simple as the normal approximation, yields confidence regions that are almost as accurate as those obtained from the log-normal approximation. The empirical corrections to the normal approximation do not yield more accurate covariances and confidence regions than the (simplified) log-normal approximation. Moreover, they fail to produce positive-semidefinite data covariance matrices in certain cases, rendering them unusable for parameter estimation.
}{
The log-normal or simplified log-normal approximation should be used in favour of the normal approximation for parameter estimation and parameter error forecasts. More generally, any approximation to the cosmic shear covariance should ensure a positive-(semi)definite data covariance matrix.
}

\maketitle

\section{Introduction}

In recent years, observations of weak lensing by large-scale structure, called cosmic shear, have become an important tool for studying the Universe. Remarkable constraints on the matter content of the Universe, its expansion history, and the amplitude of the cosmic density fluctuations have been obtained, e.g., from measurements of cosmic shear in the Canada-France-Hawaii Telescope (CFHT) Legacy Survey \citep[][]{SemboloniEtal2006, HoekstraEtal2006, FuEtal2008}, the $100\,\degt^2$ weak-lensing survey \citep[][]{BenjaminEtal2007}, and the Cosmic Evolution Survey \citep[COSMOS,][]{MasseyEtal2007_3D_WL, SchrabbackEtal2010}. Future cosmic shear surveys are expected to provide essential information on the properties of the dark matter and the dark energy, and on possible deviations of gravity from General Relativity \citep[][]{Huterer2010}.

cosmic shear surveys aim at inferring the statistical properties of the convergence and shear field from the observed image shapes of distant galaxies.
There are various statistical measures that probe the two-point statistics of the convergence and shear, e.g. the shear correlation functions \citep[][]{BlandfordEtal1991,Kaiser1992}, the shear power spectra \citep[][]{Kaiser1992}, the shear dispersion \citep[][]{MouldEtal1994}, the aperture-mass dispersion \citep[][]{SchneiderEtal1998_Map}, the ring statistics \citep[][]{SchneiderKilbinger2007}, the statistics proposed by \citet{FuKilbinger2010}, or the COSEBIs \citep[][]{SchneiderEiflerKrause2010}. In practice, however, only the shear correlation functions $\xi_+$ and $\xi_-$ are estimated directly from the data (since they can be estimated  most easily), and any other two-point statistics are computed from these estimates.
Thus, accurate knowledge of the noise properties of the shear correlation estimators is needed for reliable estimates of the errors on the cosmological parameters inferred from cosmic shear surveys.

The number of independent measurements in a cosmic shear survey is usually too small to permit an good estimate of the full covariance of the measured statistics directly from the data. Therefore, one has to resort to covariance estimates obtained by other means.
Supposedly the most accurate, but also computationally by far the most expensive approach is the direct estimation of a cosmic shear covariances from $N$-body simulations of cosmic structure formation \citep[e.g.][or this work]{SatoEtal2011}. This method requires many independent realisations of the survey field. So far, it has been successfully applied to lensing surveys covering a small fraction of the sky, e.g. the \satellitename{Chandra} Deep Field South \citep[][]{HartlapEtal2009} and the COSMOS field \citep{SchrabbackEtal2010,SemboloniEtal2011}. However, creating sufficiently many realistic mock observations to calculate covariances for future all-sky lensing surveys in various cosmologies and parameter ranges will remain a very challenging computational task in the foreseeable future.

Since reliable estimates for the covariances of cosmic shear two-point statistics from real or mock data are often out of reach, it appears desirable to have simple approximations at hand, which are easy to compute and yield sufficiently accurate covariance matrices and parameter errors. General expressions for the covariance of lensing two-point statistics, including the shear correlation functions, have been derived by \citet{SchneiderEtal2002} and \citet{JoachimiSchneiderEifler2008} under the assumption of normal (Gaussian) statistics for the shear and convergence field. This normal approximation is relatively easy to apply, and requires no additional information besides the convergence correlation itself and the basic survey parameters. Simulations show, however, that the normal approximation tends to underestimate the covariance of two-point lensing statistics \citep[e.g.][]{ScoccimarroZaldarriagaHui1999,CoorayHu2001,vanWaerbekeEtal2001,SatoEtal2009}.

There have been various attempts to improve the normal approximation by amending the resulting expressions for the cosmic shear covariance by scale-dependent correction factors that are calibrated with mock data from structure formation simulations \citep[][]{vanWaerbekeEtal2001, SemboloniEtal2007, SatoEtal2011}. A drawback of this approach is that, while any corrections to the normal approximation depend in a non-trivial way on the assumed cosmology, the correction factors can only be tuned to the particular cosmology of the simulations at hand.

Other approaches towards more accurate lensing statistics are based on halo models \citep[][]{CoorayHu2001,TakadaJain2009, PielorzEtal2010,KainulainenMarra2011}. The resulting expressions for the covariance of lensing two-point statistics require considerable more effort to compute than the corresponding expressions in the normal approximation, even if simplified by fitting formulas \citep[][]{PielorzEtal2010}. A thorough assessment of the accuracy of the halo model approach to cosmic shear covariances (e.g. by comparing its results with those from large high-resolution simulations of weak lensing) is still outstanding.

Normal random fields are arguably the most simple statistical models for the three dimensional matter distribution and the resulting weak-lensing convergence and shear fields. However, as mentioned above, they fail to accurately describe those fields' higher-order correlations such as the covariance of two-point correlations. There are few other classes of random fields that allow one to calculate their higher-order correlations analytically. These include log-normal random fields, on which we focus here.

Already \citet{Hubble1934} noted that observed galaxy densities roughly follow a log-normal distribution. \citet{ColesJones1991} suggested to describe the three-dimensional matter density distribution in our Universe by a log-normal random field. Later, numerical simulations of cosmic structure formation showed that a log-normal field provides indeed a good description of the one- and two-point statistics of the evolved matter density field, even though a point-wise exponential mapping of the initial Gaussian density field does not describe the evolved density field well \citep[][]{KofmanEtal1994,KayoTaruyaSuto2001}. The one-point probability density function (pdf) of the projected density and lensing convergence are also fitted well by a log-normal pdf \citep[][]{TaruyaEtal2002}. Even better fits to the convergence pdf, particularly in the tails of the distribution, can be obtained by generalisations of a log-normal pdf \citep[][]{DasOstriker2006,TakahashiEtal2011_arXiv,JoachimiTaylorKiessling2011_tmp}.

In this work, we derive expressions for the cosmic shear covariance under the assumption that the underlying convergence field follows log-normal statistics. Furthermore, we derive a simplified version of this log-normal approximation by only retaining the most important terms beyond normal statistics. In particular the simplified log-normal approximation meets our aim of an expression for the cosmic shear covariance that is as simple as the normal approximation.

We use numerical simulations to show that the log-normal and the simplified log-normal approximations reproduce the shear covariances for cosmic shear surveys much more accurately than the normal approximation. Furthermore, we demonstrate that, in contrast to the normal approximation, the uncertainties in the cosmological parameters inferred from the log-normal and the simplified log-normal approximation are in good agreement with the errors resulting from the covariances we estimated from the numerical simulations. We also show that the covariances and parameter errors in the log-normal approximation have accuracies comparable to or better than those derived from empirical corrections to the normal approximation proposed by \citet{SemboloniEtal2007} and \citet{SatoEtal2011}. In addition, we discuss the problem that approximations to the cosmic shear covariance, in particular those based on empirical fits, may produce covariance matrices that are not positive-semidefinite, which makes their use in parameter estimation at least questionable.

The paper is organised as follows. The theory of estimators for cosmic shear two-point statistics and their noise properties are briefly discussed in Sect.~\ref{sec:theory}. In Sect.~\ref{sec:simulations}, we describe our numerical simulations for creating mock lensing surveys and for studying the statistical properties of the convergence and shear. The results of the simulations and the comparison between the different approximations to the cosmic shear covariance are presented in Sect.~\ref{sec:results}. The main part of the paper concludes with a summary and discussion in Sect.~\ref{sec:summary}. A more detailed discussion about normal and log-normal random fields, and the covariance of estimators for the cosmic shear 2-point correlation can be found in the Appendix.

\section{Theory}
\label{sec:theory}

Here, we briefly discuss weak gravitational lensing, estimators for the cosmic shear correlation functions and their noise properties, and the estimation of cosmological parameters from cosmic shear data. 
The discussion of the statistical properties of the cosmic shear estimators is based on the real-space approach by \citet[][]{SchneiderEtal2002}, but includes log-normal convergence fields in addition to normal convergence fields.

\subsection{Gravitational lensing}
\label{sec:gravitational_lensing}

Gravitational lensing, the deflection of photons from distant sources by the gravity of intervening matter structures \citep[e.g.][]{SchneiderKochanekWambsganss_book}, causes shifts in the observed image positions relative to the sources' `true' sky positions. 
The shifts may be described by a deflection field
\begin{equation}
\label{eq:deflection_field}
 \vect{\alpha}(\vect{\theta},z) = \vect{\theta} - \vect{\beta}(\vect{\theta},z),
\end{equation}
which relates the observed image position $\vect{\theta}=(\theta_1,\theta_2)$ of a point-like source at redshift $z$ to its true angular position $\vect{\beta}=\bigl(\beta_1(\vect{\theta},z),\beta_2(\vect{\theta},z)\bigr)$.
The distortions induced by differential deflection can be quantified by the distortion matrix
\begin{equation}
\label{eq:lens_distortion}
 \left(\parder{\alpha_i(\vect{\theta},z)}{\theta_j}\right)_{i,j=1,2} =
\begin{pmatrix}
 \kappa + \gamma_1 & \gamma_2 + \omega \\
 \gamma_2 - \omega & \kappa - \gamma_1 
\end{pmatrix},
\end{equation}
which can be decomposed into a convergence $\kappa(\vect{\theta},z)$, a complex shear $\gamma(\vect{\theta},z)= \gamma_1(\vect{\theta},z) + \ii \gamma_2(\vect{\theta},z)$, and an asymmetry $\omega(\vect{\theta},z)$.

We define the effective convergence and shear for a population of sources with normalised redshift distribution $p_z(z)$ by
\begin{align}
\label{eq:df_effective_convergence}
  \kappa(\vect{\theta}) &= \int\idiff[]{z}\, p_z(z) \kappa(\vect{\theta}, z) 
  \text{ and}\\
\label{eq:df_effective_shear}
  \gamma(\vect{\theta}) &= \int\idiff[]{z}\, p_z(z) \gamma(\vect{\theta}, z).
\end{align}

The tangential component $\gammat(\vtheta,\vvartheta)$ and cross component $\gammax(\vtheta,\vvartheta)$ of the shear $\gamma(\vtheta)$ at position $\vtheta$ with respect to the direction $\vvartheta$ are defined by \citep[e.g.][]{SchneiderVanWaerbekeMellier2002}
\begin{align}
\label{eq:df_effective_shear_tangential_component}
\gammat(\vtheta,\vvartheta) &= -\Re\left(\gamma(\vtheta)\ee^{-2\ii\varphi(\vvartheta)}\right)
  \text{ and}\\
\label{eq:df_effective_shear_cross_component} 
\gammax(\vtheta,\vvartheta) &= -\Im\left(\gamma(\vtheta)\ee^{-2\ii\varphi(\vvartheta)}\right)
,
\end{align} 
where $\varphi(\vvartheta)$ denotes the polar angle of the vector $\vvartheta$.

For a statistically homogeneous and isotropic universe, the convergence correlation function $\xi_{\kappa}(\vartheta)$ can be defined by the expectation \citep[e.g.][]{BlandfordEtal1991,Kaiser1992}
\begin{subequations}
\begin{align}
  \xi_{\kappa} \bigl(\lvert \vvartheta\rvert\bigr) &= \bEV{\kappa(\vtheta) \kappa(\vtheta + \vvartheta)}
.
\end{align} 
Here, $\bEV{f}$ denotes the ensemble average of a function $f$ over all realisations of the observable universe for a fixed set of cosmological parameters. The two shear correlation functions $\xi_+(\vartheta)$ and $\xi_-(\vartheta)$ are defined by
\begin{align}
  \xi_{\pm} \bigl(\lvert\vvartheta \rvert\bigr) &=
  \EV{\gammat(\vtheta,\vvartheta)\gammat(\vtheta+\vvartheta,\vvartheta)} \pm \EV{\gammax(\vtheta,\vvartheta)\gammax(\vtheta+\vvartheta,\vvartheta)}
.
\end{align}
\end{subequations}

If one neglects higher-order lensing effects \citep[which are small, cf. ][]{HilbertEtal2009_RT,KrauseHirata2010} and contamination by observational systematics, the deflection field~\eqref{eq:deflection_field} is a gradient field, the asymmetry $\omega$ vanishes, and the convergence $\kappa$ and shear $\gamma$ are related by \citep[][]{Kaiser1995}
\begin{equation}
\label{eq:relation_kappa_gamma}
\left(\parder{}{\theta_1} + \ii \parder{}{\theta_2}\right) \kappa(\vect{\theta}, z) = \left(\parder{}{\theta_1} - \ii \parder{}{\theta_2}\right)  \gamma(\vect{\theta}, z).
\end{equation}
As a consequence \citep{CrittendenEtal2002},
\begin{align}
\label{eq:xi_p_from_xi_kappa}
  \xi_+ (\vartheta) &=  \xi_{\kappa} (\vartheta)
  \text{ and}\\
\label{eq:xi_m_from_xi_kappa}
  \xi_- (\vartheta) &= \int_0^{\infty}\idiff[]{\vartheta'}\,\vartheta' \KerM (\vartheta,\vartheta') \xi_{\kappa} (\vartheta')
,
\end{align}
where
\begin{equation}
\label{eq:kernel_xi_k_to_xi_m}
  \KerM(\vartheta,\vartheta') = \left(\frac{4}{\vartheta^2}-\frac{12\vartheta^{\prime 2}}{\vartheta^4}\right)H(\vartheta - \vartheta') + \frac{1}{\vartheta'}\DiracDelta(\vartheta - \vartheta').
\end{equation}
Here, $H$ denotes the Heaviside step function, and $\DiracDelta$ denotes the Dirac delta function.

\subsection{Estimators for the shear correlations and their covariance}
\label{sec:cosmic_shear_estimators}

A suitable estimator for the shear $\gamma(\thetagal^{(i)})$ at the sky position $\thetagal^{(i)}$ of a galaxy $i$ at redshift $\zgal^{(i)}$ from a sample with redshift distribution $p_z(z)$ is the observed image ellipticity $\gammagalobs^{(i)}$. To lowest order, the observed ellipticity is a sum,
 \begin{equation}
\gammagalobs^{(i)} = \epsilongal^{(i)} + \gamma\bigl(\thetagal^{(i)},\zgal^{(i)}\bigr),
\end{equation}
of the `true' shear  $\gamma\bigl(\thetagal^{(i)}, \zgal^{(i)}\bigr)$ and an intrinsic ellipticity $\epsilongal^{(i)}$, which vanishes on average. From a survey with $\Ngal$ galaxies providing shear estimates $\gammagalobs^{(1)},\ldots,\gammagalobs^{(N)}$ at positions  $\thetagal^{(1)},\ldots,\thetagal^{(N)}$, one can estimate the shear correlations $\xi_\pm(\vartheta)$ at separations $\vartheta$ by\footnote{see Appendix~\ref{sec:appendix_shear_correlation} for a detailed discussion of cosmic shear estimators and their noise properties, which also considers a more general estimator with non-uniform weights for the galaxies.}
\begin{equation}
\label{eq:xi_pm_estimator}
\est{\xi}_{\pm}(\vartheta) =
 \frac{\sum_{i,j=1}^{\Ngal} \,\Delta\bigl(\vartheta, \lvert \thetagal^{(j)} - \thetagal^{(i)} \rvert \bigr)
       \left( \gammagalobst^{(i,j)} \gammagalobst^{(j,i)} \pm \gammagalobsx^{(i,j)} \gammagalobsx^{(j,i)} \right)
      }{\sum_{i,j=1}^{\prime\Ngal} \,\Delta\bigl(\vartheta, \lvert \thetagal^{(j)} - \thetagal^{(i)} \rvert \bigr)}
.
\end{equation}
Here, the bin window function
\begin{equation}
\label{eq:bin_function}
  \Delta\bigl(\vartheta, \theta \bigr)
  = \begin{cases}
  1 & \text{for } \lvert \theta - \vartheta \rvert \leq \varDelta(\vartheta)/2 \text{ and}\\
  0 & \text{otherwise,}
  \end{cases}
\end{equation}
where $\varDelta(\vartheta)$ denotes the bin width, and $\gammagalobst^{(i,j)}$  and $\gammagalobsx^{(i,j)}$ denote the tangential and cross component, respectively, of the shear estimated from the shape of galaxy $i$ with respect to the line joining galaxy $i$ and $j$.

If (i) the bin width is small compared to the scales on which correlations change, (ii)  the galaxy positions are not correlated with the shear field,
(iii) the galaxy redshifts are not correlated with each other or the shear field, and (iv) the galaxies' intrinsic ellipticities are uncorrelated with each other and the shear, the estimators \eqref{eq:xi_pm_estimator} are unbiased \citep[][]{SchneiderEtal2002},
\begin{equation}
  \EV{\est{\xi}_\pm (\vartheta)} = \xi_\pm(\vartheta).
\end{equation}

The covariance of the cosmic shear estimators determines the accuracy to which one can deduce cosmological parameters from these estimators.
The covariance $c_{\pm\pm}(\vartheta_1, \vartheta_2) = \cov\bigl(\est{\xi}_\pm(\vartheta_1),\est{\xi}_\pm(\vartheta_2)\bigr)$ of the estimators \eqref{eq:xi_pm_estimator} for scales $\vartheta_1$ and $\vartheta_2$ can then be split into an ellipticity-noise part $c^{(\epsilon)}_{\pm\pm}(\vartheta_1, \vartheta_2)$, a mixed part  $c^{(\epsilon \gamma)}_{\pm\pm}(\vartheta_1, \vartheta_2)$, and a cosmic variance part $c^{(\gamma)}_{\pm\pm}(\vartheta_1, \vartheta_2)$:
\begin{equation}
\label{eq:c_pmpm_split}
c_{\pm\pm}(\vartheta_1, \vartheta_2) =  
  c^{(\epsilon)}_{\pm\pm}(\vartheta_1, \vartheta_2) +c^{(\epsilon \gamma)}_{\pm\pm}(\vartheta_1, \vartheta_2) + c^{(\gamma)}_{\pm\pm}( \vartheta_1, \vartheta_2)
 .
\end{equation}

First, we consider the covariance for a single survey field $\FOV$ with area $\AFOV$, galaxy density $\ngal = \Ngal/\AFOV$, and linear dimensions large compared to $\vartheta_1$ and $\vartheta_2$. The ellipticity-noise part $c^{(\epsilon)}_{\pm\pm}$ vanishes for $c_{+-}$, and for disjoint bins. Its contribution to the auto-variance of the bin with radius $\vartheta$ depends on the variance $\sigmaepsilongal^2 = \bEV{\lvert\epsilongal^{(i)}\rvert^{2}}$ of the intrinsic ellipticity and the expected effective number of galaxy pairs $\MeanNormSum(\vartheta)=  2 \pi \vartheta \varDelta(\vartheta) \AFOV \ngal^2$ in the bin,
\begin{equation}
\label{eq:c_epsilon}
c^{(\epsilon)}_{++}\left(\vartheta, \vartheta \right) =
c^{(\epsilon)}_{--}\left(\vartheta, \vartheta \right) =
  \frac{\sigmaepsilongal^4}{\MeanNormSum(\vartheta) }
.
\end{equation}

The mixed part $c^{(\epsilon \gamma)}_{\pm\pm}$ contributes to both the auto- and covariance of bins: 
\begin{subequations}
\label{eq:c_epsilon_gamma}
\begin{align}
c^{(\epsilon\gamma)}_{++}\left(\vartheta_1, \vartheta_2 \right)
&=
\frac{2\sigmaepsilongal^2}{\pi\ngal \AFOV} 
  \int_{0}^{\pi}\!\!\idiff[]{\varphi_1}\,
\xi_+\bigl(\lvert \vartheta_2 \uvect{0} - \vartheta_1 \uvect{\varphi_1}  \rvert\bigr)
,\\
\begin{split}
c^{(\epsilon\gamma)}_{+-}\left(\vartheta_1, \vartheta_2 \right)
&=
\frac{2\sigmaepsilongal^2}{\pi\ngal \AFOV} 
  \int_{0}^{\pi}\!\!\idiff[]{\varphi_1}\,
\xi_-\bigl(\lvert \vartheta_2 \uvect{0} - \vartheta_1 \uvect{\varphi_1}  \rvert\bigr) 
\\&\quad\times
  \cos\left[4\varphi(\vartheta_2 \uvect{0} - \vartheta_1 \uvect{\varphi_1} )\right]
\text{, and}
\end{split}
\\
\begin{split}
c^{(\epsilon\gamma)}_{--}\left(\vartheta_1, \vartheta_2 \right)
&=
\frac{2\sigmaepsilongal^2}{\pi\ngal \AFOV} 
  \!\! \int_{0}^{\pi}\!\!\idiff[]{\varphi_1} \,
\xi_+\bigl(\lvert \vartheta_2 \uvect{0} - \vartheta_1 \uvect{\varphi_1}  \rvert\bigr)
 \cos(4\varphi_1)
.
\end{split}
\end{align}
\end{subequations}
Here, $\uvect{\varphi}$ denotes a unit vector in the plane with polar angle $\varphi$.

Calculating the cosmic variance part $c^{(\gamma)}_{\pm\pm}$ requires knowledge about the 4-point statistics of the convergence field. If one assumes that the convergence field is a homogeneous and isotropic normal random field, the cosmic variance parts read \citep{SchneiderEtal2002}
\begin{align}
\label{eq:c_gamma_normal}
\begin{split}
c^{(\gamma)}_{\pm\pm}\left(\vartheta_1, \vartheta_2 \right)
   &= 
  \frac{4}{\pi\AFOV}\!\int_{0}^{\rFOV}\idiff[]{\theta_3}\, \theta_3   
  \int_{0}^{\pi}\idiff[]{\varphi_1}\, \zeta_{\pm}\bigl( \vartheta_1 \uvect{\varphi_1}, \theta_3 \uvect{0} \bigr) 
   \\&\quad \times  
  \!\int_{0}^{\pi}\idiff[]{\varphi_2}\, \zeta_{\pm}\bigl( \vartheta_2 \uvect{\varphi_2}, \theta_3 \uvect{0} \bigr)
\end{split}
\end{align}
where
\begin{subequations}
\begin{align}
  \zeta_{+}\left( \vtheta, \vtheta' \right) &= \xi_{+}(\lvert \vtheta' - \vtheta \rvert)
	,\\
	\zeta_{-}\left( \vtheta, \vtheta' \right) &= \xi_{-}(\lvert \vtheta' - \vtheta \rvert)
	 \cos\left[\varphi(\vtheta) - \varphi(\vtheta' - \vtheta) \right]
,
\end{align}
\end{subequations}
and $\rFOV\approx \sqrt{\AFOV/\pi}$ denotes  the `radius' of the survey field, a scale similar to the field's linear dimensions.\footnote{
\citet{SchneiderEtal2002} and \citet{JoachimiSchneiderEifler2008} assume $\rFOV\approx \infty$ for simplicity. This is justified for sufficiently large fields, where correlations on scales larger than the field are negligible, and extending the integration boundary of $\theta_3$ in Eq.~\eqref{eq:c_gamma_normal} to infinity does not affect the integral significantly. For small survey fields, however, this may lead to an overestimation of the cosmic covariance, which was observed by \cite{SatoEtal2011} in computer experiments with Gaussian convergence fields.
}

In this work, we also consider zero-mean shifted log-normal random fields (see Appendix for details) as models for the convergence field. Such a field $\kappa$ can be obtained from a homogeneous and isotropic normal random field $n$ with mean $\mu$ and standard deviation $\sigma$ by the point-wise transformation
\begin{equation}
	\kappa(\vtheta) = \exp\bigl[n(\vtheta)\bigr] - \kappa_0,
\end{equation}
which implies a zero-mean shifted log-normal pdf,
\begin{equation}
\label{eq:zero_mean_shifted_log_normal_kappa_pdf}
p_\kappa(\kappa) = 
\begin{cases}
 \dfrac{\exp\Biggl\{-\dfrac{\bigl[\ln(\kappa / \kappa_0 + 1) + \sigma^2/2 \bigr]^2}{2\sigma^2}\Biggl\} }{\sqrt{2\pi}(\kappa + \kappa_0)\sigma}
   & \text{for } \kappa > -\kappa_0 , \\
 0 & \text{otherwise.}	
\end{cases}
\end{equation}
Choosing the shift $\kappa_0 = \exp(\mu + \sigma^2/2)$ ensures zero mean. Since $-\kappa_0$ marks the lower limit for all possible $\kappa$, we call $\kappa_0$ minimum-convergence parameter. 

If the convergence field is well described by a zero-mean shifted log-normal random field, the cosmic variance term $c^{(\gamma)}_{++}$ of the cosmic shear covariance reads (see Appendix~\ref{sec:appendix_shear_correlation})
\begin{equation}
\label{eq:c_gamma_pp_log_normal}
\begin{split}
&
c^{(\gamma)}_{++}\left(\vartheta_1, \vartheta_2 \right)
=
 \xi_{+}(\vartheta_1) \xi_{+}(\vartheta_2)
 \frac{8 \pi}{\kappa_0^{2}\AFOV} 
  \int_{0}^{\rFOV}\iidiff[]{\theta_3}\, \theta_3\, \xi_{+}(\theta_3)
 \\&
 +
  \frac{4 \eta(\vartheta_1) \xi_{+}(\vartheta_2)}{\kappa_0^2\AFOV}  
  \int_{0}^{\rFOV}\iidiff[]{\theta_3}\, \theta_3\, \xi_{+}(\theta_3) 
   \int_{0}^{\pi}\iidiff[]{\varphi_1}\,
  \xi_{+}(\lvert \theta_3\uvect{0} - \vartheta_1\uvect{\varphi_1} \rvert)
 \\&+
 \frac{4 \xi_{+}(\vartheta_1) \eta(\vartheta_2)}{\kappa_0^2\AFOV}
  \int_{0}^{\rFOV}\iidiff[]{\theta_3}\, \theta_3\, \xi_{+}(\theta_3) 
  \int_{0}^{\pi}\iidiff[]{\varphi_2}\,
  \xi_{+}(\lvert \theta_3\uvect{0} - \vartheta_2\uvect{\varphi_2} \rvert)
 \\&+
 \frac{\eta(\vartheta_1)\eta(\vartheta_2)}{2\pi \AFOV}
 \int_{0}^{\rFOV}\idiff[]{\theta_3}\,\theta_3 \int_{0}^{2\pi}\idiff[]{\varphi_1} \int_{0}^{2\pi}\idiff[]{\varphi_2}\,
 \\&\times
 \xi_{+}(\lvert \theta_3\uvect{0} - \vartheta_1\uvect{\varphi_1} \rvert)
 \xi_{+}(\lvert \theta_3\uvect{0} - \vartheta_2\uvect{\varphi_2} \rvert)
   \\&\times 
  \Bigl[
  2 + 4 \kappa_0^{-2} \xi_{+}(\theta_3)
  +\kappa_0^{-4} 
 \xi_{+}(\theta_3)
 \xi_{+}(\lvert \theta_3\uvect{0} - \vartheta_1\uvect{\varphi_1} - \vartheta_2\uvect{\varphi_2} \rvert)
 \Bigr]
\end{split}
\end{equation}
with $\eta(\vartheta) = \kappa_0^{-2} \xi_{+}(\vartheta) +1$.
We are unable to derive equally `concise' expressions for the cosmic variance parts $c^{(\gamma)}_{+-}$ and $c^{(\gamma)}_{--}$. However, $c^{(\gamma)}_{+-}$ and $c^{(\gamma)}_{--}$ can be obtained from any given $c^{(\gamma)}_{++}$ by an integral transformation with the kernel \eqref{eq:kernel_xi_k_to_xi_m} (see Appendix~\ref{sec:appendix_shear_correlation}):
\begin{align}
\label{eq:c_pp_to_pm}
c^{(\gamma)}_{+-}(\vartheta_1,\vartheta_2) &= 
  \int_0^{\infty}\idiff[]{\vartheta'_2}\,\vartheta'_2\, \KerM (\vartheta_2,\vartheta'_2)c^{(\gamma)}_{++}(\vartheta_1,\vartheta'_2)
,\\
\begin{split}
\label{eq:c_pp_to_mm}
c^{(\gamma)}_{--}(\vartheta_1,\vartheta_2) &= 
  \int_0^{\infty}\idiff[]{\vartheta'_1}\,\vartheta'_1\, \KerM(\vartheta_1,\vartheta'_1)
  \int_0^{\infty}\idiff[]{\vartheta'_2}\,\vartheta'_2\, \KerM(\vartheta_2,\vartheta'_2)
  \\&\quad\times   
  c^{(\gamma)}_{++}(\vartheta'_1,\vartheta'_2)
.
\end{split}
\end{align}

Equation~\eqref{eq:c_gamma_pp_log_normal} comprises terms of 2nd to 6th order in $\xi_{\kappa}$. If only the quadratic terms, which are also present in the expression~\eqref{eq:c_gamma_normal}, and the simplest cubic term in $\xi_{\kappa}$ are retained, the cosmic variance contribution reduces to
\begin{equation}
\label{eq:c_gamma_simplified_log_normal}
\begin{split}
c^{(\gamma)}_{\pm\pm}\left(\vartheta_1, \vartheta_2 \right)
&= 
  \frac{4}{\pi\AFOV}\!\int_{0}^{\rFOV}\idiff[]{\theta_3}\, \theta_3\,
  \!\int_{0}^{\pi}\idiff[]{\varphi_1}\, \zeta_{\pm}\bigl( \vartheta_1 \uvect{\varphi_1}, \theta_3 \uvect{0} \bigr) 
  \\&\quad\times
  \!\int_{0}^{\pi}\idiff[]{\varphi_2}\, \zeta_{\pm}\bigl( \vartheta_2 \uvect{\varphi_2}, \theta_3 \uvect{0} \bigr)
\\&\quad
+
  \xi_{\pm}(\vartheta_1) \xi_{\pm}(\vartheta_2)\frac{8 \pi}{\kappa_0^{2}\AFOV}
  \int_{0}^{\rFOV}\idiff[]{\theta_3}\,\theta_3\, \xi_{+}(\theta_3)
.
\end{split}
\end{equation}

In the following, we call the approximation defined by Eqs.\eqref{eq:c_pmpm_split}-\eqref{eq:c_gamma_normal} the normal approximation to the cosmic shear covariance. If Eqs.~\eqref{eq:c_gamma_pp_log_normal}-\eqref{eq:c_pp_to_mm} are used instead of Eq.~\eqref{eq:c_gamma_normal}, we call the resulting approximation  to the cosmic shear covariance the log-normal approximation. If Eq.~\eqref{eq:c_gamma_simplified_log_normal} is used instead, we call the resulting approximation the simplified log-normal approximation.

In general, the convergence field is neither a normal nor a zero-mean shifted log-normal field. In case one has a large sample of $N_\text{r}$ (quasi-)independent realisations of the survey field with measured correlations $\est{\xi}_\pm^r(\vartheta)$, one can estimate the mean using the sample mean,
\begin{equation}
\label{eq:df_sample_covariance}
\est{\xi}_{\pm}(\vartheta) = \frac{1}{N_\text{r}} \sum_{r=1}^{N_\text{r}} \est{\xi}_\pm^r(\vartheta)
,
\end{equation}
and the covariance using the sample covariance,
\begin{equation}
c_{\pm\pm}\left(\vartheta_1, \vartheta_2 \right) = 
    \frac{1}{N_\text{r} - 1}\!\!\sum_{r=1}^{N_\text{r}}
    \bigl[\est{\xi}_\pm^r(\vartheta_1) - \est{\xi}_\pm(\vartheta_1)\bigr]
    \bigl[\est{\xi}_\pm^r(\vartheta_2) - \est{\xi}_\pm(\vartheta_2)\bigr]
.
\end{equation}

For many surveys, the survey area is not a single field, but consists of several small areas that are far apart from each other on the sky. In this case, the measurements in each field can be considered independent from the measurements in the other fields. If all survey fields have similar field areas and galaxy densities, the covariances for the whole survey can be obtained by computing the covariances for a single field and dividing the result by the number of fields. 

\subsection{Estimation of cosmological parameters}
\label{sec:parameter_estimation}

From a set of shear correlation functions $\xi_\pm(\vartheta_1),\ldots,\xi_\pm(\vartheta_{\NBins})$ measured at $\NBins$ angular separations, one can form a data vector 
\begin{equation}
\label{eq:df_xi_data_vector}
 \vect{d}=\transposed{\bigl(\xi_+(\vartheta_1),\ldots,\xi_+(\vartheta_{\NBins}),\xi_-(\vartheta_1),\ldots,\xi_-(\vartheta_{\NBins})\bigr)}
.
\end{equation}
In one of its simplest forms, Bayesian model-parameter estimation from a measured cosmic shear data vector $\vect{d}$ amounts to a maximum likelihood parameter estimation assuming a quadratic log-likelihood
\begin{equation}
\label{eq:quadratic_log_likelihood}
  \loglikelihood(\vect{\pi}|\vect{d}) = -\frac{1}{2}\transposed{\left[\vect{d} - \vect{\mu}(\vect{\pi}) \right]} \matrb{C}_\text{d}^{-1} \left[\vect{d} - \vect{\mu}(\vect{\pi}) \right] +\text{const.}
\end{equation}
Here, $\vect{\mu}(\vect{\pi})$ denotes the prediction of the cosmic shear correlation vector for the model characterized by a set of $N_\text{p}$ parameters $\vect{\pi}=\transposed{(\pi_1,\ldots,\pi_{N_\text{p}})}$.
If one uses the normal or log-normal approximation to the cosmic shear covariance, the inverse covariance matrix $\matrb{C}_\text{d}^{-1}$ of the data vector $\vect{d}$ reads:
\begin{align}
\label{eq:def_csdv_inv_cov_matrix}
  \matrb{C}_\text{d}^{-1} &= 
  \begin{pmatrix}
  \matrb{c}_{++} & \matrb{c}_{+-} \\
  \transposed{\matrb{c}}_{+-} & \matrb{c}_{--} \\
  \end{pmatrix}^{-1}
&\text{with}&&
  \matrb{c}_{\pm\pm} &=
  \begin{pmatrix}
  c_{\pm\pm}(\vartheta_i,\vartheta_j)
  \end{pmatrix}_{i,j=1}^{\NBins}
.
\end{align}

If the cosmic shear covariances $\matrb{c}_{\pm\pm}$ are estimated from the sample covariances of $N_\text{r}$ realisations,  the inverse covariance can be estimated by \citep[][]{Anderson2003_book,HartlapEtal2007}
\begin{equation}
\label{eq:def_csdv_inv_cov_matrix_from_sample_cov}
  \matrb{C}_\text{d}^{-1} = 
\frac{N_\text{r} - 2 \NBins - 2}{N_\text{r} - 1}
  \begin{pmatrix}
  \matrb{c}_{++} & \matrb{c}_{+-} \\
  \transposed{\matrb{c}}_{+-} & \matrb{c}_{--} \\
  \end{pmatrix}^{-1}
,
\end{equation}
where the correction factor ensures that this estimate is unbiased if the joint distribution of the measured correlation functions is well approximated by a multivariate normal distribution.
This estimate requires $N_\text{r} > (2 \NBins + 2)$. For smaller $N_\text{r}$, the sample covariance matrix is not invertible, and other methods are required to estimate the inverse covariance matrix.

The parameter set $\vect{\pi}_\text{ML}$ maximizing the (log-)likelihood satisfies
\begin{equation}
  \vect{0} = \transposed{\left.\parder{\vect{\mu}(\vect{\pi})}{\vect{\pi}}\right|_{\vect{\pi}_\text{ML}}} \matrb{C}_\text{d}^{-1} \left[\vect{d} - \vect{\mu}(\vect{\pi}_\text{ML}) \right].
\end{equation}
A Taylor expansion of $\loglikelihood(\vect{\pi}|\vect{d})$ around $\vect{\pi}_\text{ML}$ up to second order yields
\begin{equation}
\label{eq:ml_likelihood_gauss_approximation}
  \loglikelihood(\vect{\pi}|\vect{d}) \approx -\frac{1}{2}\transposed{\left(\vect{\pi} - \vect{\pi}_\text{ML}\right)} \matrb{C}^{-1}_{\pi} \left(\vect{\pi} - \vect{\pi}_\text{ML}\right) +\text{const.},
\end{equation}
where
\begin{equation}
  \matrb{C}_{\pi}^{-1} = 
  \begin{pmatrix}
   \displaystyle\sum_{k,l=1}^{\NBins} 
   \matr{C}^{-1}_{\text{d},kl}
   \left[
     \dparder{\mu_k}{\pi_i} \dparder{\mu_l}{\pi_j}
     +  (d_k - \mu_k) \dparder{^2 \mu_k}{\pi_i \partial \pi_j}
   \right]_{\vect{\pi}_\text{ML}}
   \end{pmatrix}_{i,j = 1}^{N_\text{p}}
\end{equation} 
differs from the Fisher matrix by additional contributions from second derivatives of $\vect{\mu}$ when $\vect{\mu}(\vect{\pi}_\text{ML})\neq \vect{d}$.

The equations above show how the covariances and confidence regions of the model parameters depend on the assumed covariances of the cosmic shear correlation functions in a maximum-likelihood estimation with a Gaussian likelihood.  
In a Bayesian analysis with constant priors on the model parameters, the likelihood $\likelihood(\vect{\pi}|\vect{d})$ is proportional to the posterior distribution $p(\vect{\pi}|\vect{d})$. In this case, the above equations also specify how the posterior distribution depends on the assumed covariance of the cosmic shear correlation functions. For more general prior distributions $p(\vect{\pi})$, the posterior reads
\begin{equation}
\label{eq:general_posterior}
  p(\vect{\pi}|\vect{d}) = \frac{\likelihood (\vect{\pi}|\vect{d}) p(\vect{\pi})}{ \int\diff[N_\text{p}]{\vect{\pi}}\, \likelihood (\vect{\pi}|\vect{d}) p(\vect{\pi}) }.
\end{equation}

\section{Simulations}
\label{sec:simulations}

We use a numerical approach to assess the performance of predictions for the cosmic shear covariance.
By ray-tracing through the Millennium Run (MR), a large $N$-body simulation of cosmic structure formation by \citet{SpringelEtal2005_Millennium}, we create a suite of simulated fields of view with maps of the effective convergence and shear. From these maps, we measure the convergence distribution and the convergence and shear correlations. We then estimate the covariance of the shear correlation functions in different ways: from the measured one- and two-point statistics of the convergence using the normal, log-normal, and simplified log-normal approximation, and from the sample covariance of the shear correlation in the simulated survey fields.

The MR assumes a flat $\Lambda$CDM cosmology with a matter density $\Omega_\mathrm{m} = 0.25$, a baryon density $\Omega_\mathrm{b} = 0.045$, and a cosmological-constant energy density $\Omega_\Lambda=0.75$ (in units of the critical density), a Hubble constant $h=0.73$ (in units of $100\,\mathrm{km}\,\mathrm{s}^{-1}\Mpc^{-1}$), a primordial spectral index $n_\mathrm{s}=1$ and a normalisation parameter $\sigma_8=0.9$ for the linear density power spectrum. These values also define our fiducial cosmology in the later analysis. The simulation employed a customized version of \softwarename{gadget-2} \citep[][]{Springel2005_GADGET2} with $10^{10}$ particles of $8.6\times 10^8 h^{-1}\,\Msolar$ in a cube of $500h^{-1}\,\Mpc$ comoving side length, and a comoving force softening length of $5h^{-1}\,\kpc$.

We employ the multiple-lens-plane ray-tracing algorithm described in \citet[][]{HilbertEtal2009_RT} to simulate gravitational lensing observations.
The ray-tracing algorithm takes into account the gravitational deflection by the dark matter, represented by the dark-matter particles of the MR, and the deflection by the stellar mass in galaxies \citep[for details, see][]{HilbertEtal2008_StrongLensing_II} as inferred from the galaxy-formation model of \cite{DeLuciaBlaizot2007}.

Using the ray-tracing, we generate 64 fields of view. Each field has an area of $4 \times 4\,\degt^2$ (yielding a total area of $1024\,\degt^2$) and is covered by a regular grid of $4096^2$ pixels (yielding a resolution of $3.5\,\arcsect$). For each pixel, a ray is traced through the MR up to redshift 3.2, and the convergence and shear along the ray is recorded. To study the covariance of smaller survey fields, we also create 256 fields of $2 \times 2\,\degt^2$ by splitting the $4 \times 4\,\degt^2$ fields evenly.

The convergence and shear information along the rays is then used to calculate the effective convergence and shear in the simulated fields for a source population with  median redshift $z_\text{median}=1$ and a redshift distribution \citep[][]{BrainerdBlandfordSmail1996}
\begin{equation}
\label{eq:source_redshift_distribution}
 p_{z}(z)=\frac{3z^2}{2z_0^3}\exp\left[-\left(\frac{z}{z_0}\right)^{3/2} \right],
 \text{ where }
 z_0 = \frac{z_\text{median}}{1.412}.
\end{equation}
The only source of noise (i.e. uncertainties affecting the estimation of cosmological parameters) in these ellipticity noise-free maps is the cosmic variance (each field represents only a single finite realization of the underlying cosmology). 
To create noisy lensing maps that also incorporate the uncertainties in cosmic shear surveys arising from the intrinsic shapes of source galaxies, we add Gaussian noise with standard deviation $\sigmapixel =  \sigmaepsilongal/\sqrt{2 \ngal \Apixel}$
to each pixel (with pixel area $\Apixel \approx 3.5^2\,\arcsect^2$) of the simulated shear fields. We assume an intrinsic galaxy ellipticity distribution with standard deviation $\sigmaepsilongal = 0.4$, and we consider surveys with galaxy densities $\ngal = 25\,\arcmint^{-2}$ and $\ngal = 100\,\arcmint^{-2}$.

\section{Results}
\label{sec:results}

\subsection{The convergence distribution}
\label{sec:convergence_pdf}

\begin{figure}
\centerline{\includegraphics[width=\linewidth]{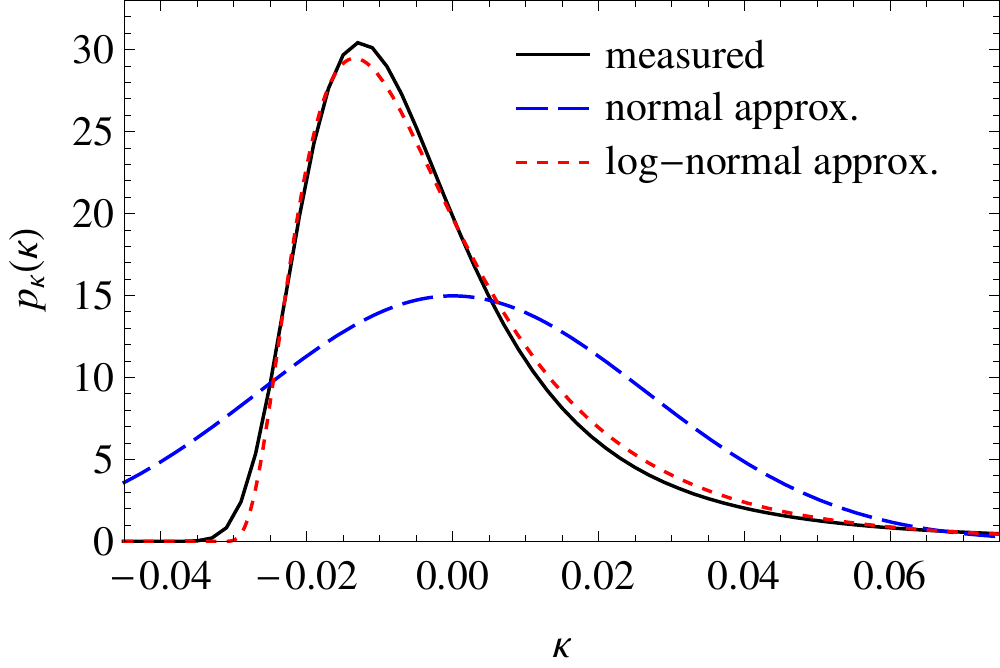}}
\caption{
\label{fig:convergence_pdf}
The pdf $p_\kappa(\kappa)$ of the convergence $\kappa$ for sources with median redshift $z_\text{median}=1$. The pdf measured from the simulations (solid line) is compared to a zero-mean normal distribution with matching variance (dashed line), and the best-fitting zero-mean shifted log-normal distribution (dotted line). 
}
\end{figure}

We determine the convergence distribution from the simulated fields by binning the ellipticity noise-free convergence values of all field pixels into bins of size $0.002$. The resulting probability density function (pdf) of the convergence distribution is shown Fig.~\ref{fig:convergence_pdf}. For comparison, we also show the pdf of a zero-mean normal distribution, whose variance equals the measured variance of the convergence, and the pdf \eqref{eq:zero_mean_shifted_log_normal_kappa_pdf} of the best-fitting zero-mean shifted log-normal distribution, whose parameters $\kappa_0$ and $\sigma$ where obtained by a simple least-squares fit.
Neither the normal  nor the zero-mean shifted log-normal distribution matches the measured convergence distribution perfectly, but (at least visually) the zero-mean shifted log-normal distribution fares far better than the normal distribution. This finding indicates that the log-normal approximation might possibly provide a better approximation to the covariance of the cosmic shear correlation than the normal approximation. 

While the normal approximation to the covariance of cosmic shear correlations only needs the convergence correlation as input, the log-normal approximation also requires the minimum-convergence parameter $\kappa_0$. One could, for example, compute the convergence for an empty beam to a fixed source redshift and use its modulus as $\kappa_0$. Simulations show, however, that there are no empty lines of sight in a $\Lambda$CDM universe \citep[][]{TaruyaEtal2002,ValeWhite2003, HilbertEtal2007_StrongLensing}. We thus use the value $\kappa_0 \approx 0.032$ obtained from the fit \eqref{eq:zero_mean_shifted_log_normal_kappa_pdf}.

\begin{figure}
\centerline{\includegraphics[width=\linewidth]{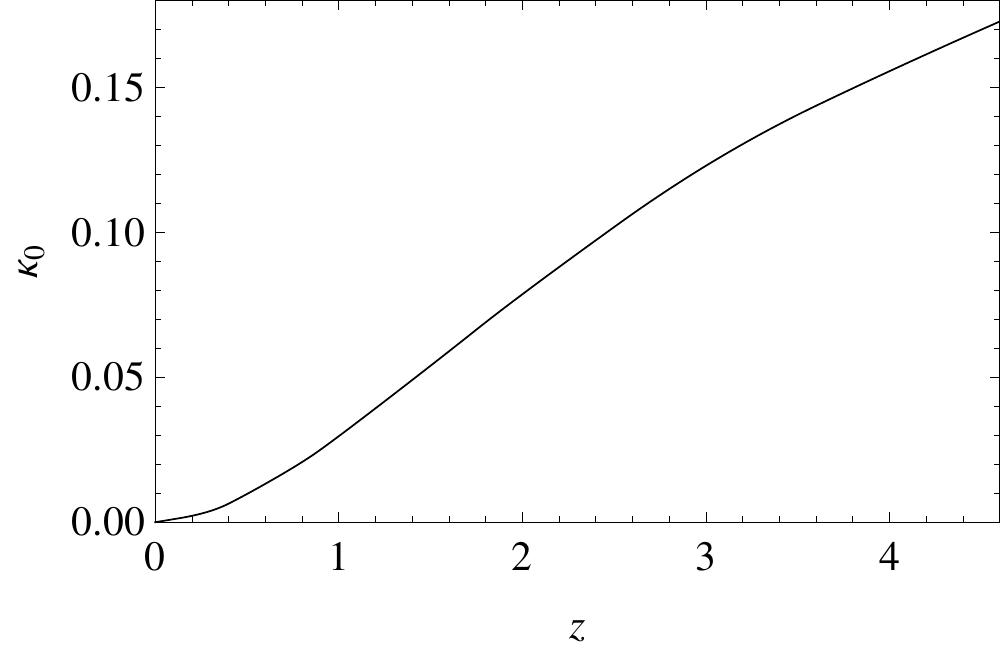}}
\caption{
\label{fig:minimum_convergence_parameter}
The minimum-convergence parameter $\kappa_0$ as a function of source redshift $z$ (obtained from fits to the pdf of the convergence for sources at redshift $z$).
}
\end{figure}
\begin{table}
\center
  \caption{
\label{tab:kappa_0_et_al}
The parameters $\kappa_0$ and $\sigma$ of the fit~\eqref{eq:zero_mean_shifted_log_normal_kappa_pdf} to the convergence pdf for sources at various redshifts $z$, compared to the measured standard deviation $\bEV{\kappa^2}^{1/2}$ of the convergence.
}
\begin{tabular}{c c c c}
\hline
$z$   & $\kappa_0$ & $\sigma$ & $\bEV{\kappa^2}^{1/2}$ \\
\hline
0.28 & 0.0035 & 0.81 & 0.0058 \\
0.51 & 0.0100 & 0.74 & 0.0124 \\
0.76 & 0.0120 & 0.70 & 0.0199 \\
0.99 & 0.0291 & 0.66 & 0.0270 \\
1.50 & 0.0542 & 0.58 & 0.0415 \\
2.07 & 0.0819 & 0.53 & 0.0547 \\
3.06 & 0.1253 & 0.47 & 0.0716 \\
4.18 & 0.1608 & 0.44 & 0.0838 \\ 
5.29 & 0.1917 & 0.41 & 0.0922 \\
\hline
\end{tabular}
\end{table}

The minimum-convergence parameter $\kappa_0(z)$ for sources at a single redshift $z$, obtained from fits to the measured pdf of the convergence $\kappa(z)$, is shown in Fig.~\ref{fig:minimum_convergence_parameter} as a function of source redshift and listed in Table~\ref{tab:kappa_0_et_al} along with the second fit parameter $\sigma$ and the standard deviation of the convergence. The parameter $\sigma$ can be interpreted as a measure of non-Gaussianity of the convergence field. The values $0.5 \lesssim \sigma \lesssim 1$ indicate a moderate degree of non-Gaussianity \citep[][]{MartinSchneiderSimon2011_arXiv}.

The redshift dependence of $\kappa_0$ in the range $0.3 \leq z \leq 4$ is well described by
\begin{equation}
  \kappa_0(z) = 0.008 z + 0.029 z^2 - 0.0079 z^3 + 0.00065 z^4.
\end{equation}
The minimum-convergence parameter for sources at redshift $z=1$ is very similar to the value measured from the source redshift distribution \eqref{eq:source_redshift_distribution} with median redshift $z_\text{median}=1$. Moreover, the minimum-convergence for $z = 1$ and $z = 2$ are in good agreement with the corresponding values found by \citet{TaruyaEtal2002}.

\subsection{The convergence and shear correlations}
\label{sec:convergence_and_shear_2PCF}
\begin{figure}
\centerline{\includegraphics[width=\linewidth]{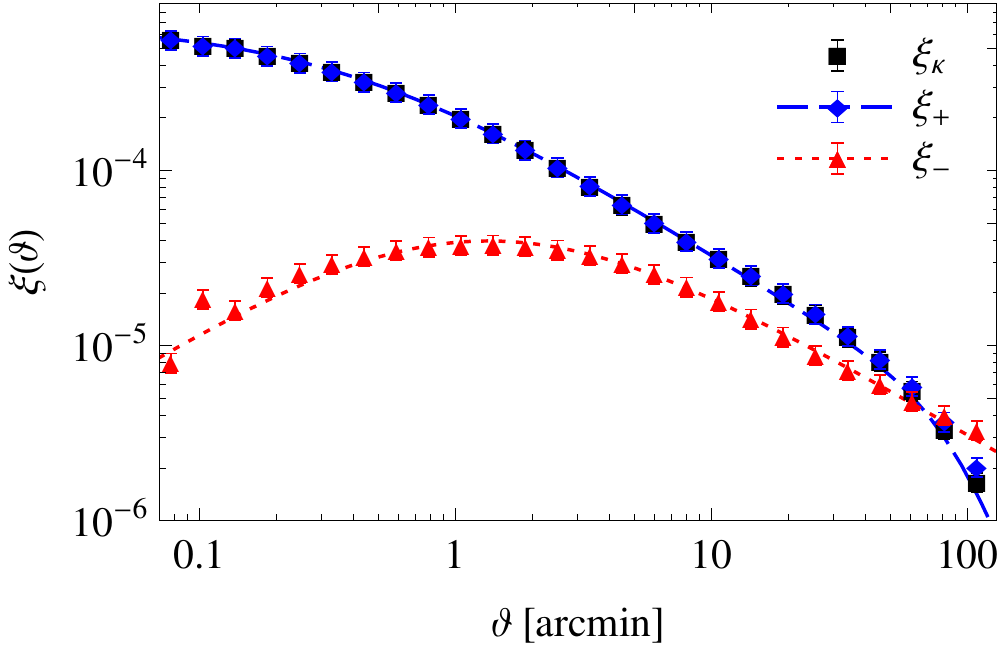}}
\caption{
\label{fig:convergence_and_shear_correlation}
The convergence correlation $\xi_\kappa(\vartheta)$ (squares) and the cosmic shear correlations $\xi_{+}(\vartheta)$ (diamonds) and $\xi_{-}(\vartheta)$ (triangles) as functions of the angular scale $\vartheta$ for sources with median redshift $z_\text{median}=1$ measured from the simulations. Error bars indicate uncertainties on the mean estimated from the field-to-field variance of the simulated fields. Also shown are the analytic fits (see text) to the correlations $\xi_{+}(\vartheta)$ (dashed line) and $\xi_{-}(\vartheta)$ (dotted line). 
}
\end{figure}

From each ellipticity noise-free simulated $4\times4\,\degt^2$ field, we estimate the convergence correlation $\xi_\kappa(\theta)$ and the cosmic shear correlations $\xi_{\pm}$ using Fast Fourier Transform (FFT) techniques employing zero-padding and book-keeping of the number of contributing pixel pairs to properly account for the non-periodic field boundaries (see Appendix \ref{sec:appendix_shear_correlation_fourier_methods} for more details).
The correlation estimates from all fields are then used to compute the mean correlations and their uncertainties from the field-to-field variance.

The resulting convergence and cosmic shear correlations are shown in Fig.~\ref{fig:convergence_and_shear_correlation}. As expected, the measured shear correlation $\xi_+$ is almost identical to the measured convergence correlation $\xi_\kappa$. Numerical tests verify furthermore that the measured correlations $\xi_\kappa$ and $\xi_-$ satisfy the relation~\eqref{eq:xi_m_from_xi_kappa} well within the error bars.

Computation of the covariance of the cosmic shear correlations in the (log-)normal approximation involves integrals over the shear correlations. To facilitate the integration, we approximate the measured correlation functions $\xi_{\pm}$ by analytic expressions
\begin{subequations}
\begin{align}
\label{eq:xi_pm_fits}
  \xi_{+}(\vartheta) &= \frac{a_1}{1+(\vartheta/\tau_1)^{b_1}}\frac{1}{1+(\vartheta/\tau_2)^{b_2}}
\text{ and}\\
   \xi_{-}(\vartheta) &= \frac{a_2 \vartheta^{b_1}}{(1+\vartheta/\tau_3)^{b_3}}
\end{align}
\end{subequations}
with parameters $a_1$, $a_2$, $b_1$, $b_2$, $b_3$, $\tau_1$, $\tau_2$, and $\tau_3$ obtained from a weighted least-squared fit. The resulting expressions fit the measured correlations very well, as Fig.~\ref{fig:convergence_and_shear_correlation} shows.

\subsection{The covariance of the cosmic shear correlation functions}
\label{sec:results_for_cosmic_shear_covariance}

\begin{figure*}
\centerline{
\includegraphics[width=0.33\linewidth]{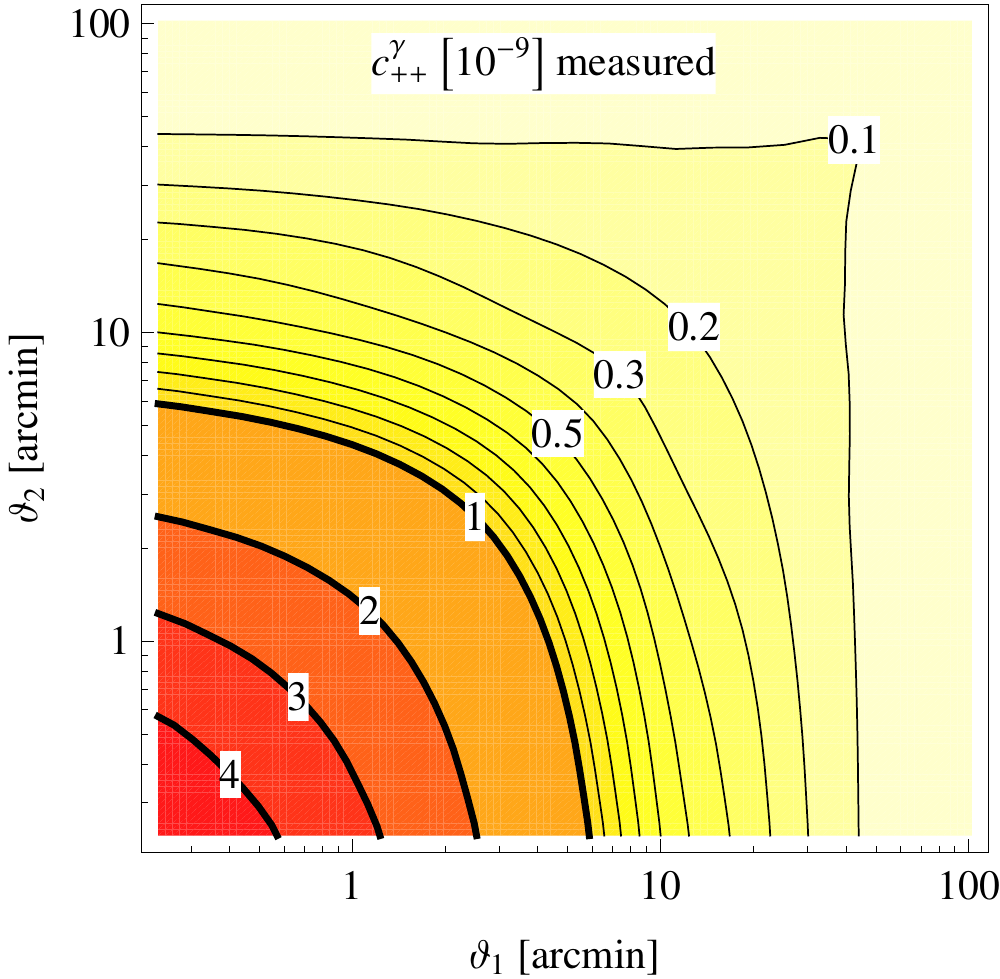}
\hfill
\includegraphics[width=0.33\linewidth]{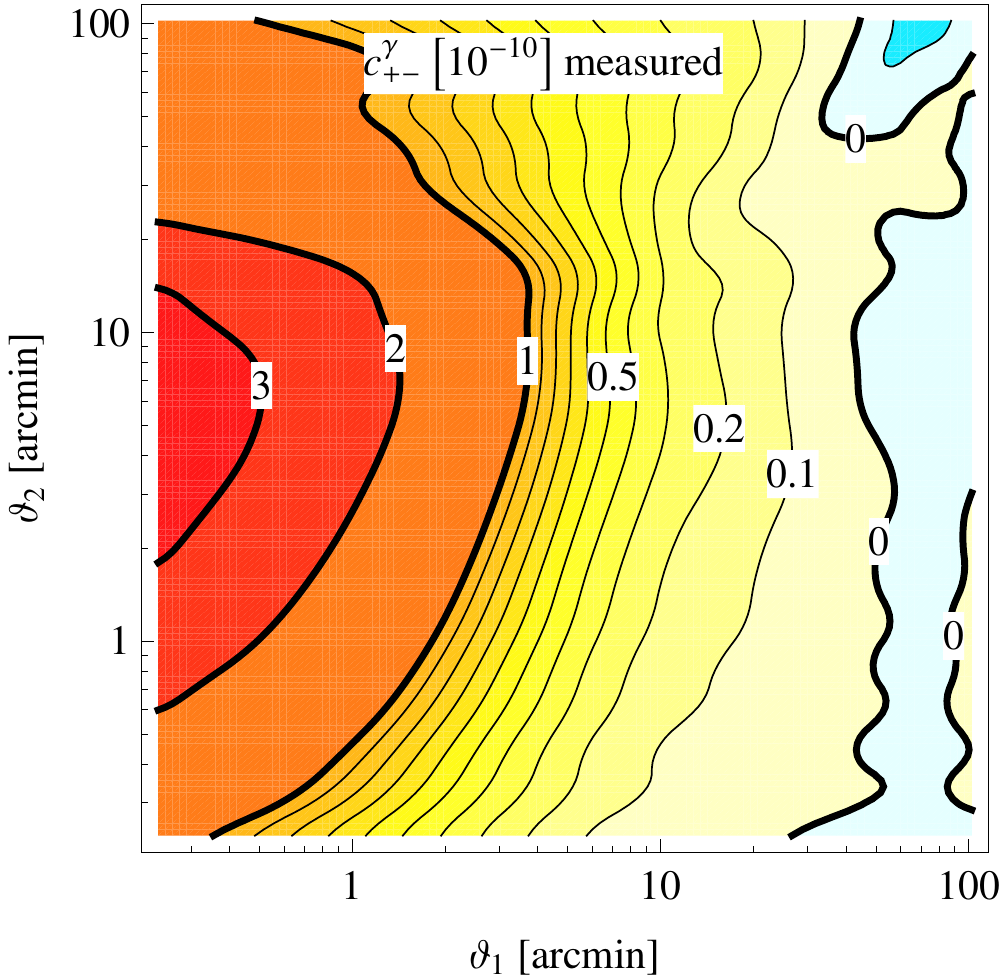}
\hfill
\includegraphics[width=0.33\linewidth]{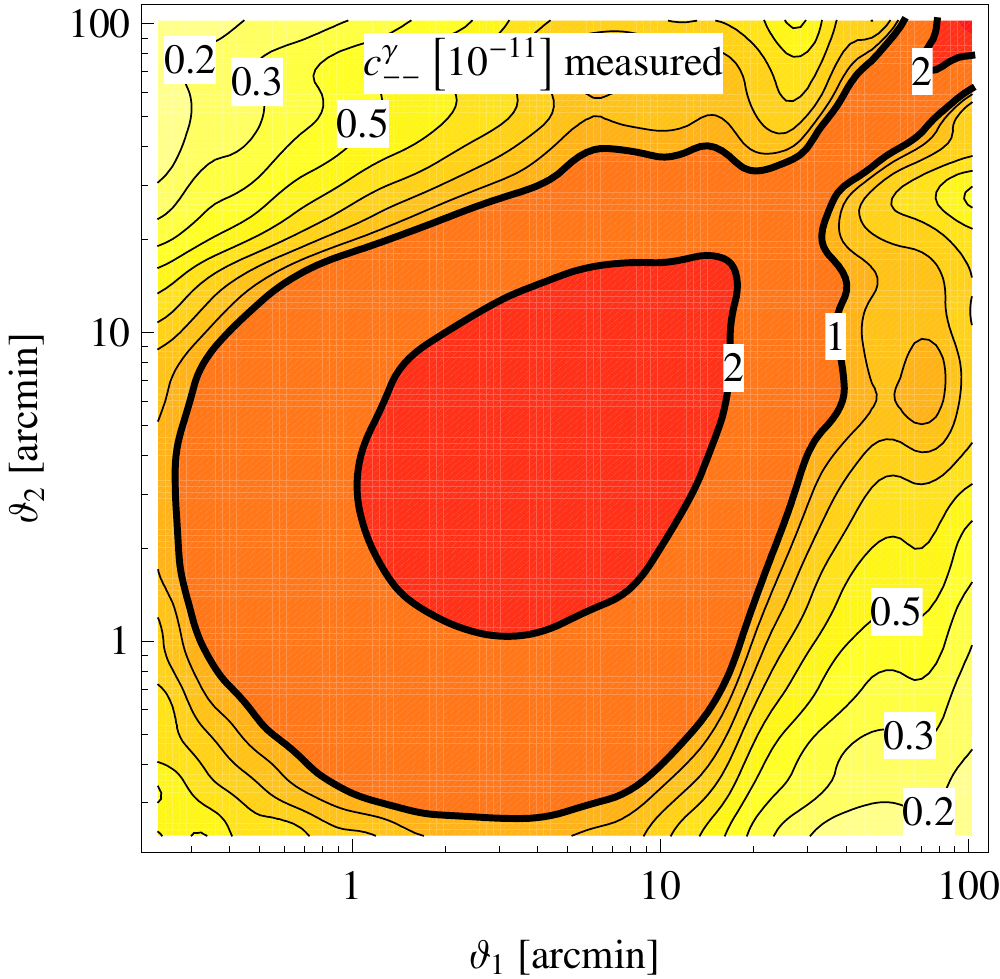}
}
\vspace{1em}
\centerline{
\includegraphics[width=0.33\linewidth]{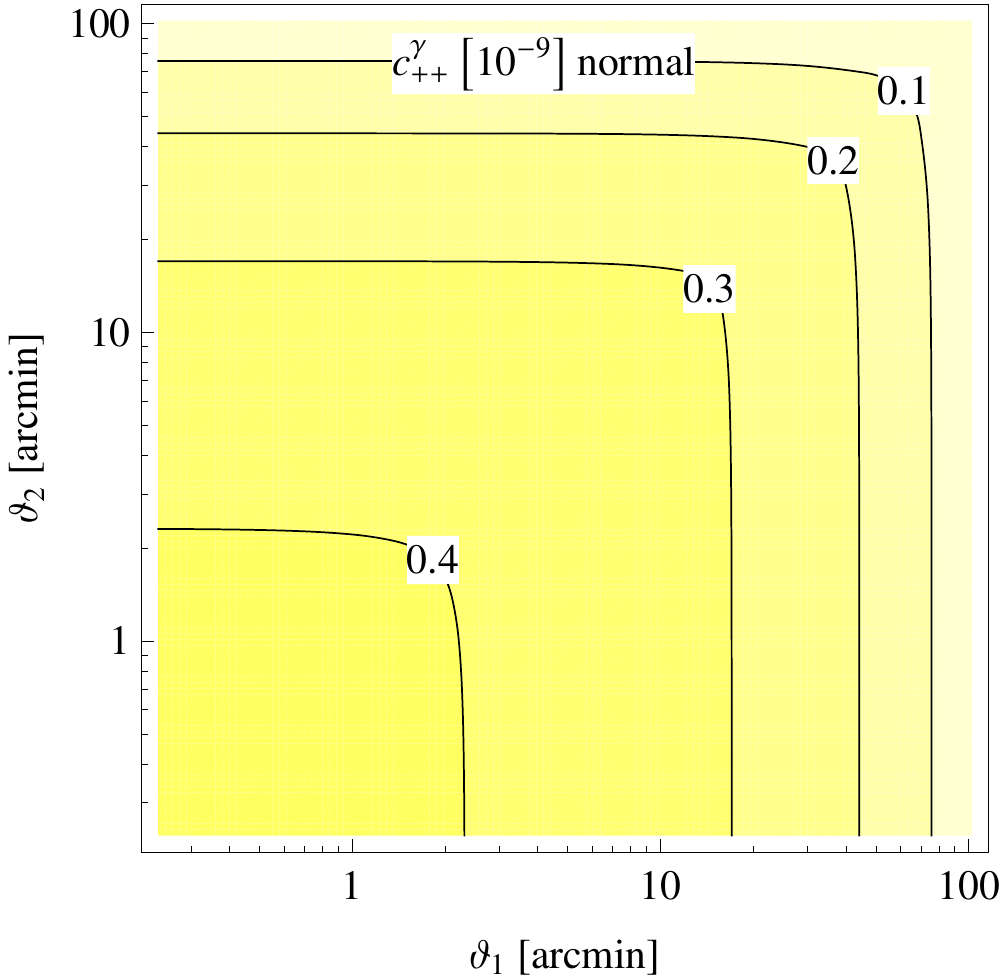}
\hfill
\includegraphics[width=0.33\linewidth]{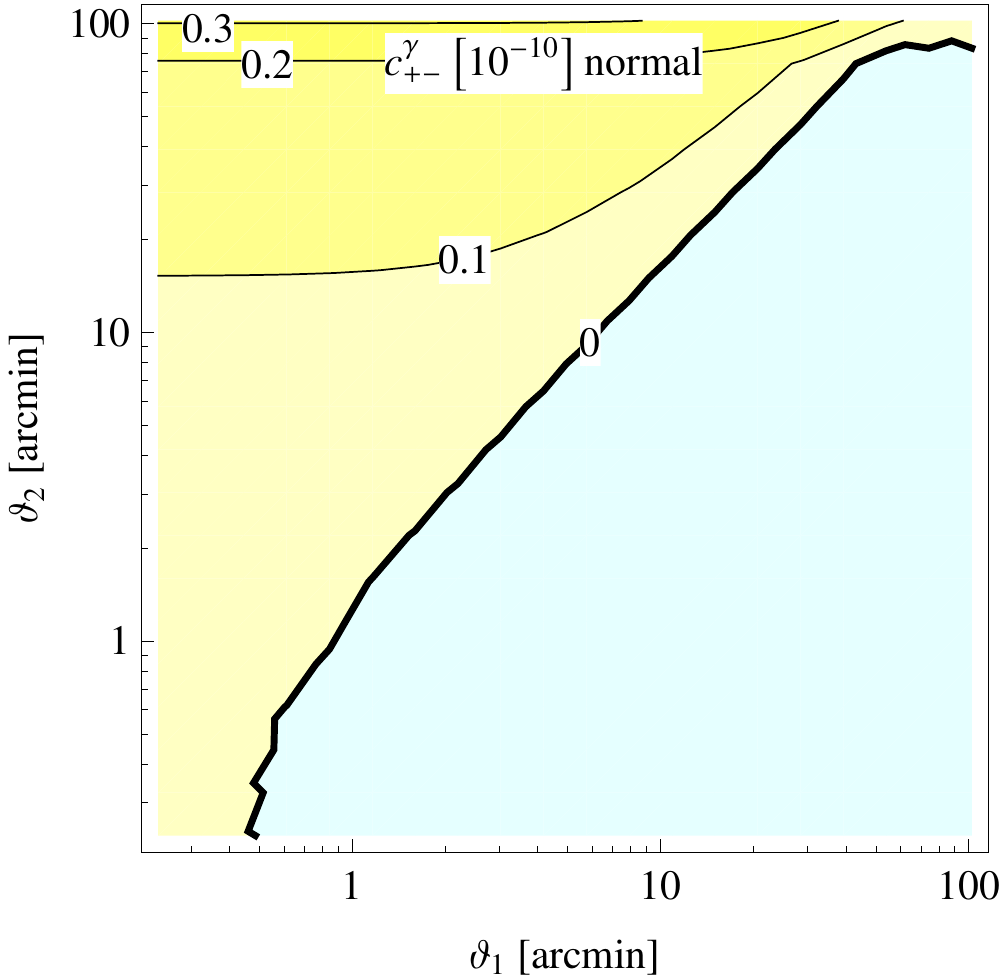}
\hfill
\includegraphics[width=0.33\linewidth]{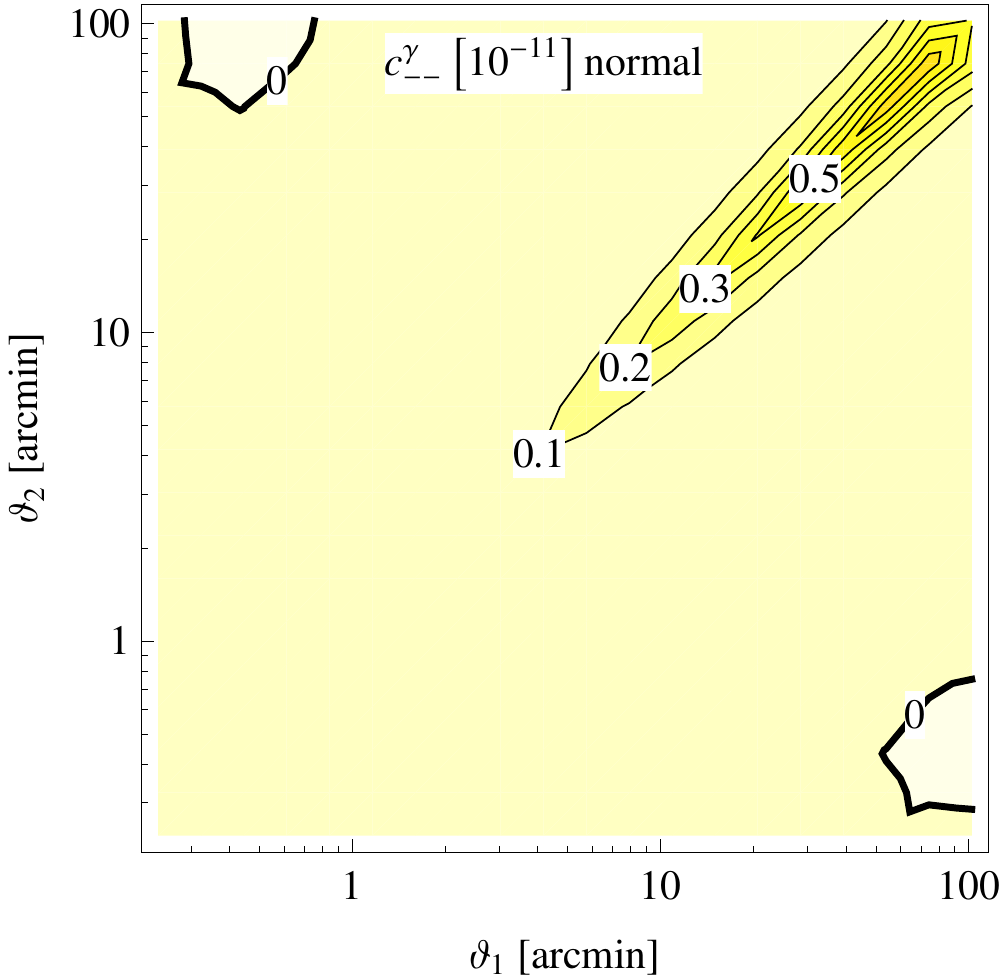}
}
\vspace{1em}
\centerline{
\includegraphics[width=0.33\linewidth]{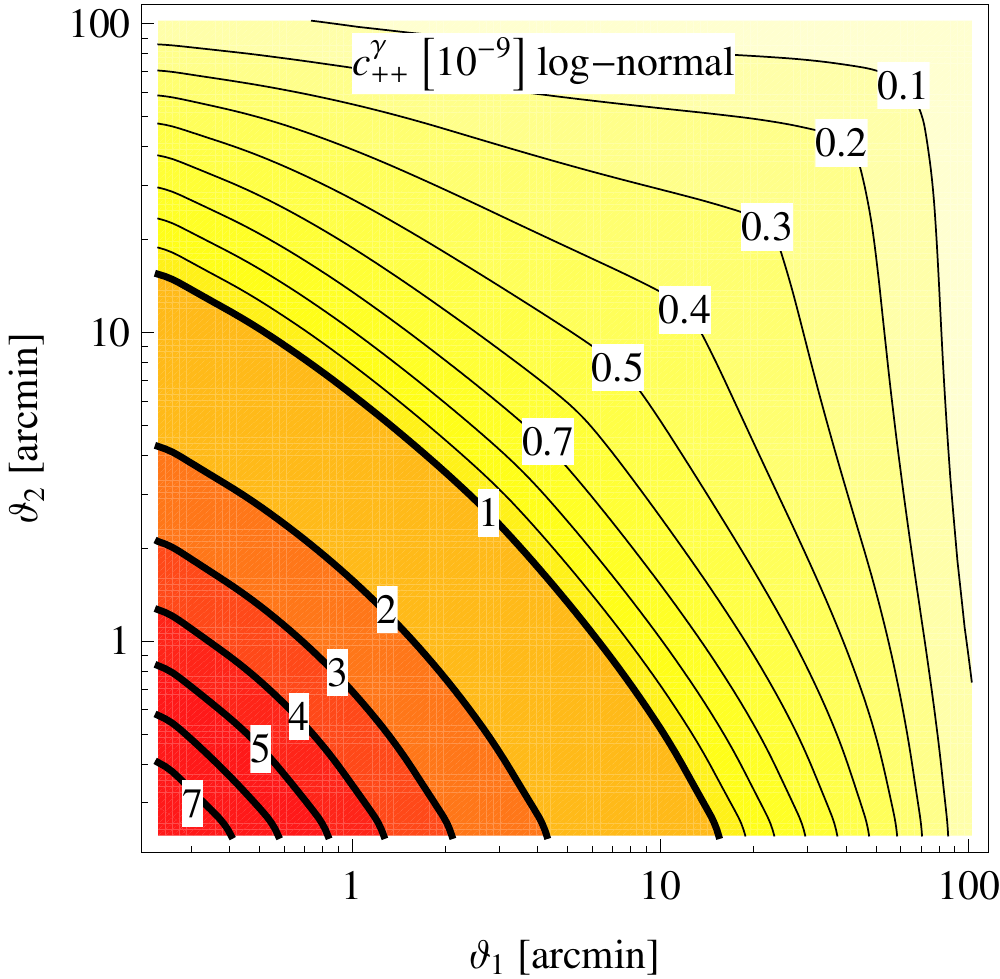}
\hfill
\includegraphics[width=0.33\linewidth]{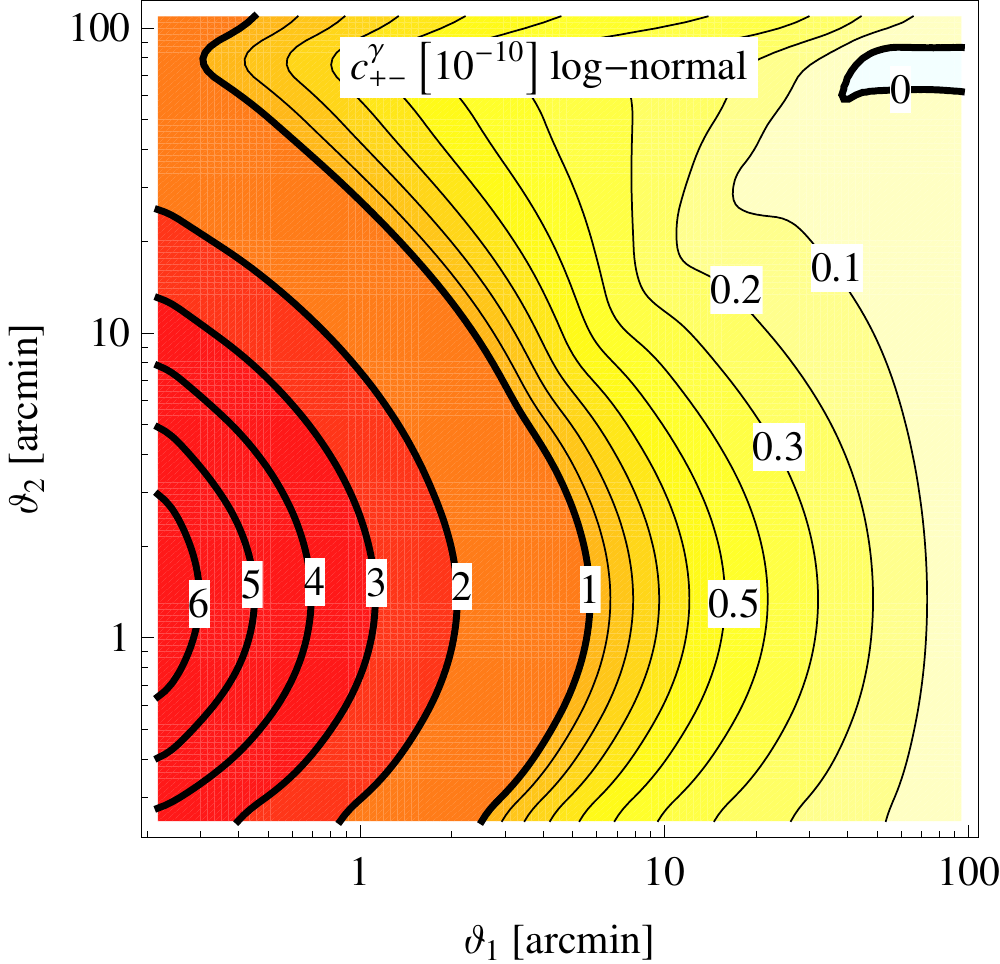}
\hfill
\includegraphics[width=0.33\linewidth]{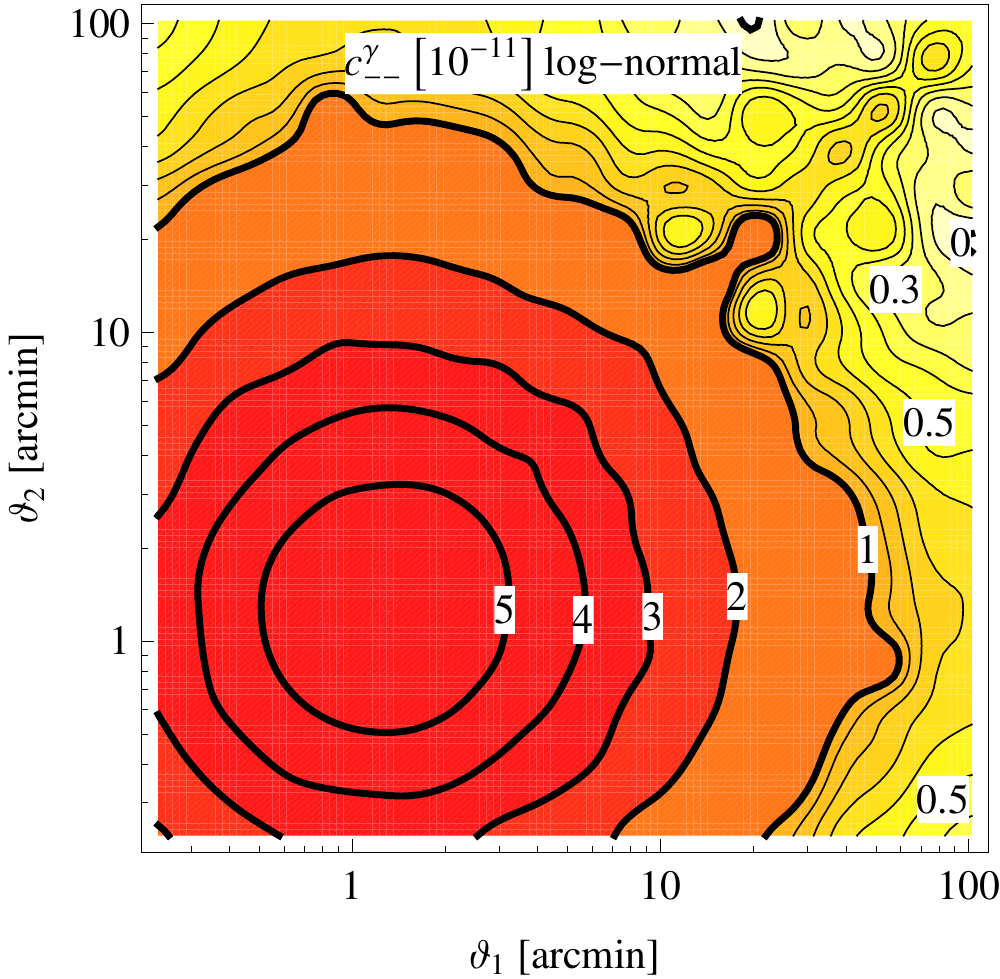}
}
\caption{
\label{fig:cosmic_covariance}
The cosmic variance contributions $c^{(\gamma)}_{\pm\pm}(\vartheta_1, \vartheta_2)$ to the cosmic shear covariance as functions of angular separations $\vartheta_1$ and $\vartheta_2$. Shown are the contributions $c^{(\gamma)}_{++}$ (left column), $c^{(\gamma)}_{+-}$ (middle column), and $c^{(\gamma)}_{--}$ (right column) for $2\times2\,\degt^2$ fields measured from the lensing simulations (top row), computed employing the normal approximation (2nd row), and the log-normal approximation (bottom row).
}
\end{figure*}

We estimate the cosmic variance contribution $c^{(\gamma)}_{\pm\pm}$ to the cosmic shear covariance from the sample covariance~\eqref{eq:df_sample_covariance} of the measured shear correlation functions $\est{\xi}_\pm^r(\vartheta)$ in the ellipticity noise-free simulated fields. Furthermore, we calculate estimates for $c^{(\gamma)}_{\pm\pm}$ within the normal approximation and the log-normal approximation, using the fits~\eqref{eq:xi_pm_fits} as inputs for the shear correlation functions.
The estimates from the sample covariance of the $2\times2\,\degt^2$ fields are compared to the corresponding estimates based on the normal and on the log-normal approximation in Fig.~\ref{fig:cosmic_covariance}.

As is already known from earlier studies \citep[e.g.][]{SemboloniEtal2007,SatoEtal2011}, the normal approximation grossly underestimates the covariance $c^{(\gamma)}_{++}$ on scales $\lesssim 10\,\arcmint$. The reason for this failure is that the convergence and shear fields are very `non-Gaussian' on such small scales, where the growth of matter structures is very non-linear.

Or simulations show furthermore that the normal approximation completely fails to reproduce the measured $c^{(\gamma)}_{+-}$ and $c^{(\gamma)}_{--}$ on all scales considered. As Eqs.~\eqref{eq:xi_m_from_xi_kappa}, \eqref{eq:c_pp_to_mm}, and \eqref{eq:c_pp_to_pm} show, the shear correlation $\xi_-$ and the cosmic variance parts $c^{(\gamma)}_{+-}$ and $c^{(\gamma)}_{--}$ at given scales depend on the statistical properties of the convergence field on all smaller scales. These scales, of course, include the non-Gaussian scales $\lesssim 10\,\arcmint$, where the normal assumption fails. 

In contrast to the normal approximation, the log-normal approximation yields estimates of the cosmic variance terms $c^{(\gamma)}_{\pm\pm}$ that are very similar in shape and magnitude to the measured cosmic variance contribution. Differences are noticeable on scales below a few $\arcmint$, where the log-normal approximation overestimates the covariance. Furthermore, the log-normal approximation underestimates the covariances involving $\xi_{-}$ on scales $> 10\,\arcmint$. The same holds for the covariances obtained from the simplified log-normal approximation (not shown), which are almost identical to the covariances from the log-normal approximation.

\subsection{Comparison to empirical fits for the cosmic shear covariance}
\label{sec:comparison_to_fits_for_the_cosmic_shear_covariance}

\begin{figure*}
 \centerline{
 \includegraphics[width=0.33\linewidth]{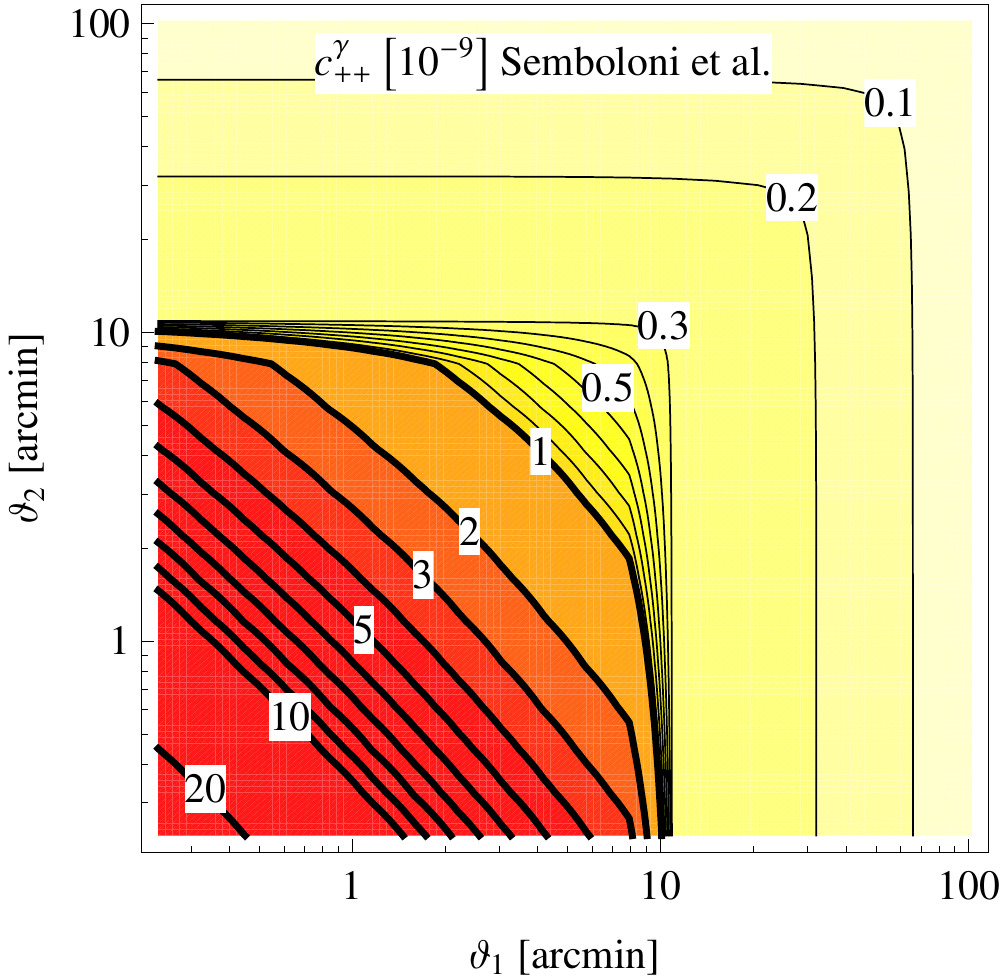}
 \hfill
 \includegraphics[width=0.33\linewidth]{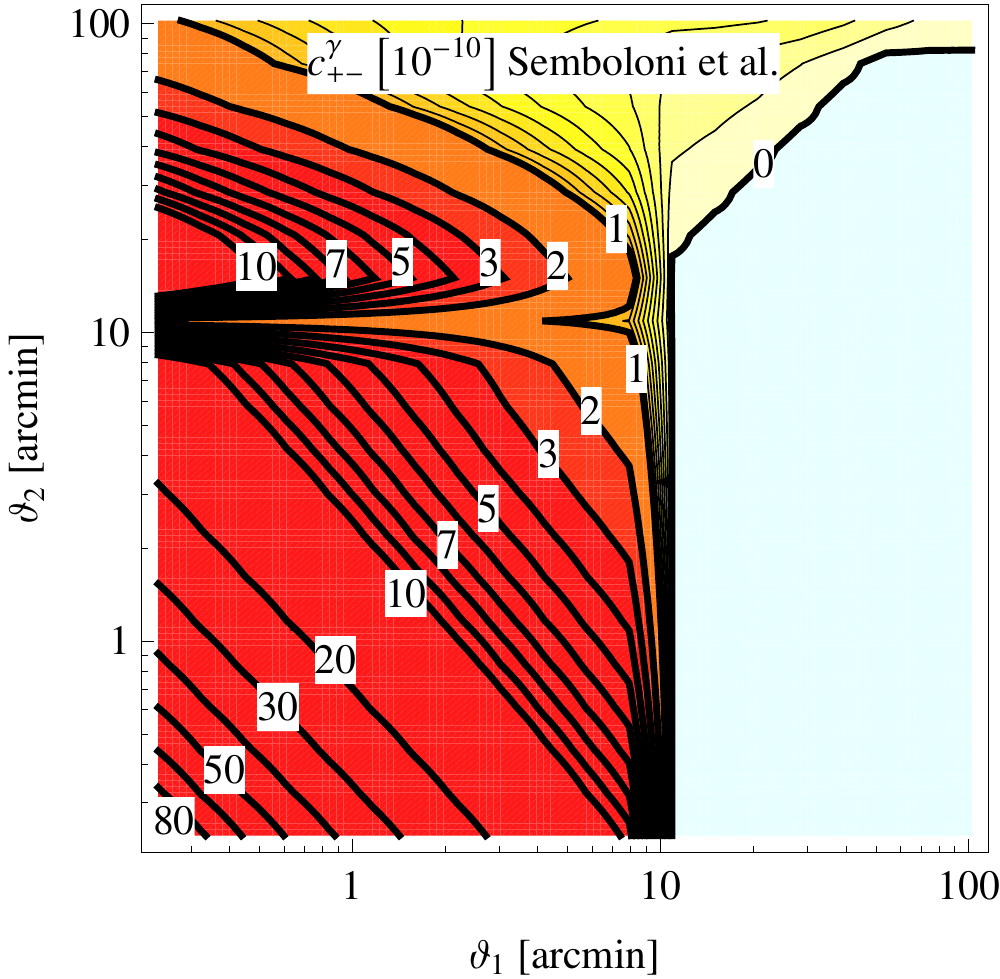}
 \hfill
 \includegraphics[width=0.33\linewidth]{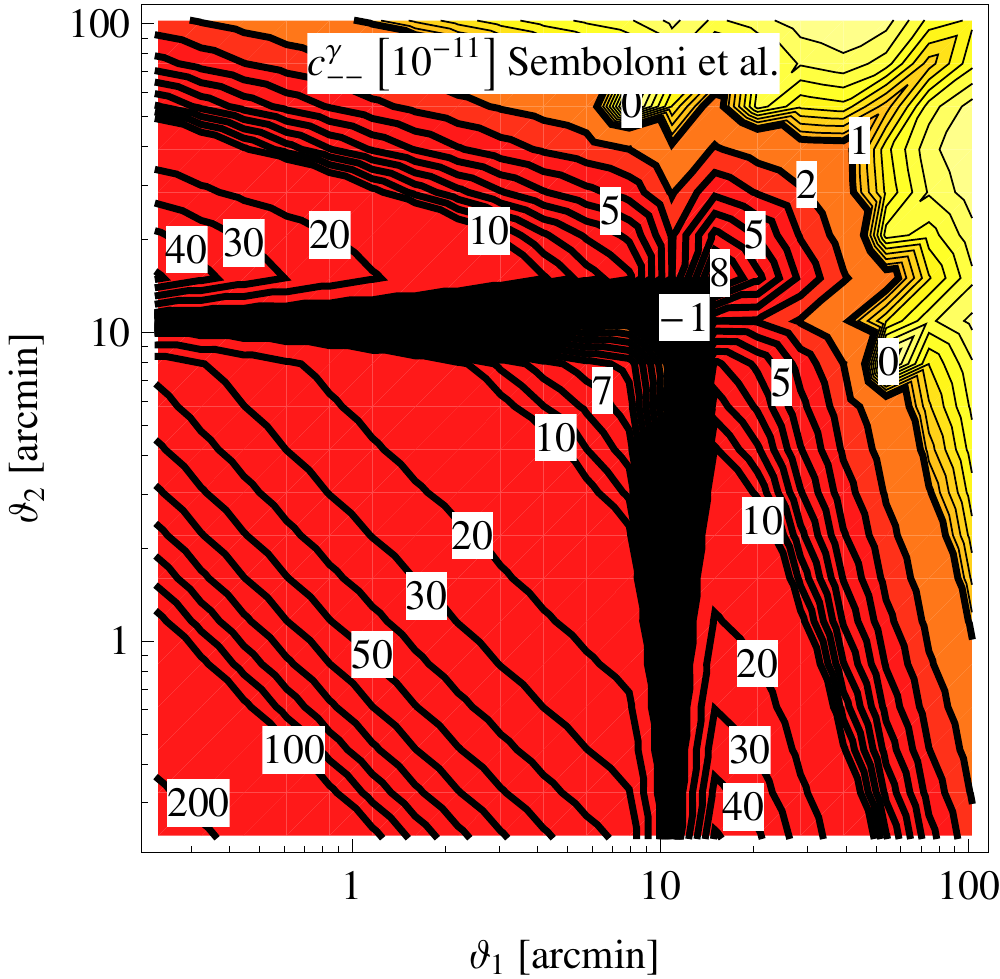}
 }
 \vspace{1em}
 \centerline{
 \includegraphics[width=0.33\linewidth]{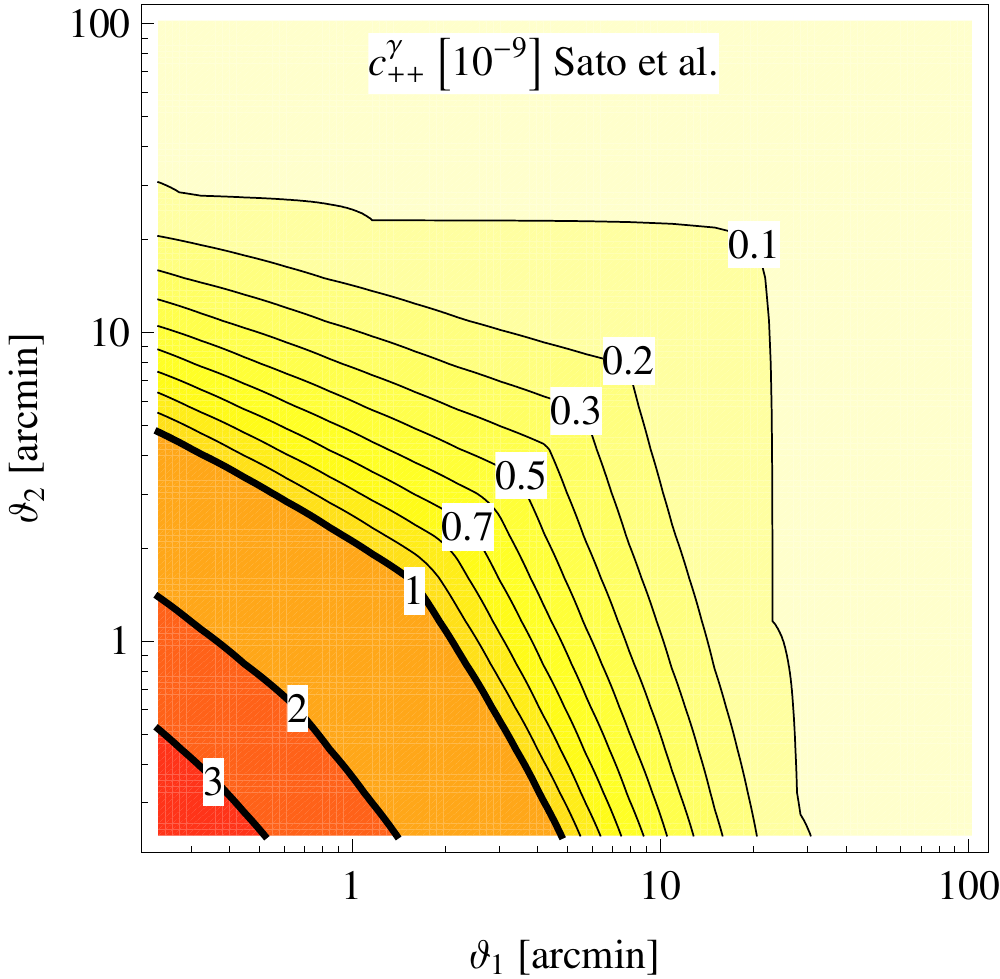}
 \hfill
 \includegraphics[width=0.33\linewidth]{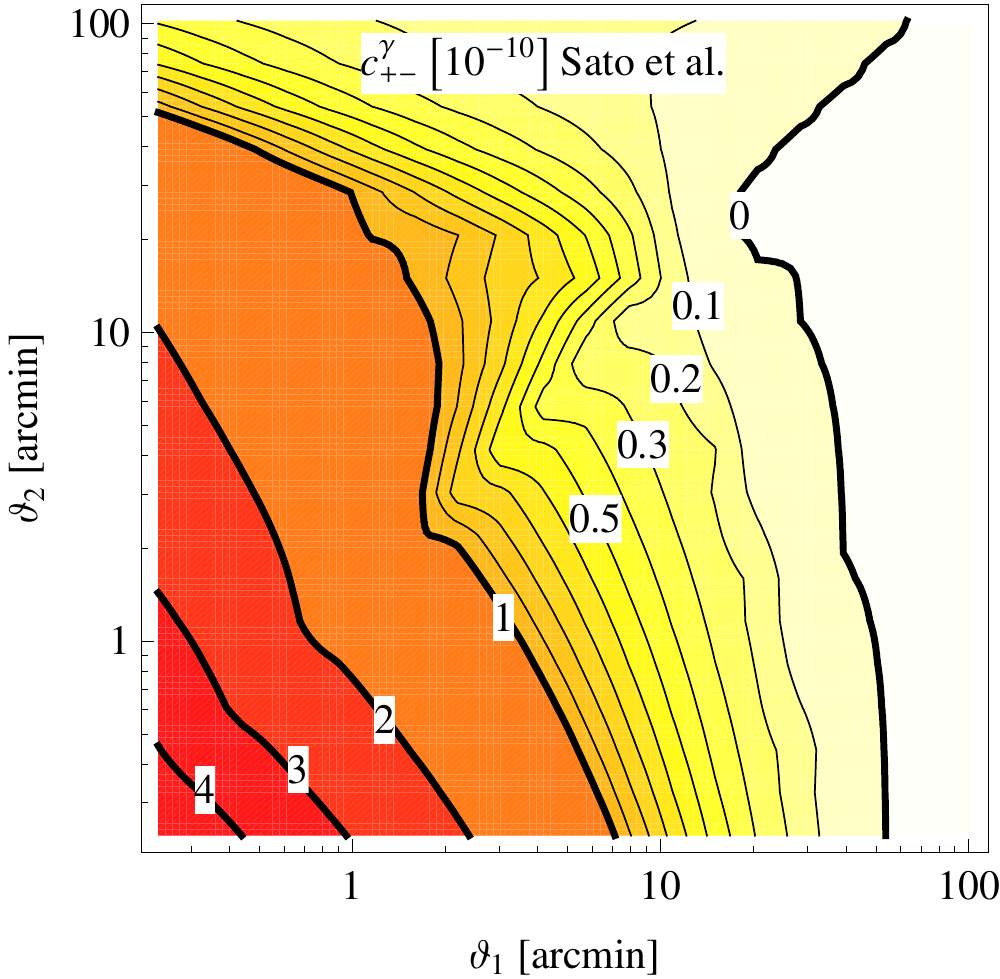}
 \hfill
 \includegraphics[width=0.33\linewidth]{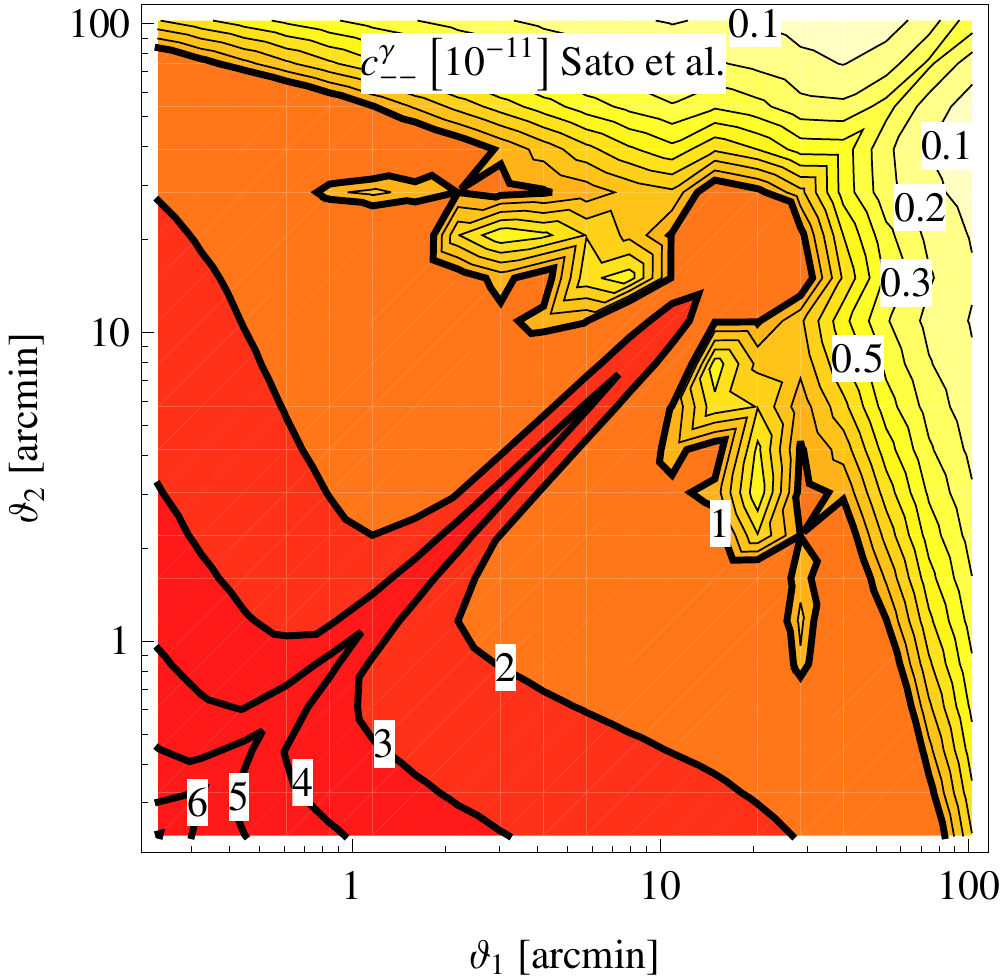}
 }
\caption{
\label{fig:cosmic_covariance_fits}
The cosmic variance contributions $c^{(\gamma)}_{++}(\vartheta_1, \vartheta_2)$ (left column), $c^{(\gamma)}_{+-}(\vartheta_1, \vartheta_2)$ (middle column), and $c^{(\gamma)}_{--}(\vartheta_1, \vartheta_2)$ (right column) as functions of angular separations $\vartheta_1$ and $\vartheta_2$ for $2\times2\,\degt^2$ fields computed from the fit by \citet[][top row]{SemboloniEtal2007} and the fit by \citet[][bottom row]{SatoEtal2011}.
}
\end{figure*}

The failure of the normal approximation on small scales has motivated \citet{SemboloniEtal2007} and \citet{SatoEtal2011} to develop empirical corrections. Figure~\ref{fig:cosmic_covariance_fits} shows the covariances resulting from applying the correction factors they proposed for sources at redshift $z = 1$ to the cosmic variance part $c^{(\gamma)}_{++}$ in the normal approximation.\footnote{
The corrections by \citet{SatoEtal2011} assume explicitly that the normal prediction Eq.~\eqref{eq:c_gamma_normal} is computed with $\rFOV = \infty$ \citep[as in][]{SchneiderEtal2002}. We thus apply their correction factor to the normal prediction computed assuming $\rFOV = \infty$ instead of $\rFOV=\sqrt{\AFOV/\pi}$. Moreover we extend the range of the fit correction $F$ defined in eq. (B1) to all scales where $F>1$ to avoid discontinuities in the covariance.
}
The estimate based on \citet{SemboloniEtal2007} overestimates $c^{(\gamma)}_{++}$ on small scales by a large factor, whereas the corrections according to \citet{SatoEtal2011} yield estimates that underestimate $c^{(\gamma)}_{++}$ by 20-50\%.

\citet{SemboloniEtal2007} and \citet{SatoEtal2011} do not provide corrections for the cosmic variance parts $c^{(\gamma)}_{+-}$ or $c^{(\gamma)}_{--}$. We thus employ Eqs.~ \eqref{eq:c_pp_to_mm} and \eqref{eq:c_pp_to_pm} to compute $c^{(\gamma)}_{+-}$ and $c^{(\gamma)}_{--}$ from the predictions for  $c^{(\gamma)}_{++}$. As  Fig.~\ref{fig:cosmic_covariance_fits} shows in comparison to Fig.~\ref{fig:cosmic_covariance}, the estimates based on \citet{SemboloniEtal2007} fail to reproduce the estimates from the lensing simulations both quantitatively and qualitatively. On most scales, the estimate is too large by orders of magnitude. Around the transition scale between the Gaussian and non-Gaussian regime at $10\,\arcmint$, the estimated covariances show artificially sharp drops. 

The estimates based on \citet{SatoEtal2011} perform better. However, they still overpredict the covariances on small scales by a factor of a few, and underpredict the covariances on large scales. Since the empirical corrections were designed to yield good estimates of $c^{(\gamma)}_{++}$ for the range of scales and cosmologies considered by \citet{SemboloniEtal2007} and \citet{SatoEtal2011}, this indicates that we have used the corrections outside their range of applicability. For the cosmology and scales considered here, the log-normal approximation appears to perform better than the empirical corrections.

\subsection{Positive-semidefiniteness of the cosmic variance contribution}
\label{sec:positive_semidefiniteness_of_the_cosmic_variance_contribution}

In the absence of other noise (e.g. ellipticity noise and other sources not considered here), the cosmic variance part of the cosmic shear covariance represents the full data covariance.  This implies that the actual cosmic variance part of the parameter covariance matrix as well as its components $c^{(\gamma)}_{++}$ and $c^{(\gamma)}_{--}$ are positive-semidefinite regardless of the statistical properties of the convergence field. Positive-semidefiniteness of the data covariance is, moreover, an essential requirement for a sound Bayesian parameter inference employing a quadratic log-likelihood. Thus, a highly desirable property of any approximation to the cosmic variance contribution is that it yields a positive-semidefinite cosmic variance part, so that even in cases of high galaxy densities and low ellipticity noise, the full data covariance remains positive-semidefinite.

In the cases studied here, we encounter a serious problem of the corrections suggested by \citet{SemboloniEtal2007} and \citet{SatoEtal2011}: Both fail to yield positive-semidefinite covariance parts $c^{(\gamma)}_{++}$ and $c^{(\gamma)}_{--}$. As a consequence, the cosmic variance part of the data covariance matrix is not positive-semidefinite, but has significantly negative eigenvalues. This may lead to a non-positive data covariance matrix $\matrb{C}_\text{d}$, a non-positive parameter covariance matrix $\matrb{C}_\pi$, and to a breakdown of the parameter estimation.
 
Apparently, the empirical corrections do not contain any mechanism ensuring positive-semidefiniteness of the data covariance. Indeed, it seems difficult to devise a valid covariance matrix of correlation functions without strong guidance from, e.g., analytic models.
 
In contrast, both the normal and the log-normal approximation `should' yield positive-semidefinite covariance matrices by construction,
since the resulting matrices are covariance matrices of correlation functions of a (log-)normal field.\footnote{
This certainly holds if the cosmic shear correlations assumed in the computation of the covariances are valid correlation functions of a \mbox{(log-)}normal field, which we checked for our calculations.
}
Within the simplified log-normal approximation, the cosmic variance part of the covariance matrix is a sum of the cosmic variance part resulting from the normal approximation and another positive-semidefinite matrix, and thus is also expected to be positive-semidefinite. 
However, due to the employed approximations for finite fields of view and numerical inaccuracies, the normal and (simplified) log-normal approximations may also yield indefinite cosmic variance parts with negative eigenvalues. These negative eigenvalues are found to be essentially consistent with zero, indicating the presence of linear constraints on the data vector [e.g. stemming from the relations \eqref{eq:xi_p_from_xi_kappa} and \eqref{eq:xi_m_from_xi_kappa}], and of much smaller magnitude than the negative eigenvalues encountered when using the empirical fits. Setting the negative eigenvalues to zero `by hand' does not change the inferred posterior distributions of the parameters significantly for the case of the normal and log-normal approximation, as shown in the following section.

\subsection{Error estimates for cosmological parameters from cosmic shear}
\label{sec:results_for_parameter_errors}

\begin{figure*}
\centerline{
\includegraphics[width=0.333\linewidth]{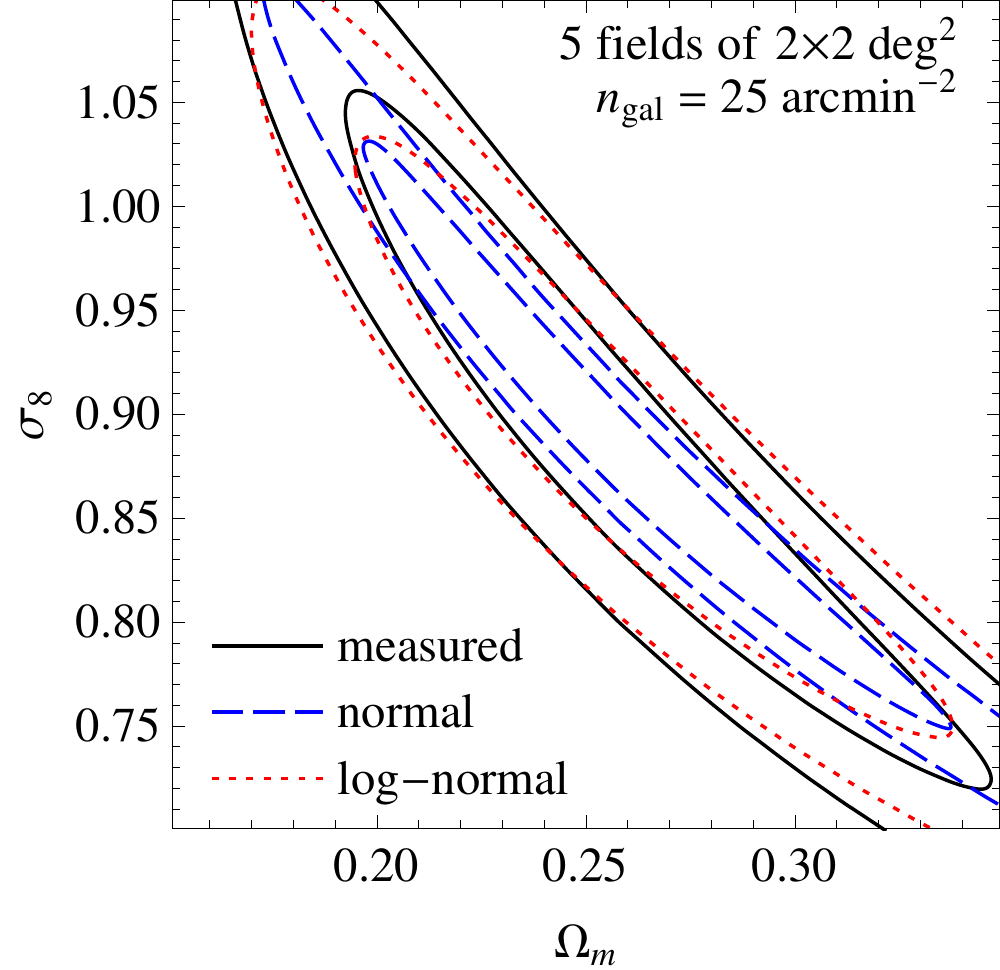}
\includegraphics[width=0.333\linewidth]{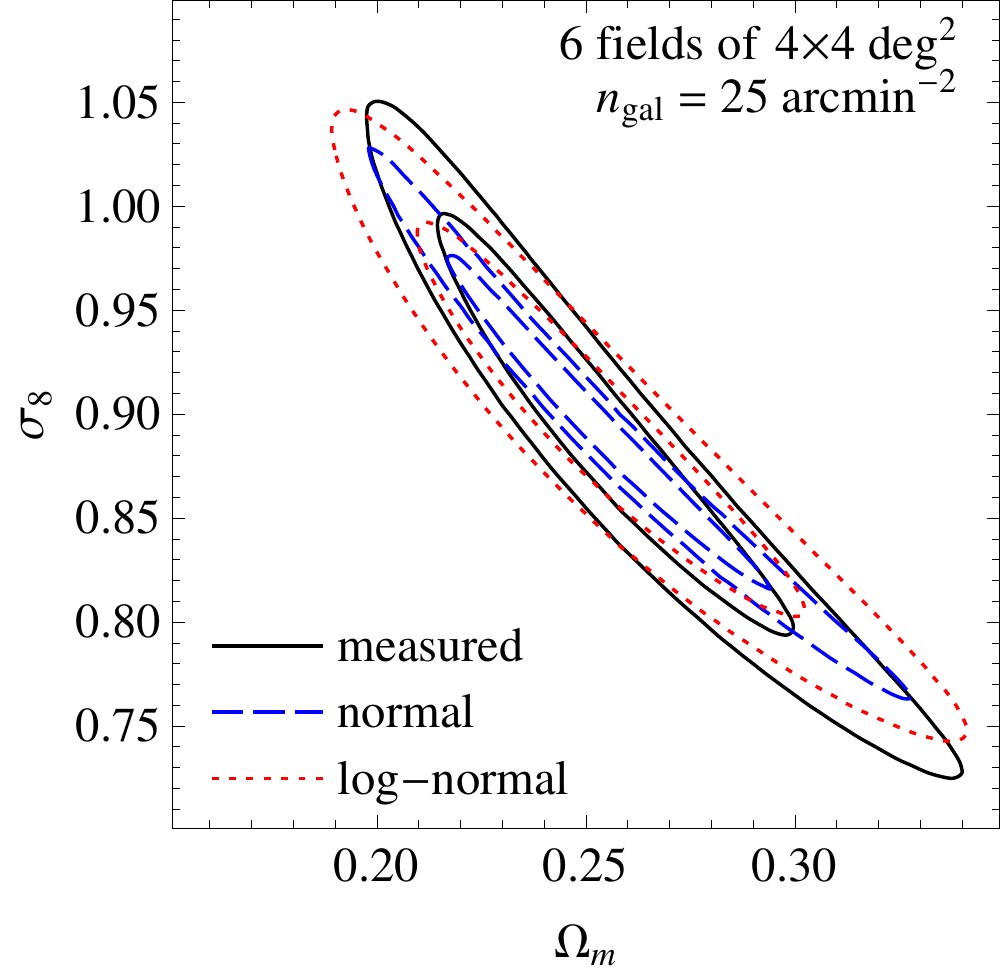}
\includegraphics[width=0.333\linewidth]{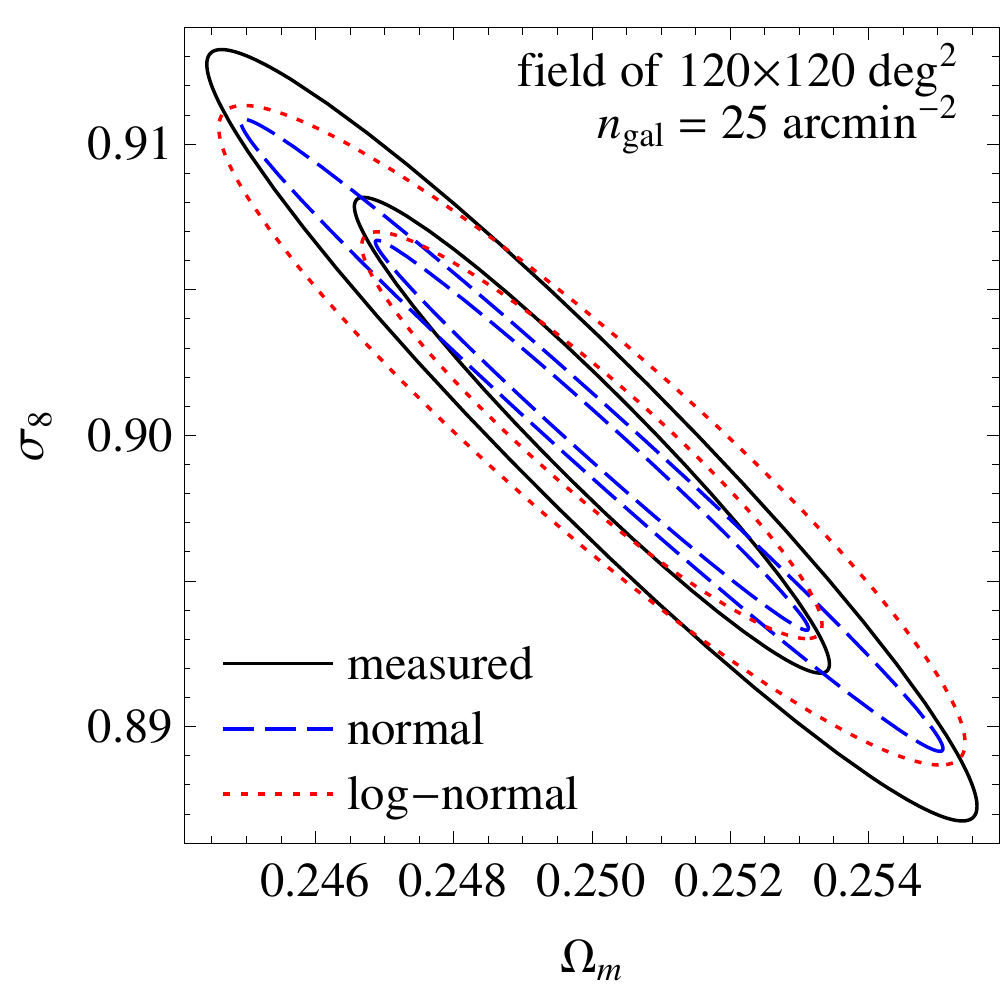}
}
\centerline{
\includegraphics[width=0.333\linewidth]{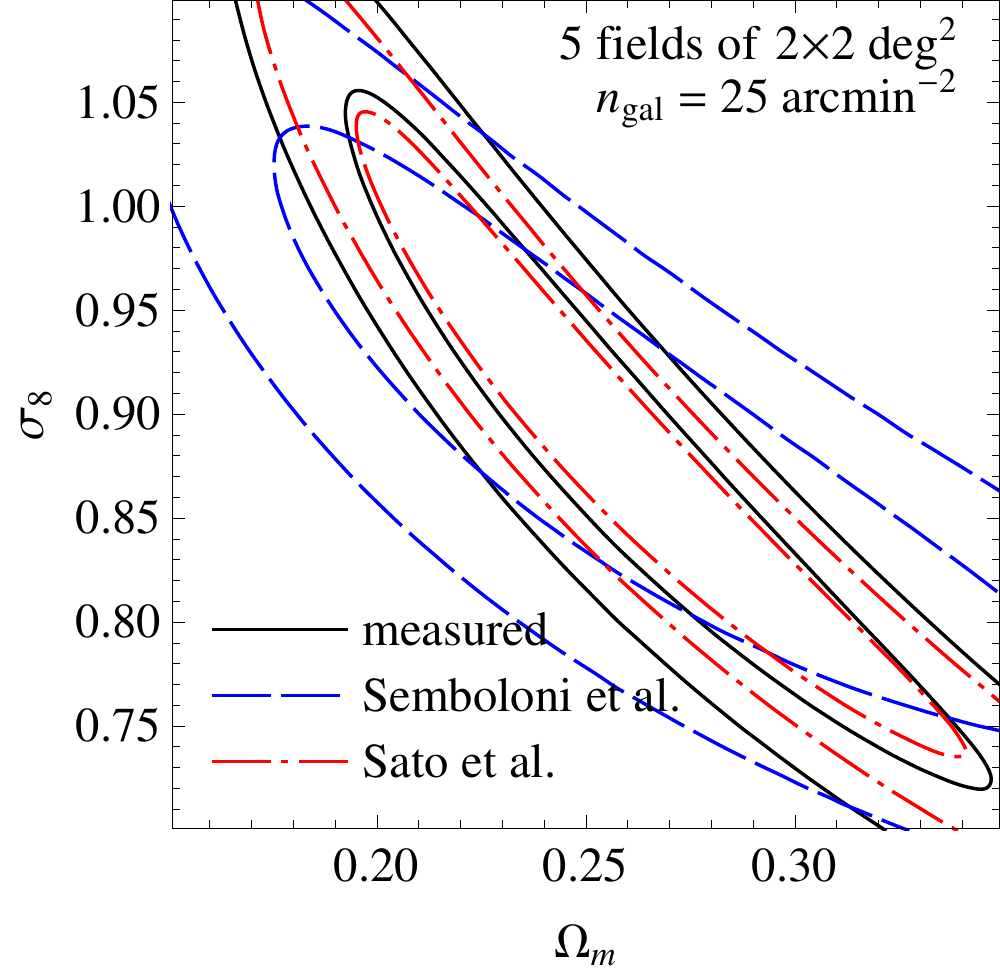}
\includegraphics[width=0.333\linewidth]{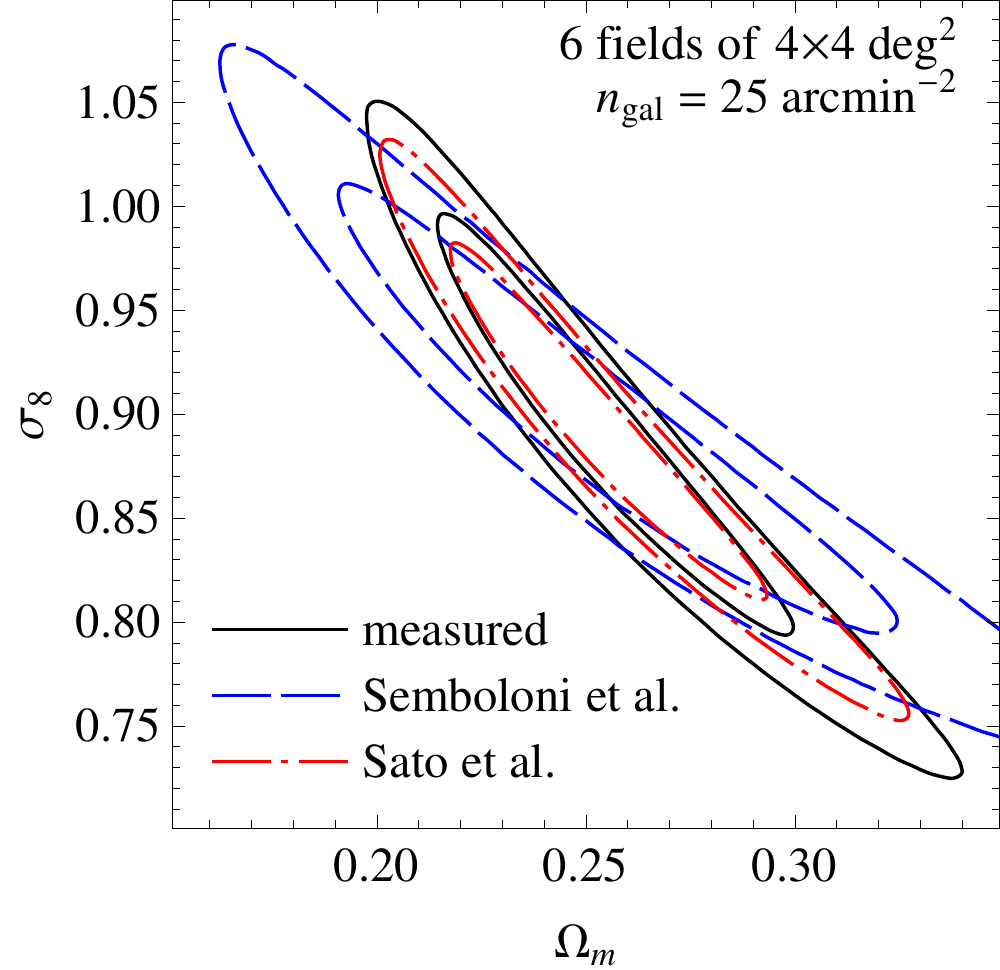}
\includegraphics[width=0.333\linewidth]{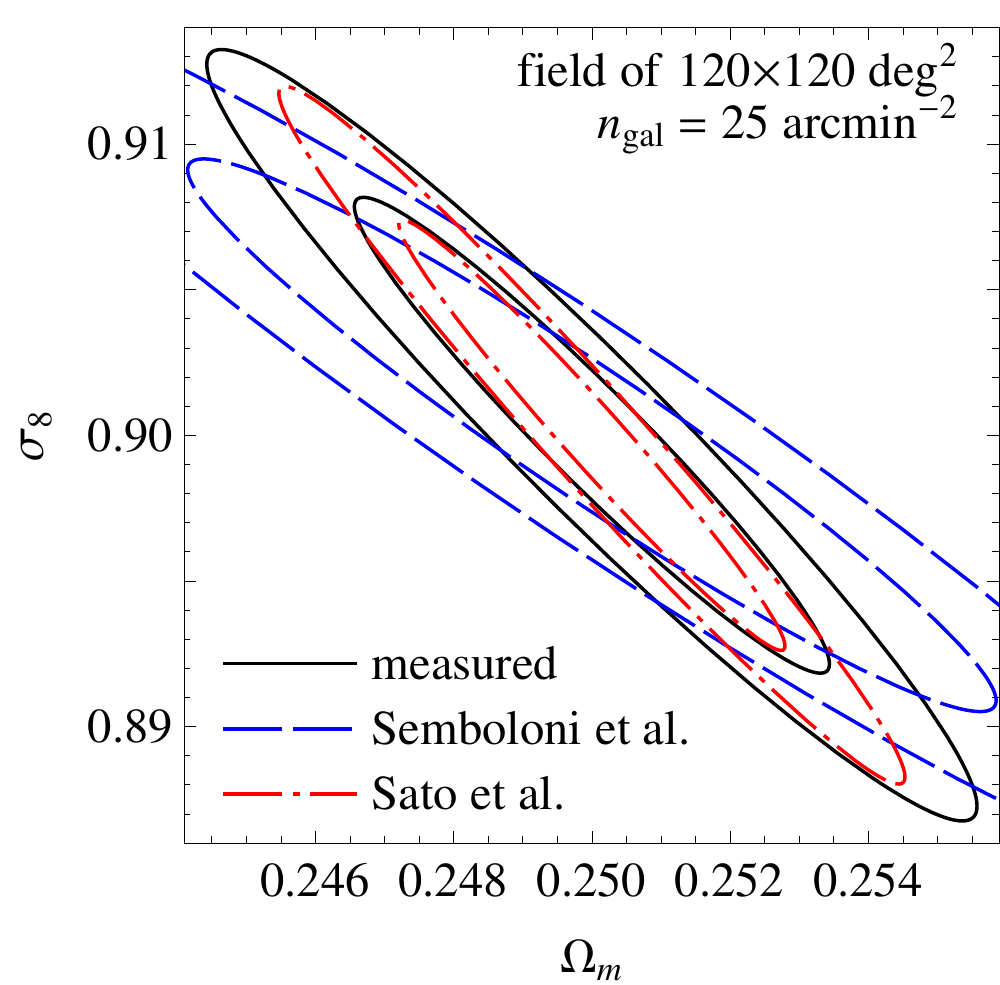}
}
\caption{
\label{fig:likelihood_contours}
Joint posterior distributions of the mean matter density $\Omega_\mathrm{m}$ and the matter power spectrum normalization $\sigma_8$ of a flat $\Lambda$CDM universe inferred from a cosmic shear survey with $\ngal = 25\,\arcmint^{-2}$ (assuming flat priors for $\Omega_\mathrm{m}$ and $\sigma_8$, and all other cosmological parameters known). Shown are the 68\% (inner) and 95\% (outer) confidence contours for a small survey in 5 fields of $2\times2\,\degt^2$ (left panels), a survey in 6 fields of $4\times4\,\degt^2$ (middle panels), and a large survey in a $120\times120\,\degt^2$ field (right panels) obtained when using the covariance matrix measured from our lensing simulations (solid lines), the normal approximation (top, dashed lines), the log-normal approximation (top, dotted lines), the empirical corrections to the normal approximation by \citet[][bottom, long dashed lines]{SemboloniEtal2007}, and the corrections by \citet[][bottom, long dash-dotted lines]{SatoEtal2011}.
}
\end{figure*}

To study how the approximations to the cosmic shear covariance affect the inferred accuracy of cosmological-parameter estimation, we consider the scenario that cosmic shear data is used to constrain the parameters $\Omega_\mathrm{m}$ and $\sigma_8$ of a flat $\Lambda$CDM cosmology. To keep the discussion simple, we choose priors that are constant for $\Omega_\mathrm{m} \in [0.15, 0.35]$ and $\sigma_8 \in [0.7,1.1]$ and zero otherwise. Furthermore, we assume the other cosmological parameters known to take the fiducial values: $h=0.73$, $\Omega_\mathrm{b}=0.045$, and $n_\mathrm{s}=1$.

We assume the cosmic shear correlation functions $\xi_+(\vartheta)$ and $\xi_-(\vartheta)$ are measured in 20 logarithmically spaced bins in the range $0.2\,\arcmint \leq \vartheta \leq 120\,\arcmint$. The model predictions $\vect{\mu}(\Omega_\mathrm{m}, \sigma_8)$ for the shear correlations are computed using \textsc{nicaea} \citep[][]{Nicaea}. The data vector $\vect{d}$ containing the measured shear correlations is assumed to coincide with the predicted values for the fiducial cosmology with $\Omega_\mathrm{m} = 0.25$ and $\sigma_8 = 0.9$.

As first example, we study the parameter constraints obtained from a small survey with $\ngal = 25\,\arcmint^{-2}$ in 5 fields of $2\times2\,\degt^2$ (i.e. a total area of  $20\,\degt^2$) similar to the Deep Lens Survey \citep[see, e.g.,][]{KuboEtal2009}. We estimate the inverse covariance matrix directly from the sample covariance of the simulated noisy shear maps using Eq.~\eqref{eq:def_csdv_inv_cov_matrix_from_sample_cov}, and combine the estimate with the \textsc{nicaea} predictions to compute the likelihood and the posterior.

In the left panels of Fig.~\ref{fig:likelihood_contours}, the resulting confidence contours in the $(\Omega_\mathrm{m}, \sigma_8)$-plane are compared to the contours obtained from the various approximations to the cosmic shear covariance. The normal approximation substantially underestimates the size of the confidence regions, in particular in the direction perpendicular to the major degeneracy. In contrast, the contours based on the log-normal approximation are very similar to the contours based on the measured covariances. The same holds for the simplified log-normal approximation (whose contours are not shown, since they are almost identical to those from the log-normal approximation). Only a slight tilt of the contours based on the (simplified) log-normal approximation against the contours based on the ray-tracing is visible.
The corrections proposed by \citet[][]{SemboloniEtal2007} yield confidence regions that are much larger than the regions inferred from the simulations, and are strongly tilted. The confidence regions based on the corrections proposed by \citet[][]{SatoEtal2011} are noticeably smaller than the regions inferred from the simulations.

As second example, we consider a survey with $\ngal = 25\,\arcmint^{-2}$ in 6 fields of $4\times4\,\degt^2$ (i.e. a total area of  $96\,\degt^2$) similar to the Canada-France-Hawaii Telescope Wide Synoptic Legacy Survey \citep[][]{HoekstraEtal2006}. The resulting confidence contours are shown in the middle panels of Fig.~\ref{fig:likelihood_contours}. As in the first case, the normal approximation underestimates the size of the confidence regions, whereas the (simplified) log-normal approximation yields confidence regions similar to those inferred from our lensing simulations.

As can be seen in the right panels of Fig.~\ref{fig:likelihood_contours}, the normal approximation substantially underestimates the size of the confidence regions also for very large surveys like the planned  Large Synoptic Survey Telescope (LSST) survey \citep[][]{AbellEtal2009_LSST_book} or the \satellitename{Euclid} survey \citep[][]{RefregierEtal2010_Euclid_book}. Even for such large surveys, much cosmological information is contained in the shear correlations on small scales, which are not well described by the normal approximation. The (simplified) log-normal approximation, describing the small scales better, yields confidence regions in much better agreement with the regions inferred from our lensing simulations (using a scaled version of the cosmic covariance measured in the $4\times4\,\degt^2$ fields in combination with analytic expressions for the mixed and ellipticity noise). The contours based on \citet[][]{SemboloniEtal2007} appear much too large and strongly tilted, whereas the contours based on \citet[][]{SatoEtal2011} appear too small.

\begin{figure*}
\centerline{
\includegraphics[width=0.333\linewidth]{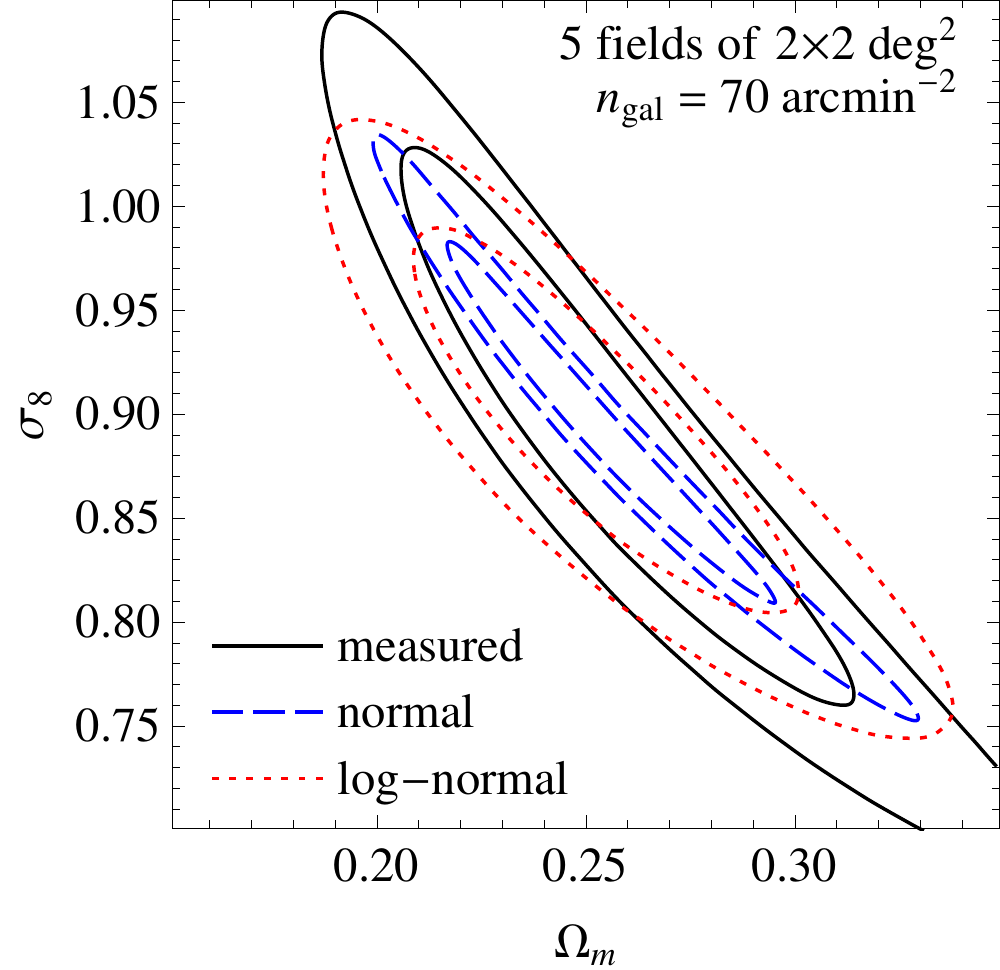}
\hspace{1em}
\includegraphics[width=0.333\linewidth]{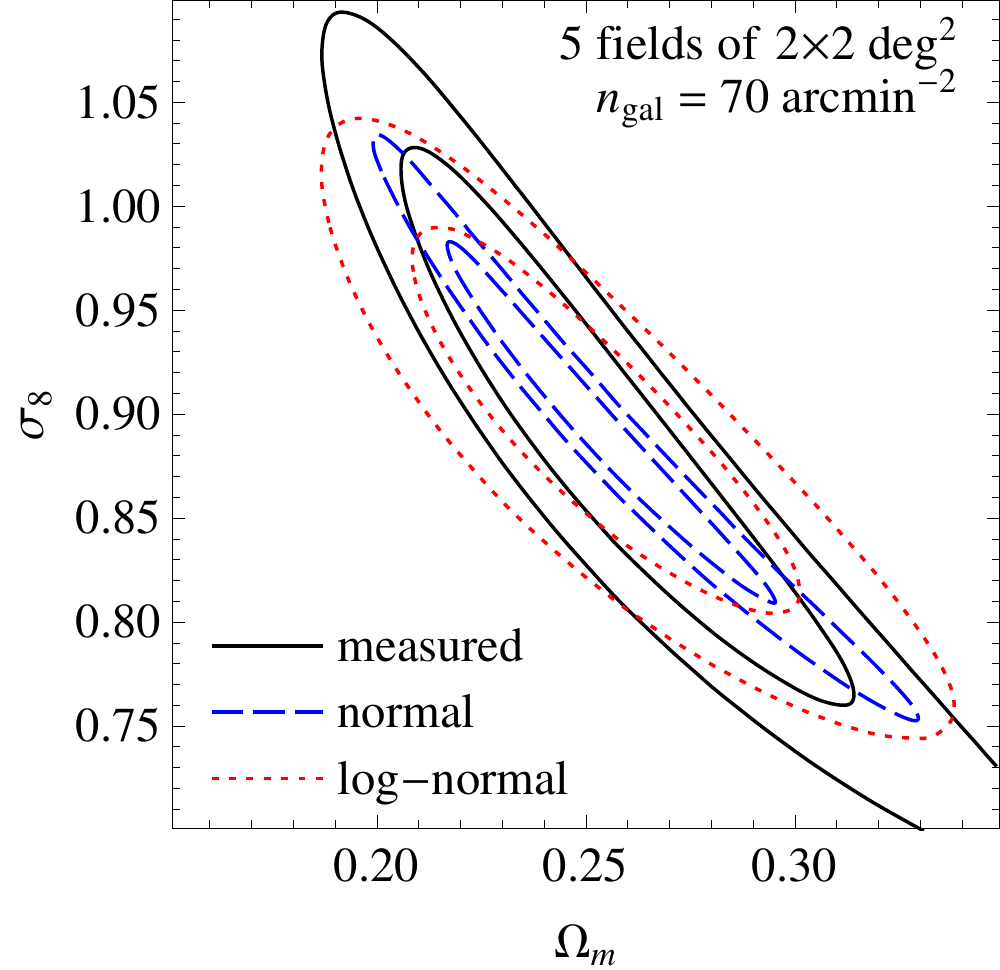}
}
\centerline{
\includegraphics[width=0.333\linewidth]{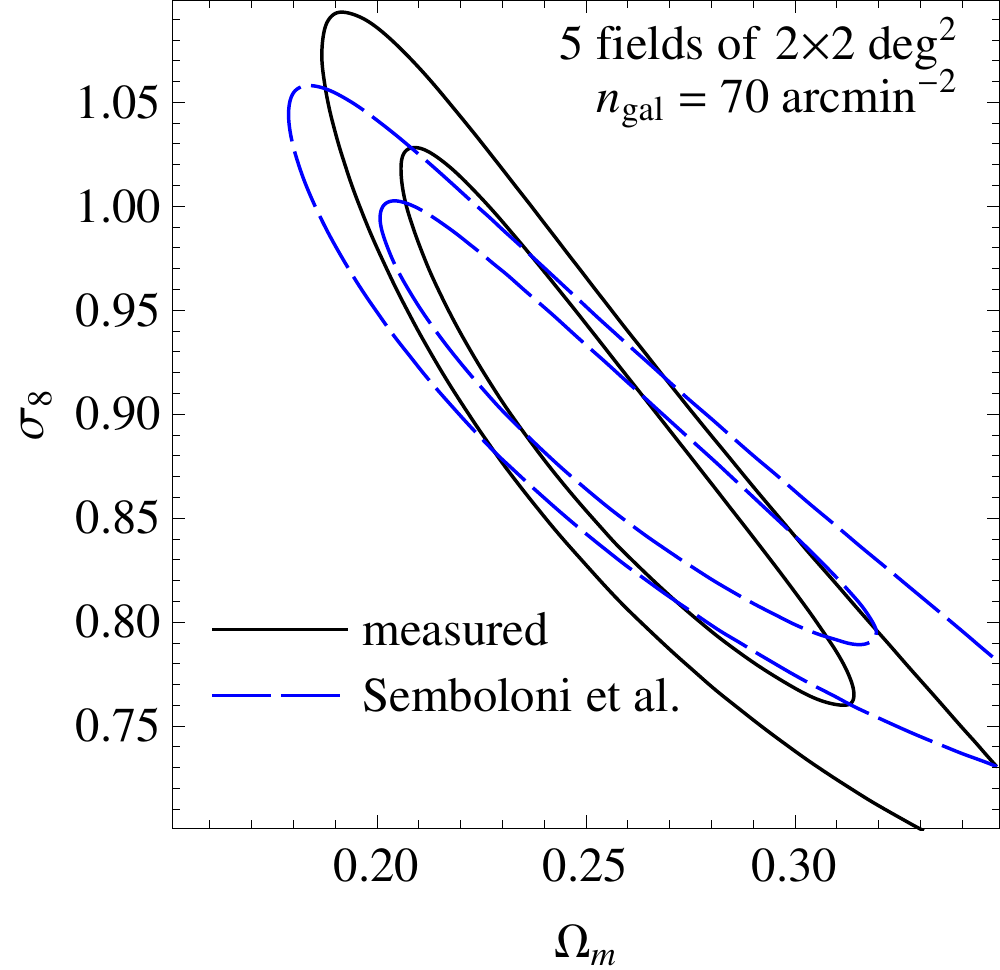}
\hspace{1em}
\includegraphics[width=0.333\linewidth]{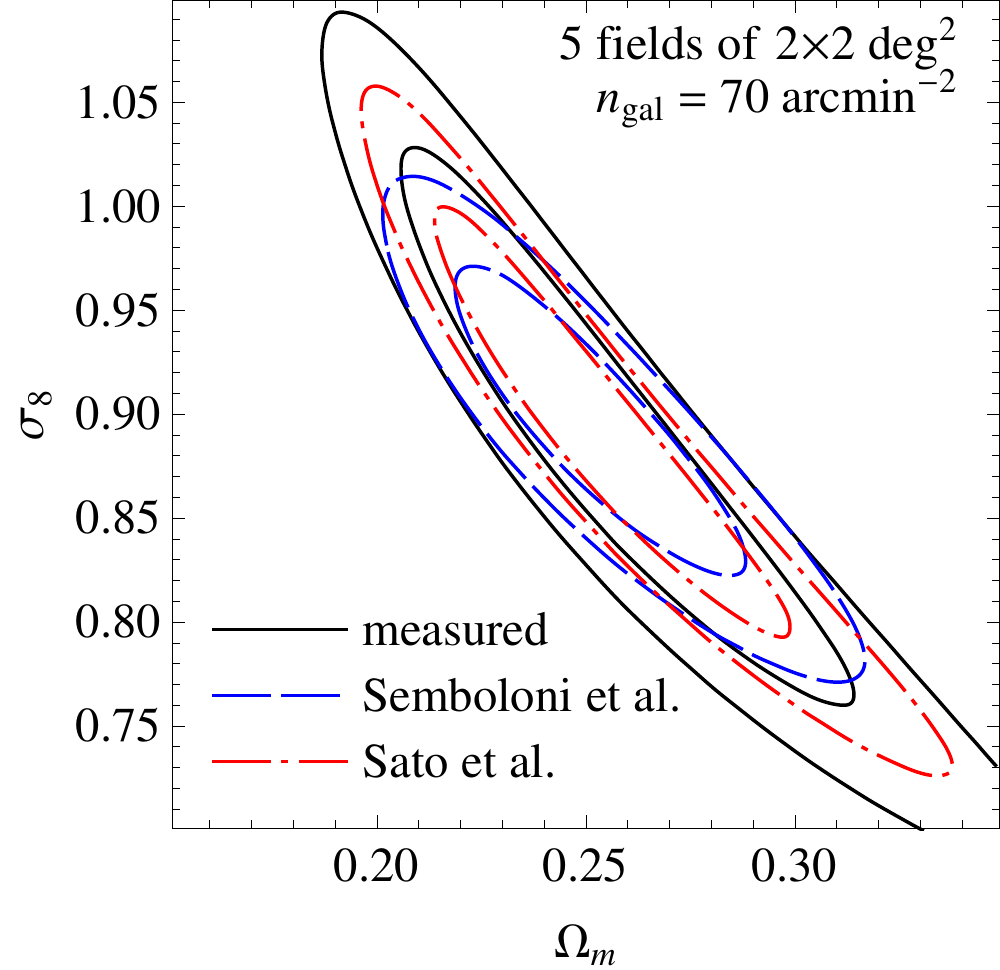}
}
\caption{
\label{fig:likelihood_contours_hi_n}
68\% (inner) and 95\% (outer) confidence contours of the joint posterior distributions of the mean matter density $\Omega_\mathrm{m}$ and the matter power spectrum normalization $\sigma_8$ of a flat $\Lambda$CDM universe inferred from a cosmic shear survey with $\ngal = 70\,\arcmint^{-2}$ in 5 fields of $2\times2\,\degt^2$, obtained using the covariance from our lensing simulations (solid lines), the normal approximation (top, dashed lines), the log-normal approximation (top, dotted lines), the empirical corrections to the normal approximation by \citet[][bottom, long dashed lines]{SemboloniEtal2007}, and the corrections by \citet[][bottom, long dash-dotted lines]{SatoEtal2011}. For the left panels, the cosmic covariance parts of the data covariance matrices have be used as given by the approximations. For the right panels, the cosmic variance parts have been modified where needed to ensure positive-semidefiniteness (see text for details).
}
\end{figure*}

In the cases just discussed, both the cosmic variance and the intrinsic ellipticity noise are important sources of error for the inferred parameters. To investigate the case that the cosmic variance is the dominant source of parameter uncertainty, we consider again a survey in 5 fields of $2\times2\,\degt^2$, but with a higher galaxy density $\ngal = 70\,\arcmint^{-2}$.
The left panels of Fig.~\ref{fig:likelihood_contours_hi_n} compare confidence contours estimated from the simulations and the contours obtained from the approximations. The normal approximation underestimates the credible parameter regions even more than in the afore discussed cases. The (simplified) log-normal approximation still yields confidence regions comparable in size to those estimated from the simulations, but the shape difference is more apparent (this might become problematic in cases where the compatibility of different cosmological experiments is judged by the overlap of their parameter confidence regions). The confidence regions based on the fit by \citet[][]{SemboloniEtal2007} are of similar size, but strongly tilted.

When using the fits proposed by \citet[][]{SatoEtal2011}, we do not obtain any reasonable confidence contours for the posterior. Both data and parameter covariance matrices are indefinite, and the likelihood features a saddle point at the fiducial parameter values instead of a maximum. The problem can be traced back to the cosmic variance part of the parameter covariance matrix, which is indefinite.

One may approach the problem of negative eigenvalues in the parameter covariance matrix stemming from an indefinite cosmic variance part as follows: Using its eigensystem decomposition, the cosmic variance part of the covariance matrix can be specified by an orthogonal set of eigenvectors, which describe the principal directions of scatter in the data vector due to cosmic variance, and the corresponding eigenvalues, which quantify the scatter in these directions. Directions with vanishing scatter indicate linear constraints, which confine all possible data vectors to a lower-dimensional subspace of the full data vector space (in the absence of other sources of scatter). Directions with negative eigenvalues, i.e. `negative scatter', do not make sense. One may presume, however, these negative eigenvalues stem from numerical inaccuracies in the employed method for computing the cosmic variance, and rather indicate directions with very small positive or vanishing scatter.

To ensure a positive-semidefinite data covariance matrix from the various approximations, we modify the cosmic variance part by replacing any of its negative eigenvalues in its eigensystem decomposition by zero.
The resulting confidence contours are shown in the right panels of Fig.~\ref{fig:likelihood_contours_hi_n}. There is no visible difference in the contours between the left and right panel for the normal and log-normal approximation, even though their cosmic variance parts have a few negative eigenvalues (which are of tiny magnitude). In contrast, the confidence contours based on the fit by \citet[][]{SemboloniEtal2007} change drastically if the data covariance matrix is modified in the described way. Only after applying the modification, we obtain confidence contours for the correction suggested by \citet[][]{SatoEtal2011}. The parameter confidence regions computed from the modified empirical corrections are much smaller than the confidence regions estimated from the simulations.

\section{Summary and discussion}
\label{sec:summary}

Accurate estimates for the covariance of cosmic shear correlation functions are essential for reliable estimates of the errors on the cosmological parameters inferred from cosmic shear surveys. In this work, we developed two approximations to the cosmic shear covariance based on the statistics of log-normal random fields. We used numerical simulations of cosmic shear surveys to assess the performance of this log-normal and simplified log-normal approximation and to compare them to the widely used normal approximation to the cosmic shear covariance \citep[][]{SchneiderEtal2002}.

We find that the normal approximation to the cosmic shear covariance substantially underestimates the inferred parameter confidence regions, in particular for surveys with small fields of view and large galaxy densities, but also for very large `all sky' surveys (like the proposed \satellitename{Euclid} or LSST lensing surveys). The log-normal approximation yields much more realistic confidence regions at the price of slightly more complicated expressions for the cosmic variance contribution to the cosmic shear covariance. In contrast, the simplified log-normal approximation is as simple as the normal approximation, yet appears as accurate as the log-normal approximation.
Moreover, the simplified log-normal approximation is simpler than several proposed approximations based on halo models \citep[e.g.][]{TakadaJain2009, PielorzEtal2010}. The simplified log-normal approximation is also more general than approximations based on empirical fits to numerical simulations \citep[][]{SemboloniEtal2007,SatoEtal2011}.

A particular advantage of the normal and (simplified) log-normal approximation over the empirical fits suggested by \citet{SemboloniEtal2007} and \citet{SatoEtal2011} is that the former yield positive-semidefinite data covariance matrices by construction, whereas the latter may fail to do so. A positive-semidefinite data covariance matrix is, however, essential for a sound parameter estimation employing a quadratic log-likelihood.

A disadvantage of the (simplified) log-normal approximation in comparison to the normal approximation is the need for providing a minimum-convergence parameter $\kappa_0$. The value of this parameter depends on the assumed cosmology as well as the source redshift distribution of the weak-lensing survey. However, computing $\kappa_0$ does not require more effort than computing the expected shear correlation functions (which are needed in any case). Estimates for $\kappa_0$ from simulations, for example, require much fewer realisations than estimates for the full cosmic shear covariance.

Because of its comparable simplicity and much better accuracy, one should consider the simplified log-normal approximation in favour of the normal approximation, e.g. for the parameter estimation from current surveys, or for parameter error forecasts for future surveys. For the analysis of observed cosmic shear data from future large and deep surveys, however, better descriptions of the cosmic shear covariance will be required.

In this work, we concentrated on the covariance of the cosmic shear correlations $\xi_{\pm}$ and the resulting errors on cosmological parameters.
In future work, one should adapt the (simplified) log-normal approximation to derive covariances for other lensing two-point statistics \citep[e.g the COSEBIs introduced by][]{SchneiderEiflerKrause2010}. The (simplified) log-normal approximation could also be generalized to provide covariances for tomographic shear surveys. Furthermore, one could consider the approximation of a log-normal convergence field for predictions of higher-order shear correlations and their covariances. 

One should also investigate cosmic shear covariances for convergence fields that are more general transformations of normal random random fields \citep[as considered, e.g., by][]{DasOstriker2006,JoachimiTaylorKiessling2011_tmp}. Such work should also take into account the non-Gaussianity and cosmology-dependence of the cosmic shear likelihood \citep[][]{EiflerSchneiderHartlap2009,HartlapEtal2009,SchneiderHartlap2009,SatoSatoIchikiTakeuchi2011,JoachimiTaylor2011}. Finally, one should investigate further to what extent the matter density field or the convergence field can be described by a transformed normal random field \citep[see, e.g,][]{NeyrinckSzapudiSzalay2009,SeoEtal2011,YuEtal2011,JoachimiTaylorKiessling2011_tmp} and what the physical reasons behind the successes and limits of such a description are.

\begin{acknowledgements}
This work was supported by the DFG within the Priority Programme 1177 under the projects SCHN 342/6 and WH 6/3 and the Transregional Collaborative Research Centre TRR 33 ``The Dark Universe''.
\end{acknowledgements}



\onecolumn

\appendix

\section{Normal random fields}
\label{sec:normal_fields}

Consider a non-degenerate homogeneous and isotropic normal random field $n$ in $D$ dimensions.\footnote{
Readers interested in a rigorous definition of a normal random field should consult the standard literature on measure theory.}
This implies that the probability density function (pdf) of the field values $n=n(\vect{x})$ at any single point $\vect{x}\in\R^D$ is given by
\begin{equation}
\label{eq:normal_pdf}
  p_n(n;\mu,\sigma) = \frac{1}{\sqrt{2\pi} \sigma}\exp\left[-\frac{\left(n - \mu\right)^2}{2\sigma^2}\right],
\end{equation}
where $\mu$ and $\sigma^2$ denote the position-independent mean and variance of the random variable $n(\vect{x})$.
Moreover, the joint pdf of the field values $\vect{n}=\transposed{(n_1,\ldots,n_N)}=\transposed{\big(n(\vect{x}_1),\ldots,n(\vect{x}_N)\big)}$ at a set of $N$ mutually distinct points $\vect{x}_1,\ldots,\vect{x}_N$ is given by the pdf of a non-degenerate multivariate normal distribution:
\begin{equation}
\label{eq:normal_field_multivariate_pdf}
 p_{\vect{n}} \big(\vect{n};\vect{\mu},\matrb{\Sigma}\big) =
 \frac{1}{(2\pi)^{N/2}\sqrt{\det\matrb{\Sigma}}} \exp\left[-\frac{1}{2}\transposed{(\vect{n} - \vect{\mu})} \matrb{\Sigma}^{-1}(\vect{n} - \vect{\mu}) \right]
.
\end{equation} 
Here, $\vect{\mu}=\EV{\vect{n}}=\mu\vect{1}$ with $\vect{1}=\transposed{(1,\ldots,1)}$ in accordance with the requirement of homogeneity. The covariance matrix $\matrb{\Sigma}$ is a positive-definite real symmetric matrix, whose elements $\matrb{\Sigma}_{ij}$ are given by the two-point correlations $\xi_{n,ij}$ of the field $n$. Since the field is homogeneous and isotropic, the two-point correlation $\xi_{\vect{n},ij}$ of field values at points $\vect{x}_i$ and $\vect{x}_j$ depends only on their separation, i.e.
\begin{equation}
\matrb{\Sigma}_{ij} = \xi_{n,ij} = \xi_n(|\vect{x}_i - \vect{x}_j|),
\end{equation}
where $\xi_n(\vartheta)$ denotes the correlation function of the field $n$ at separation $\vartheta$. In particular, the diagonal elements of $\matrb{\Sigma}$ are given by the variance $\sigma^2$, i.e. $\matrb{\Sigma}_{ii} = \xi(0) = \sigma^2$.

Using the joint pdf~\eqref{eq:normal_field_multivariate_pdf}, the expectation value $\EV{f(\vect{n})}$ of a function $f(\vect{n})$ depending on the field values $\vect{n}=\transposed{\big(n(\vect{x}_1),\ldots,n(\vect{x}_N)\big)}$ at $N$ positions $\vect{x}_1,\ldots,\vect{x}_N$ can be written as
\begin{equation}
 \EV{f(\vect{n})} = 
	 \int_{\R^N}\!\!\diff[N]{\vect{n}}\,
\frac{1}{(2\pi)^{N/2}\sqrt{\det\matrb{\Sigma}}} \exp\left[
-\frac{1}{2}\transposed{(\vect{n} - \vect{\mu})} \matrb{\Sigma}^{-1}(\vect{n} - \vect{\mu})\right] f(\vect{n})
.	
\end{equation}

The moment-generating function for the field values $\vect{n}=\transposed{(n_1,\ldots,n_N)}$ at $N$ mutually distinct points $\vect{x}_1,\ldots,\vect{x}_N$ is defined as the map 
\begin{equation}
	M_{\vect{n}}: \R^N \to \R : \vect{t} \mapsto M_{\vect{n}}( \vect{t} ) = \EV{\exp(\transposed{\vect{t}} \vect{n})}.
\end{equation}
For a normal random field,
\begin{multline}
M_{\vect{n}}( \vect{t} ) = 
  \EV{\exp(\transposed{\vect{t}} \vect{n})}
= 
	 \int_{\R^N}\!\!\diff[N]{\vect{n}}\,
  \frac{1}{(2\pi)^{N/2}\sqrt{\det\matrb{\Sigma}}} \exp\left[
  -\frac{1}{2}\transposed{(\vect{n} - \vect{\mu})} \matrb{\Sigma}^{-1}(\vect{n} - \vect{\mu})
  + \transposed{\vect{t}} \vect{n} \right]
\\
=
  \int_{\R^N}\!\!\diff[N]{\vect{n}}\,
  \frac{1}{(2\pi)^{N/2}\sqrt{\det\matrb{\Sigma}}} 
  \exp\left[
  -\frac{1}{2}\transposed{\left(\vect{n} - \vect{\mu} - \matrb{\Sigma} \vect{t} \right)} \matrb{\Sigma}^{-1}
                          \left(\vect{n} - \vect{\mu} - \matrb{\Sigma} \vect{t} \right)
  + \transposed{\vect{t}} \vect{\mu} 
  +\frac{1}{2} \transposed{\vect{t}} \matrb{\Sigma} \vect{t}  
  \right]
=
  \exp\left[\transposed{\vect{t}} \vect{\mu} + \frac{1}{2} \transposed{\vect{t}} \matrb{\Sigma} \vect{t} \right] 
.
\end{multline}
Using $M_{\vect{n}}$, the $m_1\cdots m_N$-point correlation function composed of powers $m_1,\ldots, m_N$ of the field values $n_1,\ldots,n_N$ can be computed by
\begin{equation}
\begin{split}
  \EV{n_1^{m_1}\cdots n_N^{m_N}} &=
  \left. \frac{\partial^{m_1\cdots m_N} M_{\vect{n}}(\vect{t})}{\partial t_1^{m_1}\cdots\partial t_N^{m_N}}\right|_{\vect{t}=\vect{0}}
=
\left. \frac{\partial^{m_1\cdots m_N}}{\partial t_1^{m_1} \cdots \partial t_N^{m_N}}
  \exp\left[\transposed{\vect{t}}\vect{\mu} + \frac{1}{2} \transposed{\vect{t}} \matrb{\Sigma} \vect{t} \right] 
\right|_{\vect{t}=\vect{0}}
.
\end{split}	 
\end{equation}
For example,
\begin{align}
\EV{n_1 n_2} &= \mu^2 + \xi_{n,12} ,\\
\EV{n_1 n_2 n_3} &= \mu^3 + \mu \left(\xi_{n,12} + \xi_{n,13} + \xi_{n,23} \right),\\
\EV{n_1 n_2 n_3 n_4} &= \mu^4 + \mu^2\left(\xi_{n,12} + \xi_{n,13} + \xi_{n,14} + \xi_{n,23} + \xi_{n,24} + \xi_{n,34}\right)
  + \xi_{n,12}\xi_{n,34} + \xi_{n,13}\xi_{n,24} + \xi_{n,14}\xi_{n,23}
.
\end{align}

In a similar manner, one can compute expectation values for products of powers $k_1,\ldots, k_N$ of exponentiated field values  $\exp(n_1),\ldots,\exp(n_N)$:
\begin{equation}
\begin{split}
\label{eq:ev_log_prod}
  \EV{\exp(n_1)^{k_1}\cdots \exp(n_N)^{k_N}}
&=
  \EV{\exp\left( \transposed{\vect{k}}\vect{n} \right)}
  =
  \exp\left[\transposed{\vect{k}}\vect{\mu} + \frac{1}{2} \transposed{\vect{k}} \matrb{\Sigma} \vect{k} \right] 
  =
  \exp\bigg[\sum_{i=1}^{N}\left(k_i \mu +  k_i^2 \frac{\sigma^2}{2}\right)
   + \sum_{i<j} k_i k_j \xi_n\big(|\vect{x}_i - \vect{x}_j|\big) \bigg]
\\&
 =
  \lambda^{\transposed{\vect{k}}\vect{1}} \eta_0^{\transposed{\vect{k}}(\vect{k}-\vect{1})/2}
  \prod_{i<j}   \eta_{ij}^{k_i k_j}
,
\end{split}
\end{equation}
where $\lambda = \exp\left(\mu + \sigma^2/2\right)$, $\vect{k}=\transposed{(k_1,\ldots, k_N)}$, $\eta_0 = \exp\left(\sigma^2\right)$, and $\eta_{ij} = \eta\big(|\vect{x}_i - \vect{x}_j|\big)$ with $\eta(r) = \exp\big[\xi_n(r)\big]$.
In particular,
\begin{align}
  \EV{\exp(n_1)} &= \lambda = \exp\left(\mu + \sigma^2/2\right)
.
\end{align}

\section{Zero-mean shifted log-normal random fields}
\label{sec:zero_mean_shifted_log_normal_fields}

A zero-mean shifted log-normal homogeneous and isotropic random field $z$ can be defined via
\begin{equation}
  z : \R^D \to \R : \vect{x} \mapsto z(\vect{x}) = \exp\big[n(\vect{x})\big] - \lambda,
\end{equation}
where $n$ is a homogeneous and isotropic normal field with mean $\mu$ and variance $\sigma^2$, and the shift $\lambda = \exp\left(\mu + \sigma^2/2\right)$ to ensure vanishing mean for $z$.
The one-point distribution of the random field $z$ reads
\begin{equation}
\label{eq:zero_mean_shifted_log_normal_pdf}
 p_z(z; \lambda, \sigma) = 
 \begin{cases}
 \dfrac{1}{\sqrt{2\pi}(z + \lambda)\sigma} \exp\Biggl\{-\dfrac{\bigl[\ln\left(z / \lambda + 1\right) + \sigma^2/2 \bigr]^2}{2\sigma^2}\Biggr\}
    & \text{for } z > - \lambda \text{, and}\\
  0 & \text{otherwise.}
\end{cases}
\end{equation}
The negative of the shift $\lambda$ marks the lower limit for possible values for $z$.

The $N$-point correlation of $z$ can be deduced from the two-point correlation of the underlying normal field $n$ as follows. First note that an $m_1\cdots m_N$-point correlation function of $z$ composed of powers $m_1,\ldots, m_N$ of the field values $z_1,\ldots,z_N$ at $N$ mutually distinct points $\vect{x}_1,\ldots,\vect{x}_N$ can be written as:
\begin{equation}
\label{eq:shifted_log_prod_to_sum_of_log_prod}
\begin{split}
\EV{z_1^{m_1}\cdots z_N^{m_N}}
&=
\EV{\prod_{i=1}^{N} \left[\exp(n_i) - \lambda\right]^{m_i}}
=
\EV{\prod_{i=1}^{N} \sum_{k_i = 0}^{m_i} \binom{m_i}{k_i} (-\lambda)^{m_i - k_i} \exp(n_i)^{k_i}}
=
\sum_{k_1 = 0}^{m_1} \binom{m_1}{k_1} \cdots\sum_{k_N = 0}^{m_N} \binom{m_N}{k_N} (-\lambda)^{ \transposed{(\vect{m} - \vect{k})}\vect{1}} 
\EV{\exp\bigl(\transposed{\vect{k}} \vect{n}\bigr)}
,
\end{split}
\end{equation}
where $\vect{k} = \transposed{(k_1,\ldots,k_N)}$ and $\vect{m} = \transposed{(m_1,\ldots, m_N)}$.
Combining the results of the Eqs.~\eqref{eq:shifted_log_prod_to_sum_of_log_prod} and \eqref{eq:ev_log_prod}, one obtains
\begin{equation}
\EV{z_1^{m_1}\cdots z_N^{m_N}}
 =
 \lambda^{\transposed{\vect{m}}\vect{1}} 
 \sum_{k_1 = 0}^{m_1} \binom{m_1}{k_1} \cdots\sum_{k_N = 0}^{m_N} \binom{m_N}{k_N} 
 (-1)^{ \transposed{(\vect{m} - \vect{k})}\vect{1}} 
 \eta_0^{\transposed{\vect{k}}(\vect{k}-\vect{1})/2}
 \prod_{i<j}   \eta_{ij}^{k_i k_j}
.
\end{equation} 
For example,
\begin{align}
\label{eq:zero_shifted_log_normal_2p_correlation}
\EV{z_1 z_2} &=  \lambda^2 \bigl[\eta_{12} - 1 \bigr] 
  =  \lambda^2 \bigl\{\exp\big[\xi_n\big(|\vect{x}_i - \vect{x}_j|\big)\big] - 1 \bigr\} 
  = \xi_z\big(|\vect{x}_i - \vect{x}_j|\big)
,\\
\EV{z_1 z_2 z_3} &=  \lambda^3
  \bigl[
    \eta_{12}\eta_{13}\eta_{23}
    - \eta_{12} - \eta_{13} - \eta_{23}
    + 2
  \bigr]
\text{, and}\\
\begin{split}
\label{eq:zero_shifted_log_normal_4p_correlation}
\EV{z_1 z_2 z_3 z_4} &=  \lambda^4
  \bigl[
    \eta_{12}\eta_{13}\eta_{14} \eta_{23}  \eta_{24}  \eta_{34}
    - \eta_{12}\eta_{13}\eta_{23}
    - \eta_{12}\eta_{14}\eta_{24}
    - \eta_{13}\eta_{14}\eta_{34} 
    - \eta_{23}\eta_{24}\eta_{34}
    + \eta_{12}
    + \eta_{13}
    + \eta_{14}
    + \eta_{23}
    + \eta_{24}
    + \eta_{34} 
    - 3
  \bigr]
.
\end{split}
\end{align}

From Eq.~\eqref{eq:zero_shifted_log_normal_2p_correlation} follows that the correlation functions $\xi_z(r)$ and $\xi_n(r)$ are related via
\begin{equation}
  \xi_z(r) = \lambda^2 \bigl\{\exp\big[\xi_n(r)\big] - 1 \bigr\}
  \qquad\Leftrightarrow\qquad
  \xi_n(r) = \ln\big[\lambda^{-2} \xi_z(r) + 1\big]
.
\end{equation}
The correlation function $\xi_z(r)$ and the function $\eta(r) = \exp\big[\xi_n(r)\big]$ are related via
\begin{equation}
\label{eq:xi_z_of_r_from_eta_of_r}
  \xi_z(r) = \lambda^2 \bigl[\eta(r) - 1 \bigr]
\qquad\Leftrightarrow\qquad
  \eta(r) = \lambda^{-2}\xi_z(r) + 1
.
\end{equation}
Using these relations, one obtains, e.g.
\begin{align}
\label{eq:zero_shifted_log_normal_2p_correlation_of_xi}
\EV{z_1 z_2} &=  \xi_{z,12}
,\\
\label{eq:zero_shifted_log_normal_3p_correlation_of_xi}
\EV{z_1 z_2 z_3} &=  
\lambda^{-1}\left[\xi_{z,12}\xi_{z,13} + \xi_{z,12}\xi_{z,23} + \xi_{z,13} \xi_{z,23}\right] + \lambda^{-3} \xi_{z,12} \xi_{z,13} \xi_{z,23}
\text{, and}\\
\begin{split}
\label{eq:zero_shifted_log_normal_4p_correlation_of_xi}
\EV{z_1 z_2 z_3 z_4} &= 
\xi_{z,12}\xi_{z,34} + \xi_{z,13}\xi_{z,24} + \xi_{z,14} \xi_{z,23}
+ \lambda^{-2}\left[\xi_{z,12}\xi_{z,13}\xi_{z,14} + \xi_{z,12}\xi_{z,13}\xi_{z,24} + \xi_{z,12}\xi_{z,13}\xi_{z,34} + \ldots\right]
\\&\quad
+ \lambda^{-4}\left[\xi_{z,12}\xi_{z,13}\xi_{z,14} \xi_{z,23} + \ldots\right]
+ \lambda^{-6}\left[\xi_{z,12}\xi_{z,13}\xi_{z,14} \xi_{z,23} \xi_{z,24} + \ldots\right]
+ \lambda^{-8}\xi_{z,12}\xi_{z,13}\xi_{z,14}\xi_{z,23}\xi_{z,24}\xi_{z,34}
.
\end{split}
\end{align}

\section{Statistical properties of the shear-correlation estimator}
\label{sec:appendix_shear_correlation}

In this Section, we give a detailed discussion of approximations to the noise properties of the commonly used shear-correlation estimators \eqref{eq:xi_pm_estimator}. The discussion loosely follows the real-space approach of \citet{SchneiderEtal2002}. In contrast to previous works, we do not restrict the discussion to Gaussian shear fields. Instead, we first derive expressions for the covariance of the shear-correlation estimators for the case of a general convergence field, and then discuss the special cases of normal and zero-mean shifted log-normal convergence fields.

\subsection{Basic assumptions and definitions}
\label{sec:appendix_shear_correlation_basics}

To obtain analytical expressions for the statistical properties of the shear-correlation estimator, we consider an ensemble of weak-lensing surveys with $\Ngal$ galaxies in a rectangular field $\FOV$ with area $\AFOV$, which implies a galaxy number density $\ngal = \Ngal/\AFOV$.
To keep the computations simple, we make the following additional assumptions:
\begin{itemize}
  \item The number of galaxies is large, i.e. $\Ngal \gg 1$.
  \item Periodic boundary conditions are assumed for the distortion field (i.e. we neglect effects of field boundaries and distortions caused by structures outside the survey area).
  \item The shear field $\gamma$ is generated by a convergence field $\kappa$ in the survey field (i.e. we neglect $B$-mode shear).
  \item The convergence field $\kappa$ is a realisation of a statistically homogeneous and isotropic random field (we ignore that the assumption of statistical isotropy is in conflict with the assumption of periodic boundary conditions).
  \item The galaxy positions $\thetagal^{(1)},\ldots,\thetagal^{(N)}$ are uniformly and independently distributed in the survey field (i.e. we neglect correlations between the galaxy positions).
  \item The galaxy redshifts $\zgal^{(1)},\ldots,\zgal^{(N)}$ are identically and independently distributed according to a redshift distribution $p_z(z)$ (i.e. we neglect correlations between the galaxy redshifts).
   \item The observed image ellipticity $\gammagalobs^{(i)} = \epsilongal^{(i)} + \gammagal^{(i)}$ of each galaxy is a sum of its intrinsic ellipticity  $\epsilongal^{(i)}$ and the shear $\gammagal^{(i)}=\gamma(\thetagal^{(i)},\zgal^{(i)})$ at the galaxy's position $\thetagal^{(i)}$ and redshift $\zgal^{(i)}$.
 \item The intrinsic ellipticities $\epsilongal^{(1)},\ldots,\epsilongal^{(N)}$ are identically and independently distributed and are not correlated with the shear field (i.e. we neglect intrinsic alignments). The joint distribution of the two components $\epsilongali^{(i)}$ and $\epsilongalii^{(i)}$ of each intrinsic ellipticity $\epsilongal^{(i)}$ is isotropic with vanishing mean and variance $\sigmaepsilongal^2/\sqrt{2}$ per dimension (i.e. we neglect complications due to non-isotropic intrinsic ellipticity distributions caused, e.g, by insufficient PSF corrections).
\end{itemize}

The Fourier transform $\ft{f}(\vell)$ of a real function $f(\vtheta)$ on the survey field is defined by the relations:
\begin{align}
  \ft{f}(\vell) &= \frac{1}{\AFOV}\int_{\FOV}\idiff[2]{\vtheta} \ee^{-\ii \vell \cdot \vtheta} f(\vtheta)
\qquad\Leftrightarrow\qquad
  f(\vtheta) = \sum_{\vell} \ee^{\ii \vell \cdot \vtheta} \ft{f}(\vell)
.
\end{align}

Expectation values $\bEV{\est{f}}$ of observables $\est{f}(\thetagal^{(1)},\ldots)$ are obtained by computing the following (formal) average:
\begin{equation}
\label{eq:ev_df}
 \bEV{\est{f}} = \bEV{\est{f}}_{\thetagal,\zgal,\epsilongal,\kappa} = \bEV{ \bEV{ \bEV{ \bEV{\est{f}}_{\thetagal} }_{\zgal} }_{\epsilongal} }_{\kappa}.
\end{equation}
Here, $\bEV{\est{f}}_{\kappa}$ denotes the ensemble average over the realisations of the convergence field, $\bEV{\est{f}}_{\epsilongal}$ denotes the average over the intrinsic ellipticities, 
\begin{equation}
\bEV{\est{f}}_{\zgal} = \int\idiff[]{\zgal^{(1)}} p_z\bigl(\zgal^{(1)}\bigr) \cdots \int\diff[]{\zgal^{(\Ngal)}}p_z\bigl(\zgal^{(\Ngal)}\bigr) \est{f}\bigl(\thetagal^{(1)},\ldots\bigr)
\end{equation}
denotes the average over the galaxy redshifts, and
\begin{equation}
\bEV{\est{f}}_{\thetagal} = \frac{1}{\AFOV^{\Ngal}}\int_{\FOV}\idiff[2]{\thetagal^{(1)}} \cdots \int_{\FOV}\idiff[2]{\thetagal^{(\Ngal)}} \est{f}\bigl(\thetagal^{(1)},\ldots\bigr)
\end{equation}
denotes the ensemble average over the galaxy positions. We assume that the order of the averages does not matter for all observables of interest.

The tangential component $\gammat(\vtheta,\vvartheta,z)$ and cross component $\gammax(\vtheta,\vvartheta,z)$ of the shear $\gamma(\vtheta,z)$ at position $\vtheta$ and redshift $z$ with respect to the direction $\vvartheta$ are defined by \citep[e.g.][]{SchneiderVanWaerbekeMellier2002}
\begin{align}
\gammat(\vtheta,\vvartheta,z) &= -\Re\left(\gamma(\vtheta,z)\ee^{-2\ii\varphi(\vvartheta)}\right)
 \qquad\text{and}\qquad
\gammax(\vtheta,\vvartheta,z)  = -\Im\left(\gamma(\vtheta,z)\ee^{-2\ii\varphi(\vvartheta)}\right)
,
\end{align}
where $\varphi(\vvartheta)$ denotes the polar angle of the vector $\vvartheta$.

We denote the tangential and cross component of the shear $\gammagal^{(i)} = \gamma(\thetagal^{(i)}, \zgal^{(i)})$ at position $\thetagal^{(i)}$ and redshift $\zgal^{(i)}$ of a galaxy $i$ with respect to the direction $\varphigal^{(i,j)}=\varphi(\thetagal^{(j)} - \thetagal^{(i)})$ towards a galaxy $j$ at position $\thetagal^{(j)}$ by
\begin{align}
  \gammagalt^{(i,j)} &= \gammat(\thetagal^{(i)}, \thetagal^{(j)} - \thetagal^{(i)}, \zgal^{(i)})
 \qquad\text{and}\qquad
  \gammagalx^{(i,j)}  = \gammax(\thetagal^{(i)}, \thetagal^{(j)} - \thetagal^{(i)}, \zgal^{(i)}) 
  .
\end{align}
Similar definitions are made for the tangential and cross component of the intrinsic and observed image ellipticity:
\begin{align}
  \epsilongalt^{(i,j)} &= -\Re\left(\epsilongal^{(i)}\ee^{-2\ii\varphi^{(i,j)})}\right)
,&
  \epsilongalx^{(i,j)} &= -\Im\left(\epsilongal^{(i)}\ee^{-2\ii\varphi^{(i,j)}}\right)
,
\\
\gammagalobst^{(i,j)} &= -\Re\left(\gammagalobs^{(i)} \ee^{-2\ii\varphigal^{(i,j)}}\right)
  = \gammagalt^{(i,j)} + \epsilongalt^{(i,j)}
\text{, and}&
\gammagalobsx^{(i,j)} &= -\Im\left(\gammagalobs^{(i)} \ee^{-2\ii\varphigal^{(i,j)}}\right)
  = \gammagalx^{(i,j)} + \epsilongalx^{(i,j)}
.
\end{align}

Furthermore, we define the source-redshift averaged convergence $\kappa(\vtheta)$ and shear $\gamma(\vtheta)$ as in Eqs.~\eqref{eq:df_effective_convergence} and \eqref{eq:df_effective_shear}.
From the assumption that the shear field is generated by the convergence field follows that the convergence and the Cartesian shear components are related in Fourier space by:
\begin{equation}
\label{eq:relation_ft_kappa_gamma}
  \ft{\gamma}_1(\vell) = \cos\bigl[2\varphi(\vell)\bigr] \ft{\kappa}(\vell)
\qquad\text{and}\qquad
  \ft{\gamma}_2(\vell) = \sin\bigl[2\varphi(\vell)\bigr] \ft{\kappa}(\vell).
\end{equation}
We define the tangential and cross components of the effective shear as in Eqs.~\eqref{eq:df_effective_shear_tangential_component} and \eqref{eq:df_effective_shear_cross_component}. With these definitions,
\begin{align}
\label{eq:gtgt_p_gxgx_to_cartesian}
    \gammat(\vtheta,\vvartheta)\gammat(\vtheta + \vvartheta,\vvartheta) + 
    \gammax(\vtheta,\vvartheta)\gammax(\vtheta + \vvartheta,\vvartheta)
&=
    \gamma_1(\vtheta)\gamma_1(\vtheta + \vvartheta) +  
    \gamma_1(\vtheta)\gamma_2(\vtheta + \vvartheta)
\text{, and}\\
\label{eq:gtgt_m_gxgx_to_cartesian}
\begin{split}
    \gammat(\vtheta,\vvartheta)\gammat(\vtheta + \vvartheta,\vvartheta) - 
    \gammax(\vtheta,\vvartheta)\gammax(\vtheta + \vvartheta,\vvartheta)
&=
    \cos\bigl[4\varphi(\vvartheta)\bigr]\bigl[
    \gamma_1(\vtheta)\gamma_1(\vtheta + \vvartheta) - 
    \gamma_2(\vtheta)\gamma_2(\vtheta + \vvartheta)
    \bigr] 
  \\&\quad
  +
    \sin\bigl[4\varphi(\vvartheta)\bigr]\bigl[
    \gamma_1(\vtheta)\gamma_2(\vtheta + \vvartheta) + 
    \gamma_2(\vtheta)\gamma_1(\vtheta + \vvartheta)
    \bigr]
.
\end{split}
\end{align}

\subsection{Correlation functions and their relations}
\label{sec:appendix_shear_correlation_correlation_function_relations}

There exist well-known relations between the two-point correlations of the convergence and the shear, e.g. the Eqs.~\eqref{eq:xi_p_from_xi_kappa} and \eqref{eq:xi_m_from_xi_kappa}. We wish to derive similar relations between the four-point correlations of the convergence and the four-point correlations of the shear, which appear in the covariance of the cosmic shear estimators~\eqref{eq:xi_pm_estimator}. Achieving this task for general convergence fields requires relations between the convergence and the shear that hold not only for the ensemble mean, but for every realisation of the convergence field.

For every realisation, we define `empirical' correlations based on spatial averages:
\begin{equation}
\label{eq:df_empirical_correlations}
  \spavxi_{\kappa} (\vtheta)  = \frac{1}{\AFOV} \int_{\FOV} \idiff[2]{\vtheta'} \kappa(\vtheta') \kappa(\vtheta' +\vtheta)
   \qquad\text{and}\qquad
  \spavxi_{\pm} (\vtheta) = 
  \frac{1}{\AFOV} \int_{\FOV} \idiff[2]{\vtheta'} \left[\gammat(\vtheta',\vtheta) \gammat(\vtheta' +\vtheta,\vtheta) \pm \gammax(\vtheta',\vtheta) \gammax(\vtheta' + \vtheta,\vtheta) \right]
.
\end{equation}
From the convolution theorem and the relations \eqref{eq:relation_ft_kappa_gamma}, \eqref{eq:gtgt_p_gxgx_to_cartesian}, and \eqref{eq:gtgt_m_gxgx_to_cartesian} follows that for every realisation,
\begin{align}
\begin{split}
  \spavxi_{+} (\vtheta) &= 
    \frac{1}{\AFOV} \int_{\FOV} \idiff[2]{\vtheta'} 
    \left[\gamma_1(\vtheta') \gamma_1(\vtheta' +\vtheta) + \gamma_2(\vtheta') \gamma_2(\vtheta' + \vtheta) \right]
 =
    \sum_{\vell} \ee^{\ii \vell \cdot \vtheta}
    \left\{\cos\bigl[2\varphi(\vell)\bigr]^2 + \sin\bigl[2\varphi(\vell)\bigr]^2 \right\} \ft{\kappa}^*(\vell) \ft{\kappa}(\vell)
  =
    \sum_{\vell} \ee^{\ii \vell \cdot \vtheta} \ft{\kappa}^*(\vell) \ft{\kappa}(\vell)
  \\&=
    \spavxi_{\kappa}(\vtheta)
\text{, and}
\end{split}
\\
\begin{split}
  \spavxi_{-} (\vtheta) &= 
    \frac{1}{\AFOV} \int_{\FOV} \idiff[2]{\vtheta'} \Bigl\{
    \cos\bigl[4 \varphi(\vtheta)\bigr]
    \left[\gamma_1(\vtheta') \gamma_1(\vtheta' +\vtheta) - \gamma_2(\vtheta') \gamma_2(\vtheta' + \vtheta) \right]
     + 
    \sin\bigl[4 \varphi(\vtheta)\bigr]
    \left[\gamma_1(\vtheta') \gamma_2(\vtheta' +\vtheta) - \gamma_2(\vtheta') \gamma_1(\vtheta' + \vtheta) \right]
    \Bigr\}
  \\&=
    \sum_{\vell} \ee^{\ii \vell \cdot \vtheta}
    \left\{\cos\bigl[4 \varphi(\vtheta)\bigr]\cos\bigl[2\varphi(\vell)\bigr]^2 - 
           \cos\bigl[4 \varphi(\vtheta)\bigr]\sin\bigl[2\varphi(\vell)\bigr]^2 +
         2 \sin\bigl[4 \varphi(\vtheta)\bigr]\sin\bigl[2\varphi(\vell)\bigr]\cos\bigl[2\varphi(\vell)\bigr] \right\}
   \ft{\kappa}^*(\vell)\ft{\kappa}(\vell)
  \\&=
    \sum_{\vell} \ee^{\ii \vell \cdot \vtheta} \cos\bigl[4 \varphi(\vell) - 4 \varphi(\vtheta)\bigr] \ft{\kappa}^*(\vell)\ft{\kappa}(\vell)
  =
    \sum_{\vell} \ee^{\ii \vell \cdot \vtheta} \cos\bigl[4 \varphi(\vell) - 4 \varphi(\vtheta)\bigr]
    \frac{1}{\AFOV} \int_{\FOV} \idiff[2]{\vtheta'} \ee^{- \ii \vell \cdot \vtheta'}
    \spavxi_{\kappa} (\vtheta')
.
\end{split}
\end{align}
This can be written as
\begin{equation}
\label{eq:relation_xi_xi_realisation}
  \spavxi_{\pm} (\vtheta) = \int_{\FOV} \idiff[2]{\vtheta'} \spavKerPM(\vtheta,\vtheta') \spavxi_{\kappa} (\vtheta')
\end{equation}
with the kernels
\begin{align}
  \spavKerP(\vtheta, \vtheta') &= \DiracDelta(\vtheta - \vtheta') 
 \qquad\text{and}\qquad
  \spavKerM(\vtheta, \vtheta') = \frac{1}{\AFOV} \sum_{\vell} \ee^{\ii \vell \cdot (\vtheta - \vtheta') }  \cos\bigl[4\varphi(\vell) - 4 \varphi(\vtheta) \bigr]
,
\end{align}
where $\DiracDelta$ denotes the (in this case two-dimensional) Dirac delta `function'.
The relations \eqref{eq:relation_xi_xi_realisation} between the empirical convergence and shear correlations are valid for every realisation. They will provide us with the tools to translate the four-point correlations of the shear into four-point correlations of the convergence. Of course, they can also be used to derive the familiar relations between the two-point correlations of the shear and the convergence.

The ensemble-average convergence and shear correlations are defined by \citep[e.g.][]{BlandfordEtal1991,Kaiser1992}:
\begin{equation}
\label{eq:df_ea_correlations}
  \xi_{\kappa} \bigl(\lvert\vvartheta\rvert\bigr) =
  \bEV{\spavxi_{\kappa}(\vvartheta)} = 
  \bEV{\kappa(\vtheta) \kappa(\vtheta + \vvartheta)}
   \qquad\text{and}\qquad
  \xi_{\pm}  \bigl(\lvert\vvartheta\rvert\bigr) =
  \bEV{\spavxi_{\pm} (\vvartheta)} =
  \EV{\gammat(\vtheta,\vvartheta) \gammat(\vtheta + \vvartheta, \vvartheta)} \pm \EV{\gammax(\vtheta, \vvartheta) \gammax(\vtheta + \vvartheta, \vvartheta)}
.
\end{equation}
The statistical isotropy of the convergence field implies that its correlation $\xi_{\kappa}$, and thus, the shear correlations $\xi_\pm$ are isotropic.
As a consequence of the relations \eqref{eq:relation_xi_xi_realisation},
\begin{equation}
\label{eq:relation_xi_xi}
   \xi_{\pm} (\vvartheta) = \int_{0}^{\rFOV}\idiff[]{\vartheta'}\vartheta' \KerPM(\vartheta,\vartheta') \xi_{\kappa} (\vartheta')
\end{equation}
with the survey field's `radius' $\rFOV\approx \sqrt{\AFOV/\pi}$ and the kernels \citep[e.g.][]{CrittendenEtal2002}
\begin{align}
\begin{split}
  \KerP(\vartheta,\vartheta') &= \frac{1}{\vartheta'}\DiracDelta(\vartheta - \vartheta')
\text{, and} 
\end{split}
\\
\begin{split}
  \KerM(\vartheta,\vartheta') &\approx
    \int_{0}^{2\pi}\idiff[]{\varphi'}\, \spavKerM( \vartheta \uvect{0}, \vartheta' \uvect{\varphi'})
  =
  \frac{1}{\AFOV} \sum_{\vell} \ee^{\ii \lvert\vell\rvert \vartheta \cos[\varphi(\vell)] }
  \cos\bigl[4\varphi(\vell)\bigr]
  \int_{0}^{2\pi}\idiff[]{\varphi'}\ee^{- \ii \lvert\vell\rvert \vartheta' \cos[\varphi(\vell) - \varphi']}
  \approx
  \int_{0}^{\infty}\idiff[]{\ell}\,\ell J_4(\ell \vartheta) J_0(\ell \vartheta')
 \\&=
  \frac{1}{\vartheta'}\DiracDelta(\vartheta - \vartheta') + 
  \left(\frac{4}{\vartheta^2} - \frac{12\vartheta^{\prime 2}}{\vartheta^4}\right) H(\vartheta - \vartheta') 
,
\end{split}
\end{align}
where $J$ denotes the Bessel function of the first kind, $H$ the Heaviside step function, and $\DiracDelta$ the Dirac delta function.

Combining the relations \eqref{eq:relation_xi_xi} between the convergence correlation $\xi_\kappa$ and the shear correlation functions $\xi_{\pm}$ with similar relations between $\xi_\kappa$ and the correlations of the shear components $\gamma_i$, one may derive \citep[e.g.][]{Kaiser1992,SchneiderEtal2002}
\begin{subequations}
\label{eq:relation_gg_xi}
\begin{align}
  \xi_{+} \bigl(\lvert\vvartheta\rvert\bigr) &=
  \EV{\gamma_1(\vtheta)\gamma_1(\vtheta+\vvartheta)} + \EV{\gamma_2(\vtheta)\gamma_2(\vtheta + \vvartheta)}
,\\
  \xi_{-} \bigl(\lvert\vvartheta\rvert\bigr) &=
     \cos\bigl[ 4 \varphi(\vtheta) \bigr] \bigl[ \EV{\gamma_1(\vtheta)\gamma_1(\vtheta+\vvartheta)} - \EV{\gamma_2(\vtheta)\gamma_2(\vtheta+\vvartheta)} \bigr]
  +2 \sin\bigl[ 4 \varphi(\vtheta) \bigr]  \EV{\gamma_1(\vtheta)\gamma_2(\vtheta+\vvartheta)}
,\\
  \EV{\gamma_1(\vtheta, z)\gamma_1(\vtheta+\vvartheta)} 
  &=
  \frac{1}{2}\Bigl\{ \xi_{+} \bigl(\lvert\vvartheta\rvert\bigr) + \cos\bigl[ 4 \varphi(\vtheta) \bigr] \xi_{-} \bigl(\lvert\vvartheta\rvert\bigr) \Bigr\}
,\\
  \EV{\gamma_2(\vtheta, z)\gamma_2(\vtheta+\vvartheta)} 
  &=
  \frac{1}{2}\Bigl\{ \xi_{+} \bigl(\lvert\vvartheta\rvert\bigr) - \cos\bigl[ 4 \varphi(\vtheta) \bigr] \xi_{-} \bigl(\lvert\vvartheta\rvert\bigr) \Bigr\}
\text{, and}\\
  \EV{\gamma_1(\vtheta, z)\gamma_2(\vtheta+\vvartheta)} 
 &=
  \frac{1}{2} \sin\bigl[ 4 \varphi(\vtheta) \bigr] \xi_{-} \bigl(\lvert\vvartheta\rvert\bigr)
.
\end{align}
\end{subequations}

\subsection{The shear correlation estimators}
\label{sec:appendix_shear_correlation_estimators}

We consider the following estimators $\est{\xi}_\pm(\vartheta)$ of the shear correlations $\xi_\pm(\vartheta)$ at separations $\vartheta$:
\begin{align}
  \label{eq:xi_pm_general_estimator}
  \est{\xi}_\pm(\vartheta) &= \frac{\ShearSumPM(\vartheta)}{\NormSum(\vartheta)}
\text{ with}\\\nonumber
  \ShearSumPM(\vartheta) &=
  \sum_{i,j=1}^{\Ngal} \wgal^{(i)}\wgal^{(j)}\,\Delta\bigl(\vartheta, \lvert \thetagal^{(j)} - \thetagal^{(i)} \rvert \bigr)
  \left( \gammagalobst^{(i,j)} \gammagalobst^{(j,i)} \pm \gammagalobsx^{(i,j)} \gammagalobsx^{(j,i)} \right)
\qquad\text{and}\qquad
  \NormSum(\vartheta) =
  \sum_{i,j=1}^{\Ngal} \wgal^{(i)}\wgal^{(j)}\,\Delta\bigl(\vartheta, \lvert \thetagal^{(j)} - \thetagal^{(i)} \rvert \bigr)
.
\end{align}
Here, $\wgal^{(i)}$ denotes a statistical weight for the galaxy $i$ (which we assume to be uncorrelated with any other quantity of interest), and $\Delta\bigl(\vartheta, \theta \bigr)$ denotes the bin window function \eqref{eq:bin_function} for a bin centred on $\vartheta$ with width $\varDelta(\vartheta)$ (which we assume to be small compared to the scale on which correlations change). Furthermore,  $\gammagalobst^{(i,j)}$ and $\gammagalobsx^{(i,j)}$ denote the tangential and cross component, respectively, of the shear $\gammagalobs^{(i)}$ estimated from the shape of galaxy $i$ with respect to the direction $\varphigal^{(i,j)}=\varphi(\thetagal^{(j)} - \thetagal^{(i)})$ defined by the positions $\thetagal^{(i)}$ and $\thetagal^{(j)}$ of galaxies $i$ and $j$.

Since
  $
  \bEV{\gammagalobst^{(i,j)} \gammagalobst^{(j,i)} \pm \gammagalobsx^{(i,j)} \gammagalobsx^{(j,i)} }_{\epsilongal,\zgal,\kappa}
  = \xi_\pm(\lvert\thetagal^{(j)} - \thetagal^{(i)}\rvert)
  $
for $i \neq j$, the estimator \eqref{eq:xi_pm_general_estimator} is an unbiased estimator of $\xi_{\pm}(\vartheta)$ \citep[e.g.][]{SchneiderVanWaerbekeMellier2002}:
\begin{equation}
\label{eq:xi_pm_general_estimator_ev}
\begin{split}
  \EV{\est{\xi}_\pm(\vartheta) }
 &
 =
 \EV{
 \frac{\sum_{i,j} \wgal^{(i)}\wgal^{(j)}\,\Delta\bigl(\vartheta, \lvert \thetagal^{(j)} - \thetagal^{(i)} \rvert \bigr)
       \left( \gammagalobst^{(i,j)} \gammagalobst^{(j,i)} \pm \gammagalobsx^{(i,j)} \gammagalobsx^{(j,i)} \right)
      }{\sum_{i,j} \wgal^{(i)}\wgal^{(j)}\,\Delta\bigl(\vartheta, \lvert \thetagal^{(j)} - \thetagal^{(i)} \rvert \bigr)}
  }
=
 \BiggEV{
 \frac{\sum_{i,j} \wgal^{(i)}\wgal^{(j)}\,\Delta\bigl(\vartheta, \lvert \thetagal^{(j)} - \thetagal^{(i)} \rvert \bigr)
      \xi_\pm(\lvert\thetagal^{(j)} - \thetagal^{(i)}\rvert) 
      }{\sum_{i,j} \wgal^{(i)}\wgal^{(j)}\,\Delta\bigl(\vartheta, \lvert \thetagal^{(j)} - \thetagal^{(i)} \rvert \bigr)}
  }_{\thetagal}
\!=
   \xi_\pm(\vartheta) 
 .
\end{split}
\end{equation}

The sum $\NormSum(\vartheta)$ appearing in the denominator of the shear correlation estimators \eqref{eq:xi_pm_general_estimator} can be interpreted as the effective number of galaxy pairs in the bin centred on $\vartheta$. The expected effective number of pairs $\MeanNormSum(\vartheta)$ in the bin reads 
\begin{equation}
\begin{split}
\label{eq:mean_norm_sum}
  \MeanNormSum(\vartheta) &= 
  \EV{\NormSum(\vartheta)}
=
  \sum_{i,j=1}^{\Ngal} \wgal^{(i)}\wgal^{(j)}
  \frac{1}{\AFOV^{\Ngal}}\int_{\FOV}\idiff[2]{\thetagal^{(1)}}\ldots \int_{\FOV}\idiff[2]{\thetagal^{(\Ngal)}}\,
  \Delta\bigl(\vartheta, \lvert \thetagal^{(j)} - \thetagal^{(i)} \rvert \bigr)
  \approx
  \frac{\ABin(\vartheta)}{\AFOV} \PairSum,
\end{split}
\end{equation}
where
\begin{equation}
  \ABin(\vartheta) = 
  \frac{1}{\AFOV}\int_{\FOV}\idiff[2]{\vtheta} \int_{\FOV}\idiff[2]{\vtheta'} \Delta\bigl(\vartheta, \lvert \vtheta' - \vtheta \rvert \bigr)
  = 2 \pi \vartheta \varDelta(\vartheta)
\end{equation}
denote the effective bin area and 
\begin{equation}
  \PairSum = \sum_{i,j=1}^{\Ngal} \wgal^{(i)} \wgal^{(j)} 
\end{equation} 
denotes the effective total number of pairs. In the case of uniform weights $\wgal^{(i)} = 1$ $\forall i$, the effective number of pairs $\PairSum = \Ngal^2$, and the expectation \eqref{eq:mean_norm_sum} reduces to:
\begin{equation}
  \MeanNormSum(\vartheta) = \ngal^2 \AFOV \ABin(\vartheta)
  = 2 \pi \vartheta \varDelta(\vartheta) \AFOV\,\ngal^2
  .
\end{equation}

\subsection{The covariance of the shear correlation estimators}
\label{sec:appendix_shear_correlation_estimators_covariance}

We denote the covariance of the estimators \eqref{eq:xi_pm_general_estimator} by:
\begin{equation}
\label{eq:c_pmpm_df}
c_{\pm\pm}(\vartheta_1, \vartheta_2) = 
   \EV{\est{\xi}_\pm(\vartheta_1), \est{\xi}_\pm(\vartheta_2)} - \EV{\est{\xi}_\pm(\vartheta_1)}\EV{\est{\xi}_\pm(\vartheta_2)}
 = 
 \EV{\frac{\ShearSumPM(\vartheta_1)\,\ShearSumPM(\vartheta_2)}{\NormSum(\vartheta_1)\,\NormSum(\vartheta_2)}}
   - \xi_\pm(\vartheta_1) \xi_\pm(\vartheta_2)
   .
\end{equation}
To calculate $c_{\pm\pm}$, we first neglect the randomness in the effective number of galaxy pairs $\NormSum(\vartheta)$:
\begin{equation}
\label{eq:c_pmpm_separate_numerator_and_denominator}
c_{\pm\pm}(\vartheta_1, \vartheta_2)
 \approx 
  \frac{\EV{\ShearSumPM(\vartheta_1)\,\ShearSumPM(\vartheta_2)}}{\MeanNormSum(\vartheta_1) \MeanNormSum(\vartheta_2)}
 - \xi_\pm(\vartheta_1) \xi_\pm(\vartheta_2).
\end{equation}
The term
\begin{equation}
  \EV{\ShearSumPM(\vartheta_1)\,\ShearSumPM(\vartheta_2)}
= \sum_{i,j,k,l=1}^{\Ngal} \wgal^{(i)}\wgal^{(j)} \wgal^{(k)}\wgal^{(l)}
  \EV{
  \Delta\bigl(\vartheta_1, \lvert \thetagal^{(j)} - \thetagal^{(i)} \rvert \bigr)
  \Delta\bigl(\vartheta_2, \lvert \thetagal^{(l)} - \thetagal^{(k)} \rvert \bigr)
  \left( \gammagalobst^{(i,j)} \gammagalobst^{(j,i)} \pm \gammagalobsx^{(i,j)} \gammagalobsx^{(j,i)} \right)
  \left( \gammagalobst^{(k,l)} \gammagalobst^{(l,k)} \pm \gammagalobsx^{(k,l)} \gammagalobsx^{(l,k)} \right)
  }
\end{equation} 
 is then decomposed into three sums, exploiting the symmetry of the galaxy indices and the fact that only terms with even powers of intrinsic ellipticities contribute:
\begin{equation}
\begin{split}
  \EV{\ShearSumPM(\vartheta_1)\,\ShearSumPM(\vartheta_2)}
  &=
  2 \sum_{i,j=1}^{\Ngal} \left(\wgal^{(i)}\right)^2 \left(\wgal^{(j)}\right)^2
  \Bigl\langle
  \Delta\bigl(\vartheta_1, \lvert \thetagal^{(j)} - \thetagal^{(i)} \rvert \bigr)
  \Delta\bigl(\vartheta_2, \lvert \thetagal^{(j)} - \thetagal^{(i)} \rvert \bigr)
  \left( \epsilongalt^{(i,j)} \epsilongalt^{(j,i)} \pm \epsilongalx^{(i,j)} \epsilongalx^{(j,i)} \right)
  \left( \epsilongalt^{(i,j)} \epsilongalt^{(j,i)} \pm \epsilongalx^{(i,j)} \epsilongalx^{(j,i)} \right)
  \Bigr\rangle
  \\&\quad
   + 4 \!\sum_{i,j,k=1}^{\Ngal} \left(\wgal^{(i)}\right)^2 \wgal^{(j)} \wgal^{(k)}
  \Bigl\langle
  \Delta\bigl(\vartheta_1, \lvert \thetagal^{(j)} - \thetagal^{(i)} \rvert \bigr)
  \Delta\bigl(\vartheta_2, \lvert \thetagal^{(k)} - \thetagal^{(i)} \rvert \bigr)
  \left( \epsilongalt^{(i,j)} \gammagalt^{(j,i)} \pm \epsilongalx^{(i,j)} \gammagalx^{(j,i)} \right)
  \left( \epsilongalt^{(i,k)} \gammagalt^{(k,i)} \pm \epsilongalx^{(i,k)} \gammagalx^{(k,i)} \right)
  \Bigr\rangle
  \\&\quad  
   + \!\sum_{i,j,k,l=1}^{\Ngal}\! \wgal^{(i)}\wgal^{(j)} \wgal^{(k)}\wgal^{(l)}
  \Bigl\langle
  \Delta\bigl(\vartheta_1, \lvert \thetagal^{(j)} - \thetagal^{(i)} \rvert \bigr)
  \Delta\bigl(\vartheta_2, \lvert \thetagal^{(l)} - \thetagal^{(k)} \rvert \bigr)
  \left( \gammagalt^{(i,j)} \gammagalt^{(j,i)} \pm \gammagalx^{(i,j)} \gammagalx^{(j,i)} \right)
  \left( \gammagalt^{(k,l)} \gammagalt^{(l,k)} \pm \gammagalx^{(k,l)} \gammagalx^{(l,k)} \right)
  \Bigr\rangle
.
\end{split}
\end{equation}
The second sum comprises terms involving two or three distinct galaxies. The third sum contains terms involving two to four distinct galaxies. Since all galaxies are statistically indistuingishable apart from their weights $\wgal^{(i)}$, all averages $\bEV{\Delta\bigl(\vartheta_1, \lvert \thetagal^{(j)} - \thetagal^{(i)} \rvert \bigr)\ldots}$ in the same sum involving the same number of galaxies yield the same contribution to the covariance. However, terms with different numbers of galaxies involved yield different contributions. We neglected these differences, and thereby the noise due to the finite sampling of the shear field in position and redshift by the galaxies (which is justified for large $\Ngal$):
\begin{equation}
\label{eq:c_pmpm_mean_of_shear_sum_products_approx}
\begin{split}
  \EV{\ShearSumPM(\vartheta_1)\,\ShearSumPM(\vartheta_2)}
  \approx
  \qquad 2 \PairSum^{(\epsilon)} 
  & 
  \Bigl\langle
  \Delta\bigl(\vartheta_1, \lvert \thetagal^{(2)} - \thetagal^{(1)} \rvert \bigr)
  \Delta\bigl(\vartheta_2, \lvert \thetagal^{(2)} - \thetagal^{(1)} \rvert \bigr)
  \left( \epsilongalt^{(1,2)} \epsilongalt^{(2,1)} \pm \epsilongalx^{(1,2)} \epsilongalx^{(2,1)} \right)
  \left( \epsilongalt^{(1,2)} \epsilongalt^{(2,1)} \pm \epsilongalx^{(1,2)} \epsilongalx^{(2,1)} \right)
  \Bigr\rangle
  \\
   + 4 \Ngal \PairSum^{(\epsilon\gamma)}
  &
  \Bigl\langle
  \Delta\bigl(\vartheta_1, \lvert \thetagal^{(2)} - \thetagal^{(1)} \rvert \bigr)
  \Delta\bigl(\vartheta_2, \lvert \thetagal^{(3)} - \thetagal^{(1)} \rvert \bigr)
  \left( \epsilongalt^{(1,2)} \gammagalt^{(2,1)} \pm \epsilongalx^{(1,2)} \gammagalx^{(2,1)} \right)
  \left( \epsilongalt^{(1,3)} \gammagalt^{(3,1)} \pm \epsilongalx^{(1,3)} \gammagalx^{(3,1)} \right)
  \Bigr\rangle
  \\
   + 
  \PairSum^2
  &
  \Bigl\langle
  \Delta\bigl(\vartheta_1, \lvert \thetagal^{(2)} - \thetagal^{(1)} \rvert \bigr)
  \Delta\bigl(\vartheta_2, \lvert \thetagal^{(4)} - \thetagal^{(3)} \rvert \bigr)
  \left( \gammagalt^{(1,2)} \gammagalt^{(2,1)} \pm \gammagalx^{(1,2)} \gammagalx^{(2,1)} \right)
  \left( \gammagalt^{(3,4)} \gammagalt^{(4,3)} \pm \gammagalx^{(3,4)} \gammagalx^{(4,3)} \right)
  \Bigr\rangle
.
\end{split}
\end{equation}
The effective pair numbers
\begin{equation}
  \PairSum^{(\epsilon)} = \sum_{i,j=1}^{\Ngal} \left(\wgal^{(i)}\right)^2 \left(\wgal^{(j)}\right)^2
  \qquad\text{and}\qquad
  \PairSum^{(\epsilon\gamma)} = \Ngal^{-1} \sum_{i,j,k=1}^{\Ngal} \left(\wgal^{(i)}\right)^2 \wgal^{(j)} \wgal^{(k)}
\end{equation}
introduced here as abbreviations become identical to $\Ngal^2$ for uniform weights.

The covariance obtained by combining Eqs.~\eqref{eq:c_pmpm_separate_numerator_and_denominator} and \eqref{eq:c_pmpm_mean_of_shear_sum_products_approx} is then split into three parts, with each part containing one of the three sums in the r.h.s. of Eq.~\eqref{eq:c_pmpm_mean_of_shear_sum_products_approx}:
\begin{equation}
\label{eq:c_pmpm_split_df}
c_{\pm\pm}(\vartheta_1, \vartheta_2) = 
c_{\pm\pm}^{(\epsilon)}(\vartheta_1, \vartheta_2) + c_{\pm\pm}^{(\epsilon\gamma)}(\vartheta_1, \vartheta_2) + c_{\pm\pm}^{(\gamma)}(\vartheta_1, \vartheta_2) 
.
\end{equation}
The ellipticity-noise part $c_{\pm\pm}^{(\epsilon)}$ consists of all terms containing intrinsic ellipticities, but no shear:
\begin{equation}
\label{eq:c_pmpm_epsilon_df}
\begin{split}
  c_{\pm\pm}^{(\epsilon)}(\vartheta_1, \vartheta_2)
  &=
  \frac{2 \PairSum^{(\epsilon)}}{\MeanNormSum(\vartheta_1)\,\MeanNormSum(\vartheta_2)} 
  \Bigl\langle
  \Delta\bigl(\vartheta_1, \lvert \thetagal^{(2)} - \thetagal^{(1)} \rvert \bigr)
  \Delta\bigl(\vartheta_2, \lvert \thetagal^{(2)} - \thetagal^{(1)} \rvert \bigr)
  \left( \epsilongalt^{(1,2)} \epsilongalt^{(2,1)} \pm \epsilongalx^{(1,2)} \epsilongalx^{(2,1)} \right)
  \left( \epsilongalt^{(1,2)} \epsilongalt^{(2,1)} \pm \epsilongalx^{(1,2)} \epsilongalx^{(2,1)} \right)
  \Bigr\rangle
  .
\end{split}
\end{equation}
This ellipticity-noise part obviously vanishes for disjoint bins.
The mixed part $c_{\pm\pm}^{(\epsilon\gamma)}$ comprises all terms containing products of shear and intrinsic ellipticities:
\begin{equation}
\label{eq:c_pmpm_epsilon_gamma_df}
\begin{split} 
  c_{\pm\pm}^{(\epsilon\gamma)}(\vartheta_1, \vartheta_2)
  &=
  \frac{4 \Ngal \PairSum^{(\epsilon\gamma)}}{\MeanNormSum(\vartheta_1)\,\MeanNormSum(\vartheta_2)} 
  \Bigl\langle
  \Delta\bigl(\vartheta_1, \lvert \thetagal^{(2)} - \thetagal^{(1)} \rvert \bigr)
  \Delta\bigl(\vartheta_2, \lvert \thetagal^{(3)} - \thetagal^{(1)} \rvert \bigr)
  \left( \epsilongalt^{(1,2)} \gammagalt^{(2,1)} \pm \epsilongalx^{(1,2)} \gammagalx^{(2,1)} \right)
  \left( \epsilongalt^{(1,3)} \gammagalt^{(3,1)} \pm \epsilongalx^{(1,3)} \gammagalx^{(3,1)} \right)
  \Bigr\rangle
.
\end{split}
\end{equation}
The cosmic variance part $c_{\pm\pm}^{(\gamma)}$ contains all terms involving only shear:
\begin{equation}
\label{eq:c_pmpm_gamma_df}
\begin{split} 
  c_{\pm\pm}^{(\gamma)}(\vartheta_1, \vartheta_2)
  &=
  \frac{\PairSum^2}{\MeanNormSum(\vartheta_1)\,\MeanNormSum(\vartheta_2)}
  \Bigl\langle
  \Delta\bigl(\vartheta_1, \lvert \thetagal^{(2)} - \thetagal^{(1)} \rvert \bigr)
  \Delta\bigl(\vartheta_2, \lvert \thetagal^{(4)} - \thetagal^{(3)} \rvert \bigr)
  \left( \gammagalt^{(1,2)} \gammagalt^{(2,1)} \pm \gammagalx^{(1,2)} \gammagalx^{(2,1)} \right)
  \left( \gammagalt^{(3,4)} \gammagalt^{(4,3)} \pm \gammagalx^{(3,4)} \gammagalx^{(4,3)} \right)
  \Bigr\rangle
  \\&\quad 
   - \xi_{\pm}(\vartheta_1)  \xi_{\pm}(\vartheta_2)
.
\end{split}
\end{equation}
We proceed by evaluating the three covariance parts $c_{\pm\pm}^{(\epsilon)}$, $c_{\pm\pm}^{(\epsilon\gamma)}$, and $c_{\pm\pm}^{(\gamma)}$ for particular choices of the signs under various assumptions about the statistics of the convergence field.
 
\subsubsection{The ellipticity-noise contribution}
\label{sec:appendix_shear_correlation_ellipticity_noise}

As mentioned above, $ c_{\pm\pm}^{(\epsilon)}$ vanishes for disjoint bins. The ellipticity-noise contribution to the covariance of $\xi_\pm(\vartheta)$ evaluates to:
\begin{equation}
\begin{split}
  c_{\pm\pm}^{(\epsilon)}(\vartheta, \vartheta)
  &=
  \frac{2\PairSum^{(\epsilon)}}{\MeanNormSum(\vartheta)^2} \frac{\ABin(\vartheta)}{\AFOV} 
  \Bigl\langle 
  \left( \epsilongalt^{(1,2)} \epsilongalt^{(2,1)} \pm \epsilongalx^{(1,2)} \epsilongalx^{(2,1)} \right)
  \left( \epsilongalt^{(1,2)} \epsilongalt^{(2,1)} \pm \epsilongalx^{(1,2)} \epsilongalx^{(2,1)} \right)
  \Bigr\rangle
  =
  \frac{2\PairSum^{(\epsilon)}}{\MeanNormSum(\vartheta)^2} \frac{\ABin(\vartheta)}{\AFOV} 
   \left[ \frac{\sigmaepsilongal^4}{4} + (\pm 1)(\pm 1) \frac{\sigmaepsilongal^4}{4} \right]
  .
\end{split}
\end{equation}
Hence, the ellipticity-noise part also vanishes for non-matching signs. For uniform weights,
\begin{equation}
  c_{++}^{(\epsilon)}(\vartheta, \vartheta) = c_{--}^{(\epsilon)}(\vartheta, \vartheta) = \frac{\sigmaepsilongal^4}{\MeanNormSum(\vartheta)},
  \qquad\text{ and }\qquad 
  c_{+-}^{(\epsilon)}(\vartheta, \vartheta) = c_{-+}^{(\epsilon)}(\vartheta, \vartheta) = 0
  .
\end{equation}

\subsubsection{The mixed contribution}
\label{sec:appendix_shear_correlation_mixed_noise}

The mixed part \eqref{eq:c_pmpm_epsilon_gamma_df} comprises all terms containing products of shear and intrinsic ellipticities. After using the relations \eqref{eq:relation_gg_xi} and some trigonometry, these reduce to:
\begin{align}
\begin{split} 
  \Bigl\langle
  \left( \epsilongalt^{(1,2)} \gammagalt^{(2,1)} + \epsilongalx^{(1,2)} \gammagalx^{(2,1)} \right)
  \left( \epsilongalt^{(1,3)} \gammagalt^{(3,1)} + \epsilongalx^{(1,3)} \gammagalx^{(3,1)} \right)
  \Bigr\rangle_{\epsilongal,\zgal, \kappa}
   &=
  \frac{\sigmaepsilongal^2}{2} \xi_+\bigl( \lvert \thetagal^{(3)} - \thetagal^{(2)} \rvert \bigr)
,
\end{split}
\\
\begin{split}
  \Bigl\langle
  \Bigl( \epsilongalt^{(1,2)} \gammagalt^{(2,1)} + \epsilongalx^{(1,2)} \gammagalx^{(2,1)} \Bigr)
  \Bigl( \epsilongalt^{(1,3)} \gammagalt^{(3,1)} - \epsilongalx^{(1,3)} \gammagalx^{(3,1)} \Bigr)
  \Bigr\rangle_{\epsilongal,\zgal, \kappa}
  &=
  \frac{\sigmaepsilongal^2}{2} 
  \xi_{-}\bigl(\lvert \thetagal^{(3)} - \thetagal^{(2)} \rvert \bigr)
  \cos\bigl( 4\varphi^{(1,3)} - 4\varphi^{(2,3)} \bigr)
\text{, and}
\end{split}
\\
\begin{split}
  \Bigl\langle
  \left( \epsilongalt^{(1,2)} \gammagalt^{(2,1)} - \epsilongalx^{(1,2)} \gammagalx^{(2,1)} \right)
  \left( \epsilongalt^{(1,3)} \gammagalt^{(3,1)} - \epsilongalx^{(1,3)} \gammagalx^{(3,1)} \right)
  \Bigr\rangle_{\epsilongal,\zgal, \kappa}
   &=
  \frac{\sigmaepsilongal^2}{2}
  \xi_{+}\bigl(\lvert \thetagal^{(3)} - \thetagal^{(2)} \rvert \bigr)
  \cos\bigl( 4\varphi^{(1,3)} - 4\varphi^{(1,2)} \bigr)
.
\end{split}
\end{align}
Hence,
\begin{equation}
\begin{split} 
  c_{++}^{(\epsilon\gamma)}(\vartheta_1, \vartheta_2)
  &=
  \frac{4 \Ngal \PairSum^{(\epsilon\gamma)}}{\MeanNormSum(\vartheta_1)\,\MeanNormSum(\vartheta_2)} 
  \frac{1}{\AFOV^3}\int_{\FOV}\diff[2]{\thetagal^{(1)}}\int_{\FOV}\diff[2]{\thetagal^{(2)}}\int_{\FOV}\diff[2]{\thetagal^{(3)}}
  \Delta\bigl(\vartheta_1, \lvert \thetagal^{(2)} - \thetagal^{(1)} \rvert \bigr)
  \Delta\bigl(\vartheta_2, \lvert \thetagal^{(3)} - \thetagal^{(1)} \rvert \bigr)
  \frac{\sigmaepsilongal^2}{2}
  \xi_{+}\bigl(\lvert \thetagal^{(3)} - \thetagal^{(2)} \rvert \bigr)
  \\&=
  \frac{2\sigmaepsilongal^2\Ngal \PairSum^{(\epsilon\gamma)}}{\AFOV^2\MeanNormSum(\vartheta_1)\,\MeanNormSum(\vartheta_2)} 
  \int_{\FOV}\diff[2]{\thetagal^{(2)}}\int_{\FOV}\diff[2]{\thetagal^{(3)}}
  \Delta\bigl(\vartheta_1, \lvert \thetagal^{(2)} \rvert \bigr)
  \Delta\bigl(\vartheta_2, \lvert \thetagal^{(3)} \rvert \bigr)
  \xi_{+}\bigl(\lvert \thetagal^{(3)} - \thetagal^{(2)} \rvert \bigr)
  \\&=
  \frac{ \Ngal^2 \PairSum^{(\epsilon\gamma)} }{ \PairSum^2} 
  \frac{2\sigmaepsilongal^2}{\pi \ngal \AFOV}
  \int_{0}^{\pi}\idiff[]{\varphi_1}
  \xi_{+}\bigl(\lvert \vartheta_2 \uvect{0} - \vartheta_1 \uvect{\varphi_1} \rvert \bigr)
.
\end{split}
\end{equation}
Here, $\uvect{\varphi}$ denotes a unit vector with polar angle $\varphi$. 
Similar calculations yield:
\begin{align}
\begin{split} 
  c_{+-}^{(\epsilon\gamma)}(\vartheta_1, \vartheta_2)
  &=
  \frac{ \Ngal^2 \PairSum^{(\epsilon\gamma)} }{ \PairSum^2} 
  \frac{2\sigmaepsilongal^2}{\pi \ngal \AFOV}       
  \int_{0}^{\pi}\idiff[]{\varphi_1}
  \xi_{-}\bigl(\lvert \vartheta_2 \uvect{0} - \vartheta_1 \uvect{\varphi_1} \rvert \bigr)
  \cos\bigl[4\varphi\bigl(\vartheta_2 \uvect{\varphi_2} - \vartheta_1 \uvect{\varphi_1}\bigr) \bigr]
\text{, and}
\end{split}
\\
\begin{split} 
  c_{--}^{(\epsilon\gamma)}(\vartheta_1, \vartheta_2)
  &=
  \frac{ \Ngal^2 \PairSum^{(\epsilon\gamma)} }{ \PairSum^2} 
  \frac{2\sigmaepsilongal^2}{\pi \ngal \AFOV}       
  \int_{0}^{\pi}\idiff[]{\varphi_1}
  \xi_{+}\bigl(\lvert \vartheta_2 \uvect{0} - \vartheta_1 \uvect{\varphi_1} \rvert \bigr)
  \cos\bigl(4\varphi_1 \bigr)
.
\end{split}
\end{align}
For uniform weights, this reduces to:
\begin{align}
  c_{++}^{(\epsilon\gamma)}(\vartheta_1, \vartheta_2)
  &=
  \frac{2\sigmaepsilongal^2}{\pi \ngal \AFOV}       
  \int_{0}^{\pi}\idiff[]{\varphi_1}
  \xi_{+}\bigl(\lvert \vartheta_2 \uvect{0} - \vartheta_1 \uvect{\varphi_1} \rvert \bigr)
,\\
  c_{+-}^{(\epsilon\gamma)}(\vartheta_1, \vartheta_2)
  &=
  \frac{2\sigmaepsilongal^2}{\pi \ngal \AFOV}       
  \int_{0}^{\pi}\idiff[]{\varphi_1}
  \xi_{-}\bigl(\lvert \vartheta_2 \uvect{0} - \vartheta_1 \uvect{\varphi_1} \rvert \bigr)
  \cos\bigl[4\varphi\bigl(\vartheta_2 \uvect{\varphi_2} - \vartheta_1 \uvect{\varphi_1}\bigr) \bigr]
\text{, and}\\
  c_{--}^{(\epsilon\gamma)}(\vartheta_1, \vartheta_2)
  &=
  \frac{2\sigmaepsilongal^2}{\pi \ngal \AFOV}       
  \int_{0}^{\pi}\idiff[]{\varphi_1}
  \xi_{+}\bigl(\lvert \vartheta_2 \uvect{0} - \vartheta_1 \uvect{\varphi_1} \rvert \bigr)
  \cos\bigl(4\varphi_1 \bigr)
.
\end{align}

\subsubsection{The cosmic variance contribution for general convergence fields}
\label{sec:appendix_shear_correlation_cosmic_noise_general}

To calculate the cosmic variance contribution to the covariance, we exploit the relations \eqref{eq:relation_xi_xi_realisation} between the empirical correlation functions \eqref{eq:df_empirical_correlations} to transform the four-point correlations of the shear in the cosmic variance part \eqref{eq:c_pmpm_gamma_df} into four-point correlations of the convergence:
\begin{equation}
\label{eq:c_pmpm_gamma_general_derivation}
\begin{split} 
  c_{\pm\pm}^{(\gamma)}(\vartheta_1, \vartheta_2)
&=
  \frac{1}{\AFOV^2\ABin(\vartheta_1)\ABin(\vartheta_2)}
  \int_{\FOV}\idiff[2]{\thetagal^{(1)}} \int_{\FOV}\idiff[2]{\thetagal^{(2)}} \int_{\FOV}\idiff[2]{\thetagal^{(3)}} \int_{\FOV}\idiff[2]{\thetagal^{(4)}}
  \Delta\bigl(\vartheta_1, \lvert \thetagal^{(2)} - \thetagal^{(1)} \rvert \bigr)
  \Delta\bigl(\vartheta_2, \lvert \thetagal^{(4)} - \thetagal^{(3)} \rvert \bigr)
  \\&\quad\times  
  \Bigl\langle
  \Bigl[
      \gammat\bigl(\thetagal^{(1)}, \thetagal^{(2)} - \thetagal^{(1)}\bigr) \gammat\bigl(\thetagal^{(2)}, \thetagal^{(2)} - \thetagal^{(1)}\bigr)
  \pm \gammax\bigl(\thetagal^{(1)}, \thetagal^{(2)} - \thetagal^{(1)}\bigr) \gammax\bigl(\thetagal^{(2)}, \thetagal^{(2)} - \thetagal^{(1)}\bigr) 
  \Bigr]
  \\&\quad\times\phantom{\Bigl\langle}
  \Bigl[
      \gammat\bigl(\thetagal^{(3)}, \thetagal^{(4)} - \thetagal^{(3)}\bigr) \gammat\bigl(\thetagal^{(4)}, \thetagal^{(4)} - \thetagal^{(3)}\bigr)
  \pm \gammax\bigl(\thetagal^{(3)}, \thetagal^{(4)} - \thetagal^{(3)}\bigr) \gammax\bigl(\thetagal^{(4)}, \thetagal^{(4)} - \thetagal^{(3)}\bigr) 
  \Bigr]
  \Bigr\rangle_{\kappa}
  \, - \, \xi_{\pm}(\vartheta_1)\xi_{\pm}(\vartheta_2)
\\&=
  \frac{1}{\ABin(\vartheta_1)\ABin(\vartheta_2)}
  \int_{\FOV}\idiff[2]{\thetagal^{(2)}} \int_{\FOV}\idiff[2]{\thetagal^{(4)}}
  \Delta\bigl(\vartheta_1, \lvert \thetagal^{(2)} \rvert \bigr)
  \Delta\bigl(\vartheta_2, \lvert \thetagal^{(4)} \rvert \bigr)
  \\&\quad\times
  \Biggl\langle
  \frac{1}{\AFOV}\int_{\FOV}\idiff[2]{\thetagal^{(1)}}
  \Bigl[
      \gammat\bigl(\thetagal^{(1)}, \thetagal^{(2)}\bigr) \gammat\bigl(\thetagal^{(1)} + \thetagal^{(2)}, \thetagal^{(2)}\bigr)
  \pm \gammax\bigl(\thetagal^{(1)}, \thetagal^{(2)}\bigr) \gammax\bigl(\thetagal^{(1)} + \thetagal^{(2)}, \thetagal^{(2)}\bigr) 
  \Bigr]
  \\&\quad\times\phantom{\Biggl\langle}
  \frac{1}{\AFOV}\int_{\FOV}\idiff[2]{\thetagal^{(3)}} 
  \Bigl[
      \gammat\bigl(\thetagal^{(3)}, \thetagal^{(4)}\bigr) \gammat\bigl(\thetagal^{(3)} + \thetagal^{(4)}, \thetagal^{(4)}\bigr)
  \pm \gammax\bigl(\thetagal^{(3)}, \thetagal^{(4)}\bigr) \gammax\bigl(\thetagal^{(3)} + \thetagal^{(4)}, \thetagal^{(4)}\bigr) 
  \Bigr]
  \Biggr\rangle_{\kappa} 
  \,-\, \xi_{\pm}(\vartheta_1)\xi_{\pm}(\vartheta_2)
\\&=
  \frac{1}{\AFOV^2\ABin(\vartheta_1)\ABin(\vartheta_2)}
  \int_{\FOV}\idiff[2]{\thetagal^{(1)}} \int_{\FOV}\idiff[2]{\thetagal^{(2)}} \int_{\FOV}\idiff[2]{\thetagal^{(3)}} \int_{\FOV}\idiff[2]{\thetagal^{(4)}}
  \Delta\left(\vartheta_1, \lvert \thetagal^{(2)} \rvert \right)
  \Delta\left(\vartheta_2, \lvert \thetagal^{(4)} \rvert \right)
  \\&\quad\times
  \int_{\FOV}\idiff[2]{\thetagal^{(2)\prime}} \spavKerPM\bigl(\thetagal^{(2)}, \thetagal^{(2)\prime}\bigr)
  \int_{\FOV}\idiff[2]{\thetagal^{(4)\prime}} \spavKerPM\bigl(\thetagal^{(4)}, \thetagal^{(4)\prime}\bigr)
  \EV{  
  \kappa\bigl(\thetagal^{(1)}\bigr)\kappa\bigl(\thetagal^{(1)} + \thetagal^{(2)\prime} \bigr)
  \kappa\bigl(\thetagal^{(3)}\bigr)\kappa\bigl(\thetagal^{(3)} + \thetagal^{(4)\prime} \bigr)
  }_{\kappa} 
 \,-\, \xi_{\pm}(\vartheta_1)\xi_{\pm}(\vartheta_2)
.
\end{split}
\end{equation}

From the properties of the kernels $\spavKerPM$ and $\KerPM$ follows that
\begin{equation}
\begin{split}
&
  \frac{1}{\ABin(\vartheta_1)}
  \int_{\FOV}\idiff[2]{\thetagal^{(2)}}
  \Delta\bigl(\vartheta_1, \lvert \thetagal^{(2)} \rvert \bigr)
  \int_{\FOV}\idiff[2]{\thetagal^{(2)\prime}}
  \spavKerPM\bigl(\thetagal^{(2)}, \thetagal^{(2)\prime}\bigr)
  f\bigl(\thetagal^{(2)\prime},\ldots\bigr) 
\\&=
  \frac{1}{\ABin(\vartheta_1)}
  \int_{0}^{\rFOV}\idiff{\varthetagal^{(2)}}       \varthetagal^{(2)}       \int_{0}^{2\pi}\idiff{\varphigal^{(2)}}
  \int_{0}^{\rFOV}\idiff{\varthetagal^{(2)\prime}} \varthetagal^{(2)\prime} \int_{0}^{2\pi}\idiff{\varphigal^{(2)\prime}}
  \Delta\bigl(\vartheta_1, \varthetagal^{(2)} \bigr)
  \spavKerPM\bigl(\varthetagal^{(2)} \uvect{\varphigal^{(2)}}, \varthetagal^{(2)\prime} \uvect{\varphigal^{(2)\prime}}\bigr)
  f\bigl(\varthetagal^{(2)\prime} \uvect{\varphigal^{(2)\prime}}, \ldots\bigr)
\\&=
  \frac{1}{2 \pi}
  \int_{0}^{2\pi}\idiff{\varphigal^{(2)}}
  \int_{0}^{\rFOV}\idiff{\varthetagal^{(2)\prime}} \varthetagal^{(2)\prime} \int_{0}^{2\pi}\idiff{\varphigal^{(2)\prime}}
  \spavKerPM\bigl(\vartheta_1 \uvect{0}, \varthetagal^{(2)\prime} \uvect{\varphigal^{(2)}}\bigr)
  f\bigl(\varthetagal^{(2)\prime} \uvect{\varphigal^{(2)\prime}}, \ldots\bigr)
\\&=
  \frac{1}{2 \pi}
  \int_{0}^{\rFOV}\idiff{\varthetagal^{(2)\prime}} \varthetagal^{(2)\prime}
  \KerPM\bigl(\vartheta_1, \varthetagal^{(2)\prime}\bigr)
  \int_{0}^{2\pi}\idiff{\varphigal^{(2)\prime}}
  f\bigl(\varthetagal^{(2)\prime} \uvect{\varphigal^{(2)\prime}}, \ldots\bigr)
\\&=
  \int_{0}^{\rFOV}\idiff{\vartheta'_1} \vartheta'_1
  \KerPM \bigl(\vartheta_1, \vartheta'_1\bigr)
  \frac{1}{\ABin(\vartheta'_1)}  
  \int_{\FOV}\idiff{\thetagal^{(2)\prime}}
  \Delta\bigl(\vartheta'_1, \lvert \thetagal^{(2)\prime} \rvert \bigr)
  f\bigl(\thetagal^{(2)\prime}, \ldots\bigr)
\end{split}
\end{equation}
Thus, any contribution $c_{\pm\pm}^{(\gamma)}$ can be computed from a given $c_{++}^{(\gamma)}$ via a simple integral transform:
\begin{equation}
\begin{split}
\label{eq:c_pmpm_gamma_general}
  c_{\pm\pm}^{(\gamma)}(\vartheta_1, \vartheta_2)
  &=
    \int_{0}^{\rFOV}\idiff{\vartheta'_1} \vartheta'_1 \KerPM(\vartheta_1, \vartheta'_1) 
    \int_{0}^{\rFOV}\idiff{\vartheta'_2} \vartheta'_2 \KerPM(\vartheta_2, \vartheta'_2) 
    c_{++}^{(\gamma)}(\vartheta'_1, \vartheta'_2)
.
\end{split}
\end{equation}

\subsubsection{The cosmic variance contribution for normal convergence fields}
\label{sec:appendix_shear_correlation_cosmic_noise_normal_approximation}

For a normal convergence field,
\begin{equation}
\begin{split}
\label{eq:kappa_covariance_normal}
  \Bigl\langle  
  \kappa\bigl(\thetagal^{(1)}\bigr) \kappa\bigl(\thetagal^{(1)} + \thetagal^{(2)} \bigr)
  \kappa\bigl(\thetagal^{(3)}\bigr) \kappa\bigl(\thetagal^{(3)} + \thetagal^{(4)} \bigr)
  \Bigr\rangle_{\kappa}
&=
  \xi_\kappa\bigl(\lvert \thetagal^{(2)} \rvert\bigr)
  \xi_\kappa\bigl(\lvert \thetagal^{(4)} \rvert\bigr)
+
  \xi_\kappa\bigl(\lvert \thetagal^{(3)} - \thetagal^{(1)} \rvert\bigr)
  \xi_\kappa\bigl(\lvert \thetagal^{(4)} + \thetagal^{(3)} - \thetagal^{(2)} - \thetagal^{(1)} \rvert\bigr)
\\&\quad
+
  \xi_\kappa\bigl(\lvert \thetagal^{(4)} + \thetagal^{(3)} - \thetagal^{(1)} \rvert\bigr)
  \xi_\kappa\bigl(\lvert \thetagal^{(3)} - \thetagal^{(2)} - \thetagal^{(1)} \rvert\bigr)
,
\end{split}
\end{equation}
and thus (exploiting the symmetries of the integrands),
\begin{equation}
\label{eq:c_pp_gamma_normal}
\begin{split} 
  c_{\pm\pm}^{(\gamma)}(\vartheta_1, \vartheta_2)
&=
  \frac{1}{\AFOV^2\ABin(\vartheta_1)\ABin(\vartheta_2)}
  \int_{\FOV}\idiff[2]{\thetagal^{(1)}} \int_{\FOV}\idiff[2]{\thetagal^{(2)}} \int_{\FOV}\idiff[2]{\thetagal^{(3)}} \int_{\FOV}\idiff[2]{\thetagal^{(4)}}
  \Delta\left(\vartheta_1, \lvert \thetagal^{(2)} \rvert \right)
  \Delta\left(\vartheta_2, \lvert \thetagal^{(4)} \rvert \right)
  \\&\quad\times
  \int_{\FOV}\idiff[2]{\thetagal^{(2)\prime}} \spavKerPM\bigl(\thetagal^{(2)}, \thetagal^{(2)\prime}\bigr)
  \int_{\FOV}\idiff[2]{\thetagal^{(4)\prime}} \spavKerPM\bigl(\thetagal^{(4)}, \thetagal^{(4)\prime}\bigr)
  \Bigl[
  \xi_\kappa\bigl(\lvert \thetagal^{(2)\prime} \rvert\bigr)
  \xi_\kappa\bigl(\lvert \thetagal^{(4)\prime} \rvert\bigr)
\\&\quad
+
  \xi_\kappa\bigl(\lvert \thetagal^{(3)} - \thetagal^{(1)} \rvert\bigr)
  \xi_\kappa\bigl(\lvert \thetagal^{(4)\prime} + \thetagal^{(3)} - \thetagal^{(2)\prime} - \thetagal^{(1)} \rvert\bigr)
+
  \xi_\kappa\bigl(\lvert \thetagal^{(4\prime)} + \thetagal^{(3)} - \thetagal^{(1)} \rvert\bigr)
  \xi_\kappa\bigl(\lvert \thetagal^{(3)} - \thetagal^{(2\prime)} - \thetagal^{(1)} \rvert\bigr)
\Bigr]
 \,-\, \xi_{\pm}(\vartheta_1)\xi_{\pm}(\vartheta_2)
\\&=
  \frac{2}{\AFOV\ABin(\vartheta_1)\ABin(\vartheta_2)}
  \int_{\FOV}\idiff[2]{\thetagal^{(2)}} \int_{\FOV}\idiff[2]{\thetagal^{(3)}} \int_{\FOV}\idiff[2]{\thetagal^{(4)}}
  \Delta\bigl(\vartheta_1, \lvert \thetagal^{(2)} \rvert \bigr)
  \Delta\bigl(\vartheta_2, \lvert \thetagal^{(4)} \rvert \bigr)
\\&\quad\times
  \int_{\FOV}\idiff[2]{\thetagal^{(2)\prime}} \spavKerPM\bigl(\thetagal^{(2)}, \thetagal^{(2)\prime}\bigr)
  \int_{\FOV}\idiff[2]{\thetagal^{(4)\prime}} \spavKerPM\bigl(\thetagal^{(4)}, \thetagal^{(4)\prime}\bigr)
  \xi_\kappa\bigl(\lvert \thetagal^{(3)} - \thetagal^{(2\prime)} \rvert\bigr)
  \xi_\kappa\bigl(\lvert \thetagal^{(3)} - \thetagal^{(4\prime)} \rvert\bigr)
.
\end{split}
\end{equation}
Now, we make use of the following identity:
\begin{equation}
\label{eq:spav_xi_kappa_to_xi_m_shifted_argument}
\begin{split}
\int_{\FOV}\idiff[2]{\vtheta'}\,\spavKerM(\vtheta, \vtheta') \xi_{\kappa}(|\vtheta'' - \vtheta'|)
&=
  \int_{\FOV}\idiff[2]{\vtheta'}\,\spavKerM(\vtheta, \vtheta'' - \vtheta') \xi_{\kappa}(|\vtheta'|)
\\&= 
  \int_{\FOV}\idiff[2]{\vtheta'}\,  
  \frac{1}{\AFOV} \sum_{\vell} \ee^{\ii \vell \cdot (\vtheta - \vtheta'' + \vtheta')} \cos\bigl[4\varphi(\vell) - 4 \varphi(\vtheta)\bigr]
   \xi_{\kappa}(|\vtheta'|)
\\&= {\hphantom+}
  \cos\bigl[4\varphi(\vtheta) - 4 \varphi(\vtheta'' - \vtheta)\bigr]
  \int_{\FOV}\idiff[2]{\vtheta'}\,  
  \frac{1}{\AFOV} \sum_{\vell} \ee^{\ii \vell \cdot (\vtheta'' - \vtheta - \vtheta')}
  \cos\bigl[4\varphi(\vell) - 4 \varphi(\vtheta'' - \vtheta)\bigr] 
  \xi_{\kappa}(|\vtheta'|)
  \\&\quad+
  \sin\bigl[4\varphi(\vtheta) - 4 \varphi(\vtheta'' - \vtheta)\bigr]
  \int_{\FOV}\idiff[2]{\vtheta'}\,  
  \frac{1}{\AFOV} \sum_{\vell} \ee^{\ii \vell \cdot (\vtheta'' - \vtheta - \vtheta')}
  \sin\bigl[4\varphi(\vell) - 4 \varphi(\vtheta'' - \vtheta)\bigr] 
  \xi_{\kappa}(|\vtheta'|)
\\&= 
  \cos\bigl[4\varphi(\vtheta) - 4 \varphi(\vtheta'' - \vtheta)\bigr]
  \xi_{-}(|\vtheta'' - \vtheta|)
.
\end{split}
\end{equation}
This can be generalized to:
\begin{equation}
\label{eq:spav_xi_to_xi_shifted_argument}
\int_{\FOV}\idiff[2]{\vtheta'}\,\spavKerPM(\vtheta, \vtheta') \xi_{\kappa}(|\vtheta'' - \vtheta'|) = \zeta_{\pm}(\vtheta, \vtheta'')
\text{, where}
\end{equation}
\begin{equation}
	 \zeta_{+}(\vtheta, \vtheta'') =  \xi_{+}(|\vtheta'' - \vtheta|)
	 \quad\text{and}\qquad
	 \zeta_{-}(\vtheta, \vtheta'') =  \xi_{-}(|\vtheta'' - \vtheta|)  \cos\bigl[4\varphi(\vtheta) - 4 \varphi(\vtheta'' - \vtheta)\bigr]
.	 
\end{equation}
Hence,
\begin{equation}
\begin{split} 
  c_{\pm\pm}^{(\gamma)}(\vartheta_1, \vartheta_2)
&=
  \frac{2}{\AFOV\ABin(\vartheta_1)\ABin(\vartheta_2)}
  \int_{\FOV}\idiff[2]{\thetagal^{(2)}} \int_{\FOV}\idiff[2]{\thetagal^{(3)}} \int_{\FOV}\idiff[2]{\thetagal^{(4)}}
  \Delta\bigl(\vartheta_1, \lvert \thetagal^{(2)} \rvert \bigr)
  \Delta\bigl(\vartheta_2, \lvert \thetagal^{(4)} \rvert \bigr)
  \zeta_{\pm}\bigl(\thetagal^{(2\prime)}, \thetagal^{(3)} \bigr)
  \zeta_{\pm}\bigl(\thetagal^{(4\prime)}, \thetagal^{(3)} \bigr)
\\&=
  \frac{4}{\pi\AFOV}
  \int_{0}^{\rFOV}\idiff{\theta_3} \theta_3
  \int_{0}^{\pi}\idiff{\varphi_1}
  \zeta_\pm\bigl(\lvert \vartheta_1 \uvect{\varphi_1}, \theta_3 \uvect{0} \rvert\bigr)
  \int_{0}^{\pi}\idiff{\varphi_2}  
  \zeta_\pm\bigl(\lvert \vartheta_2 \uvect{\varphi_2}, \theta_3 \uvect{0} \rvert\bigr)
.
\end{split}
\end{equation}

\subsubsection{The cosmic variance contribution for zero-mean shifted log-normal convergence fields}
\label{sec:appendix_shear_correlation_cosmic_noise_log_normal_approximation}

For zero-mean shifted log-normal convergence fields, one can use Eq.~\eqref{eq:zero_shifted_log_normal_4p_correlation_of_xi} to obtain
\begin{equation}
\label{eq:kappa_covariance_log_normal}
\begin{split}
&
  \Bigl\langle  
  \kappa\bigl(\thetagal^{(1)}\bigr) \kappa\bigl(\thetagal^{(1)} + \thetagal^{(2)} \bigr)
  \kappa\bigl(\thetagal^{(3)}\bigr) \kappa\bigl(\thetagal^{(3)} + \thetagal^{(4)} \bigr)
  \Bigr\rangle_{\kappa} 
\\&=
  \xi_\kappa\bigl(\lvert \thetagal^{(2)} \rvert\bigr)
  \xi_\kappa\bigl(\lvert \thetagal^{(4)} \rvert\bigr)
+
  \xi_\kappa\bigl(\lvert \thetagal^{(3)} - \thetagal^{(1)} \rvert\bigr)
  \xi_\kappa\bigl(\lvert \thetagal^{(4)} + \thetagal^{(3)} - \thetagal^{(2)} - \thetagal^{(1)} \rvert\bigr)
+
  \xi_\kappa\bigl(\lvert \thetagal^{(4)} + \thetagal^{(3)} - \thetagal^{(1)} \rvert\bigr)
  \xi_\kappa\bigl(\lvert \thetagal^{(3)} - \thetagal^{(2)} - \thetagal^{(1)} \rvert\bigr)
\\&\quad +
  \kappa_0^{-2}\left[
  \xi_\kappa\bigl(\lvert \thetagal^{(2)} \rvert\bigr)
  \xi_\kappa\bigl(\lvert \thetagal^{(3)} - \thetagal^{(1)} \rvert\bigr)
  \xi_\kappa\bigl(\lvert \thetagal^{(4)} + \thetagal^{(3)} - \thetagal^{(1)} \rvert\bigr)
  + \ldots \right]
\\&\quad +
  \kappa_0^{-4}\left[
  \xi_\kappa\bigl(\lvert \thetagal^{(2)} \rvert\bigr)
  \xi_\kappa\bigl(\lvert \thetagal^{(3)} - \thetagal^{(1)} \rvert\bigr)
  \xi_\kappa\bigl(\lvert \thetagal^{(4)} + \thetagal^{(3)} - \thetagal^{(1)} \rvert\bigr)
  \xi_\kappa\bigl(\lvert \thetagal^{(3)} - \thetagal^{(2)} - \thetagal^{(1)} \rvert\bigr)
  + \ldots \right]
\\&\quad +
  \kappa_0^{-6}\left[
  \xi_\kappa\bigl(\lvert \thetagal^{(2)} \rvert\bigr)
  \xi_\kappa\bigl(\lvert \thetagal^{(3)} - \thetagal^{(1)} \rvert\bigr)
  \xi_\kappa\bigl(\lvert \thetagal^{(4)} + \thetagal^{(3)} - \thetagal^{(1)} \rvert\bigr)
  \xi_\kappa\bigl(\lvert \thetagal^{(3)} - \thetagal^{(2)} - \thetagal^{(1)} \rvert\bigr)
  \xi_\kappa\bigl(\lvert \thetagal^{(4)} + \thetagal^{(3)} - \thetagal^{(2)} - \thetagal^{(1)} \rvert\bigr)
  + \ldots \right]
\\&\quad +
  \kappa_0^{-8}
  \xi_\kappa\bigl(\lvert \thetagal^{(2)} \rvert\bigr)
  \xi_\kappa\bigl(\lvert \thetagal^{(3)} - \thetagal^{(1)} \rvert\bigr)
  \xi_\kappa\bigl(\lvert \thetagal^{(4)} + \thetagal^{(3)} - \thetagal^{(1)} \rvert\bigr)
  \xi_\kappa\bigl(\lvert \thetagal^{(3)} - \thetagal^{(2)} - \thetagal^{(1)} \rvert\bigr)
  \xi_\kappa\bigl(\lvert \thetagal^{(4)} + \thetagal^{(3)} - \thetagal^{(2)} - \thetagal^{(1)} \rvert\bigr)
  \xi_\kappa\bigl(\lvert \thetagal^{(4)} \rvert\bigr)
.
\end{split}
\end{equation}
This reveals that the covariance for a zero-mean shifted log-normal convergence field equals the covariance for a normal convergence field (which consists of products of up to two correlation functions) plus `corrections' involving products of three to six correlation functions.
Inserting the correlation \eqref{eq:kappa_covariance_log_normal} into the general expression \eqref{eq:c_pmpm_gamma_general} for the cosmic variance contribution yields (again exploiting symmetries of the integrands)
\begin{equation}
\label{eq:c_pp_gamma_log_normal}
\begin{split} 
  c_{++}^{(\gamma)}(\vartheta_1, \vartheta_2)
&=
  \frac{1}{\AFOV^2\ABin(\vartheta_1)\ABin(\vartheta_2)}
  \int_{\FOV}\idiff[2]{\thetagal^{(1)}} \int_{\FOV}\idiff[2]{\thetagal^{(2)}} \int_{\FOV}\idiff[2]{\thetagal^{(3)}} \int_{\FOV}\idiff[2]{\thetagal^{(4)}}
  \Delta\bigl(\vartheta_1, \lvert \thetagal^{(2)} \rvert \bigr)
  \Delta\bigl(\vartheta_2, \lvert \thetagal^{(4)} \rvert \bigr)
  \\&\quad\times  
  \Biggl[     
  \xi_\kappa\bigl(\lvert \thetagal^{(3)} - \thetagal^{(1)} \rvert\bigr)
  \xi_\kappa\bigl(\lvert \thetagal^{(4)} + \thetagal^{(3)} - \thetagal^{(2)} - \thetagal^{(1)} \rvert\bigr)
+
  \xi_\kappa\bigl(\lvert \thetagal^{(4)} + \thetagal^{(3)} - \thetagal^{(1)} \rvert\bigr)
  \xi_\kappa\bigl(\lvert \thetagal^{(3)} - \thetagal^{(2)} - \thetagal^{(1)} \rvert\bigr)
+ \ldots
\Biggr]
\\&=
  \frac{1}{\AFOV \ABin(\vartheta_1)\ABin(\vartheta_2)}
  \int_{\FOV}\idiff[2]{\thetagal^{(2)}}\int_{\FOV}\idiff[2]{\thetagal^{(4)}} \int_{\FOV}\idiff[2]{\thetagal^{(3)}}\,
  \Delta\bigl(\vartheta_1, \lvert \thetagal^{(2)} \rvert \bigr) \Delta\bigl(\vartheta_2, \lvert \thetagal^{(4)} \rvert \bigr) 
  \biggl\{
  4 \kappa_0^{-2} \xi_{\kappa}(\lvert \thetagal^{(2)}\rvert) \xi_{\kappa}(\lvert \thetagal^{(4)}\rvert) \xi_{\kappa}(\lvert \thetagal^{(3)}\rvert)
  \\&\quad\quad
  + 2 \kappa_0^{-4} \bigl[\kappa_0^2 + \xi_{\kappa}(\lvert \thetagal^{(2)}\rvert)\bigr]
  \xi_{\kappa}(\lvert \thetagal^{(4)}\rvert) \xi_{\kappa}(\lvert \thetagal^{(3)}\rvert) \xi_{\kappa}(\lvert \thetagal^{(3)} - \thetagal^{(2)} \rvert)
  \\&\quad\quad
  + 2 \kappa_0^{-4} \bigl[\kappa_0^2 + \xi_{\kappa}(\lvert \thetagal^{(4)}\rvert)\bigr] 
  \xi_{\kappa}(\lvert \thetagal^{(2)}\rvert) \xi_{\kappa}(\lvert \thetagal^{(3)}\rvert) \xi_{\kappa}(\lvert \thetagal^{(3)} - \thetagal^{(4)} \rvert)
  \\&\quad\quad
  + 2 \kappa_0^{-4}
  \bigl[\kappa_0^2 + \xi_{\kappa}(\lvert \thetagal^{(2)}\rvert)\bigr]
  \bigl[\kappa_0^2 + \xi_{\kappa}(\lvert \thetagal^{(4)}\rvert)\bigr]
  \xi_{\kappa}(\lvert \thetagal^{(3)} - \thetagal^{(2)} \rvert) \xi_{\kappa}(\lvert \thetagal^{(3)} - \thetagal^{(4)} \rvert)
   \\&\quad\quad      
  + 4 \kappa_0^{-6}
  \bigl[\kappa_0^2 + \xi_{\kappa}(\lvert \thetagal^{(2)}\rvert)\bigr]
  \bigl[\kappa_0^2 + \xi_{\kappa}(\lvert \thetagal^{(4)}\rvert)\bigr]
  \xi_{\kappa}(\lvert \thetagal^{(3)}\rvert) \xi_{\kappa}(\lvert \thetagal^{(3)} - \thetagal^{(2)} \rvert) \xi_{\kappa}(\lvert \thetagal^{(3)} - \thetagal^{(4)} \rvert)
  \\&\quad\quad    
  + \kappa_0^{-8}
  \bigl[\kappa_0^2 + \xi_{\kappa}(\lvert \thetagal^{(2)}\rvert)\bigr]
  \bigl[\kappa_0^2 + \xi_{\kappa}(\lvert \thetagal^{(4)}\rvert)\bigr]
  \xi_{\kappa}(\lvert \thetagal^{(3)} \rvert) \xi_{\kappa}(\lvert \thetagal^{(3)} - \thetagal^{(2)}\rvert) \xi_{\kappa}(\lvert \thetagal^{(3)} - \thetagal^{(4)} \rvert)
  \xi_{\kappa}(\lvert \thetagal^{(3)} - \thetagal^{(2)} - \thetagal^{(4)} \rvert)
  \biggr\}
\\&=
  \quad4 \kappa_0^{-2}\xi_{+}(\vartheta_1) \xi_{+}(\vartheta_2)
  \frac{2 \pi}{\AFOV} \int_{0}^{\rFOV}\idiff[]{\theta_3}\,\theta_3 \xi_{+}(\theta_3)
  \\&\quad
  + 2 \kappa_0^{-4}
  \bigl[\kappa_0^2 + \xi_{+}(\vartheta_1)\bigr]
  \xi_{+}(\vartheta_2)
  \frac{2}{\AFOV} \int_{0}^{\rFOV}\idiff[]{\theta_3}\,
  \theta_3 \xi_{+}(\theta_3) 
  \int_{0}^{\pi}\idiff[]{\varphi_1}\,
  \xi_{+}(\lvert \theta_3\uvect{0} - \vartheta_1\uvect{\varphi_1} \rvert)
  \\&\quad
  + 2 \kappa_0^{-4}
  \xi_{+}(\vartheta_1) 
  \bigl[\kappa_0^2 + \xi_{+}(\vartheta_2)\bigr]
  \frac{2}{\AFOV} \int_{0}^{\rFOV}\idiff[]{\theta_3}\, 
   \theta_3 \xi_{+}(\theta_3) 
  \int_{0}^{\pi}\idiff[]{\varphi_2}\,
   \xi_{+}(\lvert \theta_3\uvect{0} - \vartheta_2\uvect{\varphi_2} \rvert)
  \\&\quad
  + 2 \kappa_0^{-4} 
  \bigl[\kappa_0^2 + \xi_{+}(\vartheta_1)\bigr]
  \bigl[\kappa_0^2 + \xi_{+}(\vartheta_2)\bigr]
  \frac{2}{\pi \AFOV} \int_{0}^{\rFOV}\idiff[]{\theta_3}\,
   \theta_3
  \int_{0}^{\pi}\idiff[]{\varphi_1}\,
  \xi_{+}(\lvert \theta_3\uvect{0} - \vartheta_1\uvect{\varphi_1} \rvert)
  \int_{0}^{\pi}\idiff[]{\varphi_2}\,
  \xi_{+}(\lvert \theta_3\uvect{0} - \vartheta_2\uvect{\varphi_2} \rvert)
   \\&\quad
  + 4 \kappa_0^{-6}  
  \bigl[\kappa_0^2 + \xi_{+}(\vartheta_1)\bigr]
  \bigl[\kappa_0^2 + \xi_{+}(\vartheta_2)\bigr]
  \frac{2}{\pi \AFOV} \int_{0}^{\rFOV}\idiff[]{\theta_3} \,
  \theta_3 \xi_{+}(\theta_3)
  \int_{0}^{\pi}\idiff[]{\varphi_1} \,
  \xi_{+}(\lvert \theta_3\uvect{0} - \vartheta_1\uvect{\varphi_1} \rvert)
  \int_{0}^{\pi}\idiff[]{\varphi_2}\,
  \xi_{+}(\lvert \theta_3\uvect{0} - \vartheta_2\uvect{\varphi_2} \rvert)
  \\&\quad 
  +\kappa_0^{-8}
  \bigl[\kappa_0^2 + \xi_{+}(\vartheta_1)\bigr]
  \bigl[\kappa_0^2 + \xi_{+}(\vartheta_2)\bigr]
  \frac{1}{2\pi \AFOV} \int_{0}^{\rFOV}\idiff[]{\theta_3} \,
  \theta_3  \xi_{+}(\theta_3) \,
  \int_{0}^{2\pi}\idiff[]{\varphi_1} \,
  \xi_{+}(\lvert \theta_3\uvect{0} - \vartheta_1\uvect{\varphi_1} \rvert)
  \int_{0}^{2\pi}\idiff[]{\varphi_2}\,
  \xi_{+}(\lvert \theta_3\uvect{0} - \vartheta_2\uvect{\varphi_2} \rvert)
  \\&\quad\qquad\times   
  \xi_{+}(\lvert \theta_3\uvect{0} - \vartheta_1\uvect{\varphi_1} - \vartheta_2\uvect{\varphi_2} \rvert)
.
\end{split}
\end{equation}
The expressions for $c_{+-}^{(\gamma)}$ and $c_{--}^{(\gamma)}$ can then be computed by combining Eqs.~\eqref{eq:c_pmpm_gamma_general} and \eqref{eq:c_pp_gamma_log_normal}.

The expression \eqref{eq:c_pp_gamma_log_normal} for a zero-mean shifted log-normal convergence field is considerably more complicated than the corresponding expression \eqref{eq:c_pp_gamma_normal} for a normal convergence field. If one only takes into account those terms in Eq.~\eqref{eq:c_pp_gamma_log_normal} that are also present in the case of a normal convergence field, and the simplest term $\propto\kappa_0^{-2}$, one obtains
\begin{equation}
\begin{split}
c^{(\gamma)}_{\pm\pm}\left(\vartheta_1, \vartheta_2 \right)
  &=
  \frac{4}{\pi\AFOV}
  \int_{0}^{\rFOV}\idiff{\theta_3}\, \theta_3
  \int_{0}^{\pi}\idiff{\varphi_1}\,
  \zeta_\pm\bigl(\lvert \vartheta_1 \uvect{\varphi_1}, \theta_3 \uvect{0} \rvert\bigr)
  \int_{0}^{\pi}\idiff{\varphi_2}\, 
  \zeta_\pm\bigl(\lvert \vartheta_2 \uvect{\varphi_2}, \theta_3 \uvect{0} \rvert\bigr)
  \,+\,
  \xi_{\pm}(\vartheta_1) \xi_{\pm}(\vartheta_2)\frac{8 \pi}{\kappa_0^{2}\AFOV}
  \int_{0}^{\rFOV}\idiff[]{\theta_3}\,\theta_3\, \xi_{+}(\theta_3)
.
\end{split}
\end{equation}

\subsection{Shear correlation covariances for surveys comprised of several fields}
\label{sec:appendix_shear_correlation_estimators_covariance_for_other_survey_geometries}

So far, we have discussed the case of a survey with a single rectangular field of view. If instead the survey consists of several independent fields of the same shape and size, and with the same galaxy density, the covariances for the whole survey can be obtained by (i) computing the covariances for a single field, and (ii) dividing the result by the number of fields.

The expressions derived here for the cosmic shear covariance depend on the survey geometry via the survey field area $\AFOV$. Its inverse $\AFOV^{-1}$ acts as an overall factor. Furthermore, the linear dimension $\rFOV$ of the survey field enters as upper limit on integrals of correlation functions. Thus, the covariances for a survey do not only depend on the total survey area, but also on the shapes, sizes, and relative separations of the individual fields comprising the survey \citep[in contrast to similar expressions derived by][]{SchneiderEtal2002, JoachimiSchneiderEifler2008}. For example, the covariances for a survey with one $4\,\degt^2$ field usually differ from the covariances for a survey with 4 independent fields of $1\,\degt^2$.

\section{Computing shear correlations with Fast Fourier Transform methods}
\label{sec:appendix_shear_correlation_fourier_methods}

Our ray-tracing simulations provide us with the convergence and shear at positions forming a square mesh. Thus, Fast Fourier Transform (FFT) methods to compute their correlations are the obvious choice. Such FFT methods are so simple and efficient that they may be the best choice even for shear data sampled at irregular positions in fields with gaps and otherwise complicated geometry (as in real shear surveys), where projection of the data onto a regular mesh is required as an intermediate step.

As an illustrative example, we first revise how to use FFT methods to compute the two-point correlation of the projected galaxy density. Assume, the density contrast $\deltagal(\vtheta) =\bigl(\ngal(\vtheta) - \meanngal\bigr)/\meanngal$, with $\ngal$ denoting the projected galaxy density and $\meanngal$ denoting the mean projected galaxy density, is known on a subset of points $\vtheta_i$ of a rectangular non-periodic mesh $(\vtheta_i)$, e.g. from a counts-in-cells. Some of the mesh points are allowed to have no data. To avoid aliasing, the mesh is first extended to a mesh twice as large as the original mesh in each dimension by adding mesh points around the original mesh. Let $\Npixel$ denote the number of mesh points of the padded mesh.

An array $\vect{w}=(w_1,\ldots,w_{\Npixel})$ with a layout commensurate with the padded mesh geometry is used to store the window function $w$, which holds the information on which array entries contain valid density data and what their statistical weights (e.g. due to varying completeness) are. For example, 
\begin{equation}
  w_i = w(\vtheta_i) =
  \begin{cases}
    1 &  \text{if data for $\vtheta_i$ is valid, and} \\
    0 & \text{otherwise}
  \end{cases}
\end{equation}
for uniform weights.
Furthermore, the weighted density contrast data  
\begin{equation}
  (w\deltagal)_i = w(\vtheta_i) \deltagal(\vtheta_i)
\end{equation} 
on the mesh points $\vtheta_i$ of the padded mesh are stored in an array $\vect{w\deltagal}$ with the same geometry as the window array $\vect{w}$.

An FFT algorithm \citep[such as, e.g., provided by FFTW, ][]{FrigoJohnson2005_FFTW3} is used to compute the discrete Fourier transform (DFT) $\vect{\ft{w}}$ of the window array:\footnote{The actual prefactor depends on the particular FFT implementation used.}
\begin{equation}
 \ft{w}_i = \ft{w}(\vect{k}_i) = \frac{1}{N_{\text{pix}}} \sum_{p} \ee^{\ii \vect{k}_i \cdot \vtheta_p} w(\vtheta_p).
\end{equation}
The squared modulus of resulting array elements is then computed, 
\begin{equation}
\ft{\bar{S}}_{\vect{w},\vect{w},i} =
\ft{\bar{S}}_{\vect{w},\vect{w}} (\vect{k}_i) = \ft{w}(\vect{k}_i)^* \ft{w}(\vect{k}_i)
= \frac{1}{\Npixel^2} \sum_{p,q} \ee^{\ii \vect{k}_i \cdot (\vtheta_p - \vtheta_q)} w(\vtheta_p) w(\vtheta_q)
,
\end{equation}
and transformed back to configuration space to obtain the spatial auto-correlation $\bar{S}_{\vect{w},\vect{w}}$ of the window function array:
\begin{equation}
\bar{S}_{\vect{w},\vect{w},i}
=
\bar{S}_{\vect{w}, \vect{w}} (\vtheta_i) 
= 
  \frac{1}{\Npixel^2}\sum_{p,q,r} \ee^{\ii \vect{k}_r \cdot (\vtheta_p - \vtheta_q -  \vtheta_i)} w(\vtheta_p) w(\vtheta_q)
=
  \frac{1}{\Npixel}\sum_{p,q} w(\vtheta_p) w(\vtheta_q)
  \delta(\vtheta_i, \vtheta_p - \vtheta_q) 
. 
\end{equation}
Here, 
\begin{equation}
 \delta(\vtheta, \vtheta')   =
  \begin{cases}
    1 &  \text{if } \vtheta = \vtheta' \text{, and} \\
    0 & \text{otherwise.}
  \end{cases}
\end{equation}
The auto-correlation $\bar{S}_{\vect{w\deltagal}, \vect{w\deltagal}}$ of the weighted density contrast is computed in the same way with FFTs.
Finally, the correlations are binned into radial bins to obtain the text-book counts-in-cells Landy-\&-Szalay estimator for the angular two-point correlation $\xi_{\text{gal}}$ of galaxies:
\begin{equation}
\est{\xi}_{\text{gal}}(\vartheta) = 
  \frac{\sum_i \Delta\bigl(\vartheta, \lvert \vtheta_i\rvert \bigr) \bar{S}_{\vect{w\deltagal}, \vect{w\deltagal}, i}}
       {\sum_i \Delta\bigl(\vartheta, \lvert \vtheta_i\rvert \bigr) \bar{S}_{\vect{w}, \vect{w}, i}}
=
  \frac{\sum_{i,j} \Delta\bigl(\vartheta, \lvert  \vtheta_j - \vtheta_i \rvert \bigr) w(\vtheta_i) w(\vtheta_j) \deltagal(\vtheta_i) \deltagal(\vtheta_j)}
       {\sum_{i,j} \Delta\bigl(\vartheta, \lvert  \vtheta_j - \vtheta_i \rvert \bigr) w(\vtheta_i) w(\vtheta_j)}
,
\end{equation}
where $\Delta\bigl(\vartheta, \theta \bigr)$ denotes the bin function \eqref{eq:bin_function}.

With the same approach, one can calculate estimates for convergence and shear correlations. One creates arrays $\vect{w}$, $\vect{w\kappa}$, $\vect{w\gamma_1}$, and $\vect{w\gamma_2}$ storing the values of the window function and the weighted convergence and shear components on the points $\vtheta_i$ of the padded mesh.
One then uses FFTs to compute the  auto-correlation $\bar{S}_{\vect{w}, \vect{w}}$ of the window function, the auto-correlation $\bar{S}_{\vect{w\kappa}, \vect{w\kappa}}$ of the weighted convergence, and the auto- and cross-correlations $\bar{S}_{\vect{w\gamma_p}, \vect{w\gamma_q}}$  of the weighted shear components.
Finally, the correlations are binned into radial bins to estimate the isotropic correlations:
\begin{align}
  \est{\xi}_\kappa(\vartheta) &= 
    \frac{\sum_i \Delta\bigl(\vartheta, \lvert \vtheta_i\rvert \bigr) \bar{S}_{\vect{w\kappa}, \vect{w\kappa}, i}}
         {\sum_i \Delta\bigl(\vartheta, \lvert \vtheta_i\rvert \bigr) \bar{S}_{\vect{w}, \vect{w}, i}}
    =
    \frac{\sum_{ij} \Delta\bigl(\vartheta, \lvert \vtheta_j - \vtheta_i \rvert \bigr) w(\vtheta_i) w(\vtheta_j) \kappa(\vtheta_i) \kappa(\vtheta_j)}
         {\sum_{ij} \Delta\bigl(\vartheta, \lvert \vtheta_j - \vtheta_i \rvert \bigr) w(\vtheta_i) w(\vtheta_j)}
,\\
  \est{\xi}_+(\vartheta) &= 
    \frac{\sum_i \Delta\bigl(\vartheta, \lvert \vtheta_i\rvert \bigr) \left[\bar{S}_{\vect{w\gamma_1}, \vect{w\gamma_1}, i} + \bar{S}_{\vect{w\gamma_2}, \vect{w\gamma_2}, i} \right]}
         {\sum_i \Delta\bigl(\vartheta, \lvert \vtheta_i\rvert \bigr) \bar{S}_{\vect{w}, \vect{w}, i}}
    \\\nonumber& =
    \frac{\sum_{ij} \Delta\bigl(\vartheta, \lvert \vtheta_j - \vtheta_i \rvert \bigr) w(\vtheta_i) w(\vtheta_j)
    \left[ \gammat(\vtheta_i, \vtheta_j - \vtheta_i) \gammat(\vtheta_j, \vtheta_j - \vtheta_i) + \gammax(\vtheta_i, \vtheta_j - \vtheta_i) \gammax(\vtheta_j, \vtheta_j - \vtheta_i) \right]}
         {\sum_{ij} \Delta\bigl(\vartheta, \lvert \vtheta_j - \vtheta_i \rvert \bigr) w(\vtheta_i) w(\vtheta_j)}
,\\
  \est{\xi}_-(\vartheta) &= 
    \frac{\sum_i \Delta\bigl(\vartheta, \lvert \vtheta_i\rvert \bigr) 
    \left\{
      \cos\bigl[4 \varphi(\vtheta_i)\bigr] \bar{S}_{\vect{w\gamma_1}, \vect{w\gamma_1}, i}
    - \cos\bigl[4 \varphi(\vtheta_i)\bigr] \bar{S}_{\vect{w\gamma_2}, \vect{w\gamma_2}, i}
    +2\sin\bigl[4 \varphi(\vtheta_i)\bigr] \bar{S}_{\vect{w\gamma_1}, \vect{w\gamma_2}, i}
    \right\}}
         {\sum_i \Delta\bigl(\vartheta, \lvert \vtheta_i\rvert \bigr) \bar{S}_{\vect{w}, \vect{w}}(\vtheta_i)}
    \\\nonumber& =
    \frac{\sum_{ij} \Delta\bigl(\vartheta, \lvert \vtheta_j - \vtheta_i \rvert \bigr) w(\vtheta_i) w(\vtheta_j)
    \left[ \gammat(\vtheta_i, \vtheta_j - \vtheta_i) \gammat(\vtheta_j, \vtheta_j - \vtheta_i) - \gammax(\vtheta_i, \vtheta_j - \vtheta_i) \gammax(\vtheta_j, \vtheta_j - \vtheta_i) \right]}
         {\sum_{ij} \Delta\bigl(\vartheta, \lvert \vtheta_j - \vtheta_i \rvert \bigr) w(\vtheta_i) w(\vtheta_j)}
.
\end{align}

FFT methods can also be used to compute galaxy-galaxy lensing statistics. For example, an estimator for the average tangential shear profile $\bEV{\gammat}(\vartheta)$ around a sample of lens galaxies as measured in a sample of source galaxies reads:
\begin{equation}
   \bEV{\est{\gammat}}(\vartheta) = -
    \frac{\sum_i \Delta\bigl(\vartheta, \lvert \vtheta_i\rvert \bigr)\left\{
      \cos\bigl[2 \varphi(\vtheta_i)\bigr] \bar{S}_{\vect{w_{\text{l}} n_{\text{l}}}, \vect{w_{\text{s}} n_{\text{s}} \gamma_1}, i}
    + \sin\bigl[2 \varphi(\vtheta_i)\bigr] \bar{S}_{\vect{w_{\text{l}} n_{\text{l}}}, \vect{w_{\text{s}} n_{\text{s}} \gamma_2}, i}
    \right\}}
         {\sum_i \Delta\bigl(\vartheta, \lvert \vtheta_i\rvert \bigr) \bar{S}_{\vect{w_{\text{l}} n_{\text{l}}}, \vect{w_{\text{s}} n_{\text{s}}}, i}}
,
\end{equation} 
where $w_{\text{l}}$, $n_{\text{l}}$, $w_{\text{s}}$, and $n_{\text{s}}$ denote the weights and densities of the lens and source galaxy samples.

\end{document}